\documentclass[preprint,preprintnumbers,amssymb,amsmath, superscriptaddress,letterpaper]{revtex4}[10pt]

\usepackage{graphicx}
\usepackage{dcolumn}
\usepackage{bm}
\usepackage{pstricks}

\newcommand{\be}{\begin{equation}}
\newcommand{\ee}{\end{equation}}
\newcommand{\bea}{\begin{eqnarray}}
\newcommand{\eea}{\end{eqnarray}}

\newcommand{\gsim}{ \mathop{}_{\textstyle \sim}^{\textstyle >} }
\newcommand{\lsim}{ \mathop{}_{\textstyle \sim}^{\textstyle <}}

\newcommand{\s}{{\rm ~s}}

\newcommand{\cm}{{\rm ~cm}}
\newcommand{\cmcubps}{{\rm ~cm^3/s}}

\newcommand{\GHz}{{\rm ~GHz}}

\newcommand{\microG}{\mu{\rm G}}
\newcommand{\eV}{{\rm ~eV}}

\newcommand{\GeV}{{\rm ~GeV}}
\newcommand{\gev}{{\rm ~GeV}}

\newcommand{\degree}{^\circ}

\newcommand{\kpc}{{\rm ~kpc}}

\newcommand{\sigmav}{\langle\sigma_Av\rangle}

\newcommand{\epp}{e^+e^-}

\begin{document}

\title{The Case for a 700+ GeV WIMP: Cosmic Ray Spectra from PAMELA, Fermi and ATIC}

\author{Ilias Cholis}
\affiliation{Center for Cosmology and Particle Physics, Department of Physics, New York University, 
New York, NY 10003}

\author{Gregory Dobler}
\affiliation{Harvard-Smithsonian Center for Astrophysics, 60 Garden St., Cambridge, MA 02138}

\author{Douglas P. Finkbeiner}
\affiliation{Harvard-Smithsonian Center for Astrophysics, 60 Garden St., Cambridge, MA 02138}

\author{Lisa Goodenough}
\affiliation{Center for Cosmology and Particle Physics, Department of Physics, New York University, 
New York, NY 10003}

\author{Neal Weiner}
\affiliation{Center for Cosmology and Particle Physics, Department of Physics, New York University, 
New York, NY 10003}

\date{\today}

\begin{abstract}
Multiple lines of evidence indicate an anomalous injection of
high-energy $\epp$ in the Galactic halo.  The recent $e^+$ fraction spectrum from
the Payload for Antimatter Matter Exploration and Light-nuclei
Astrophysics (PAMELA) shows a sharp rise up to 100 GeV.  The Fermi Gamma-ray Space Telescope has found a significant hardening of the $e^+e^-$ cosmic ray spectrum above 100 GeV, with a break, confirmed by HESS at around 1 TeV. The Advanced Thin Ionization
Calorimeter (ATIC) has also detected detected a similar excess, falling back to the expected 
spectrum at 1 TeV and above.  Excess
microwaves towards the galactic center in the WMAP data are consistent
with hard synchrotron radiation from a population of 10-100 GeV $\epp$
(the WMAP ``Haze'').  We argue that dark matter annihilations can
provide a consistent explanation of all of these data, focusing on
dominantly leptonic modes, either directly or through a new light
boson.  Normalizing the signal to the highest energy evidence (Fermi and HESS),
we find that similar cross sections provide good fits to PAMELA and
the Haze, and that both the required cross section and annihilation
modes are achievable in models with Sommerfeld-enhanced annihilation.
These models naturally predict significant production of gamma rays in
the galactic center via a variety of mechanisms. Most notably, there
is a robust inverse-Compton scattered (ICS) gamma-ray signal arising
from the energetic electrons and positrons, detectable at Fermi/GLAST
energies, which should provide smoking gun evidence for this
production.
\end{abstract}


\maketitle

\section{Introduction}
The search for astrophysical signatures of WIMP annihilation has
blossomed in recent months with tantalizing announcements of electron
and positron cosmic-ray spectra by three projects.  Unexpected features
in such spectra are of great interest for particle physics, because of
the possibility that WIMP annihilation products could be observed this
way.

The first project, ATIC, has measured the spectrum of $e^++e^-$ (ATIC
cannot distinguish $e^+$ from $e^-$) from 20 to 2000 GeV, finding a hard spectrum ($\sim E^{-3}$) concluding with a
broad bump at $300-800$ GeV \cite{ATIClatest}.  ATIC discriminates
protons from electrons by comparing the shapes of the hadronic
versus electromagnetic showers within its bismuth germanate (BGO)
calorimeter.  The electron energy is measured two ways: by the total
amount of energy deposited in the calorimeter (usually about 85\% of
the incoming electron energy) and by the position of the shower maximum.
Both techniques give consistent results \footnote{see Figure s8 in the Supplement
to \cite{ATIClatest}}, finding about 210 events at $300-800$ GeV,
compared to the 140 expected.  This result confirms similar excesses
seen previously by ATIC \cite{ATIC2005} and PPB-BETS
\cite{Torii:2008xu}.  Though suggestive, this alone would not be taken
as strong evidence for WIMP annihilation, because there are
astrophysical ways to accelerate electrons (SN shocks, pulsars, etc.).
Dark matter annihilation would produce equal numbers of $e^+$ and
$e^-$, so it is crucial to determine what fraction of the ATIC excess
is positrons.

The second project, PAMELA \cite{Picozza:2006nm}, \emph{can}
distinguish between $e^+$ and $e^-$ and finds a sharp rise in the flux
ratio $\phi(e^+)/(\phi(e^+)+\phi(e^-))$ above 10 GeV, continuing up to
$\sim 80\, \gev$ \cite{Adriani:2008zr}.  This confirms previous indications of
an excess by HEAT \cite{Barwick:1997ig} and AMS-01 \cite{ams}, but at
much higher confidence.  Although lower energies ($\lsim 10\, \gev$)
are significantly affected by solar modulation, these high energy
particles are not.  The ordinary secondary positron spectrum,
originating from cosmic ray (CR) interactions with the interstellar gas, is softer
than the primary electron spectrum, so the positron fraction was
expected to drop with energy.  The surprising upturn in the positron
fraction, therefore, is very strong evidence for a primary source of
positrons.  While astrophysical sources such as pulsars offer
alternative explanations \cite{pulsars,2001A&A...368.1063Z,pulsars3,Hooper:2008kg,
Yuksel:2008rf}, dark matter remains a compelling explanation, especially, as we shall discuss, in light of the evidence for excess electronic production in the galactic center as well.

Finally, the Fermi Gamma-ray Space Telescope has recently published its spectrum of the sum of electrons and positrons \cite{Abdo:2009zk}. While they also see a significant hardening of the spectrum, they do not see a broad peak. Rather, they see a smooth spectrum with a break at $\sim$ TeV, which is also seen by HESS \cite{Collaboration:2008aaa,Aharonian:2009ah}. Although the feature of ATIC is absent, the Fermi results are generally taken to support the PAMELA excess, and give a lower bound on the dark matter mass scale at around 1 TeV.

\begin{figure*}
\begin{center}
a) \includegraphics[width=.45\textwidth]{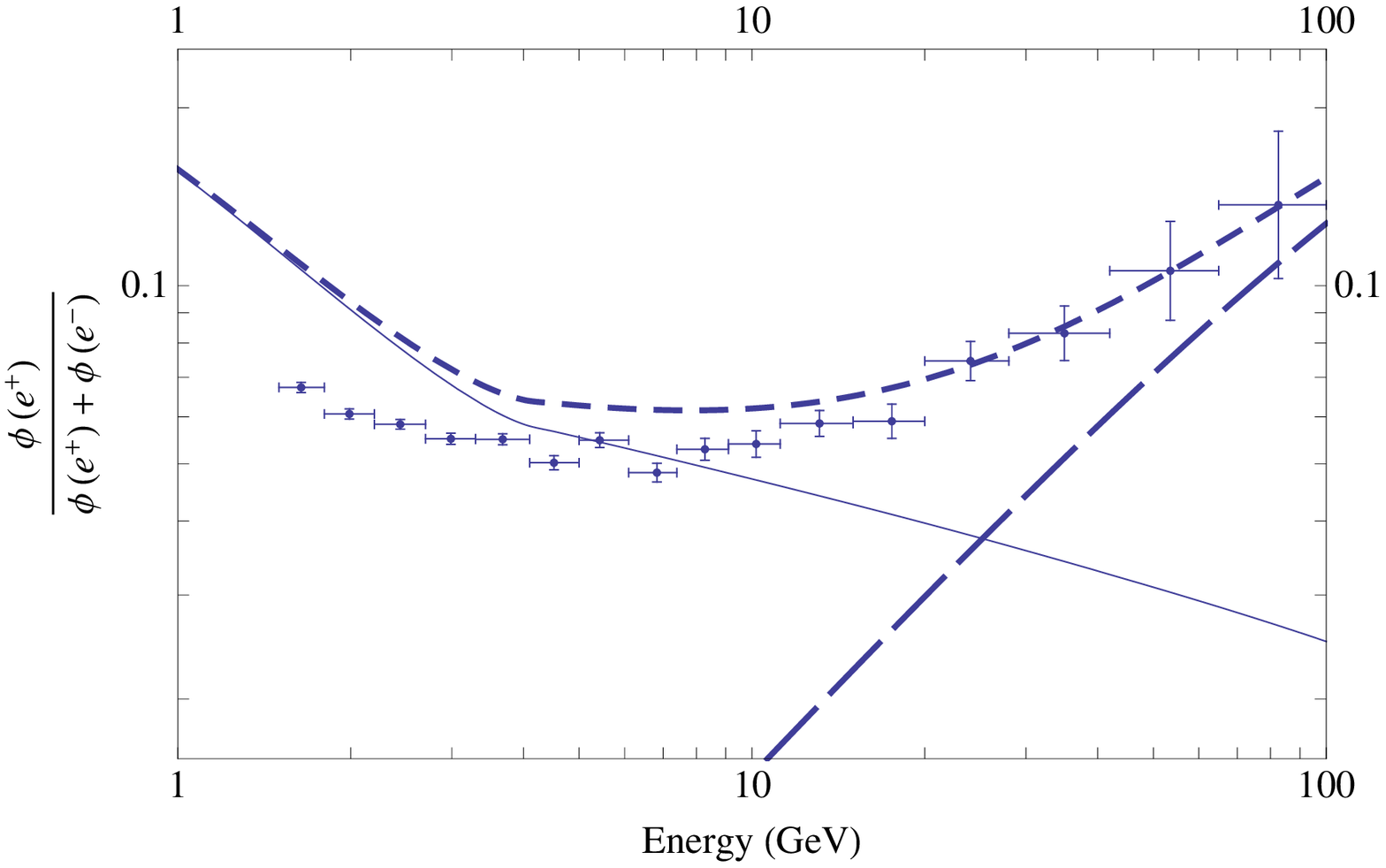}\hskip 0.2in
b) \includegraphics[width=.46\textwidth]{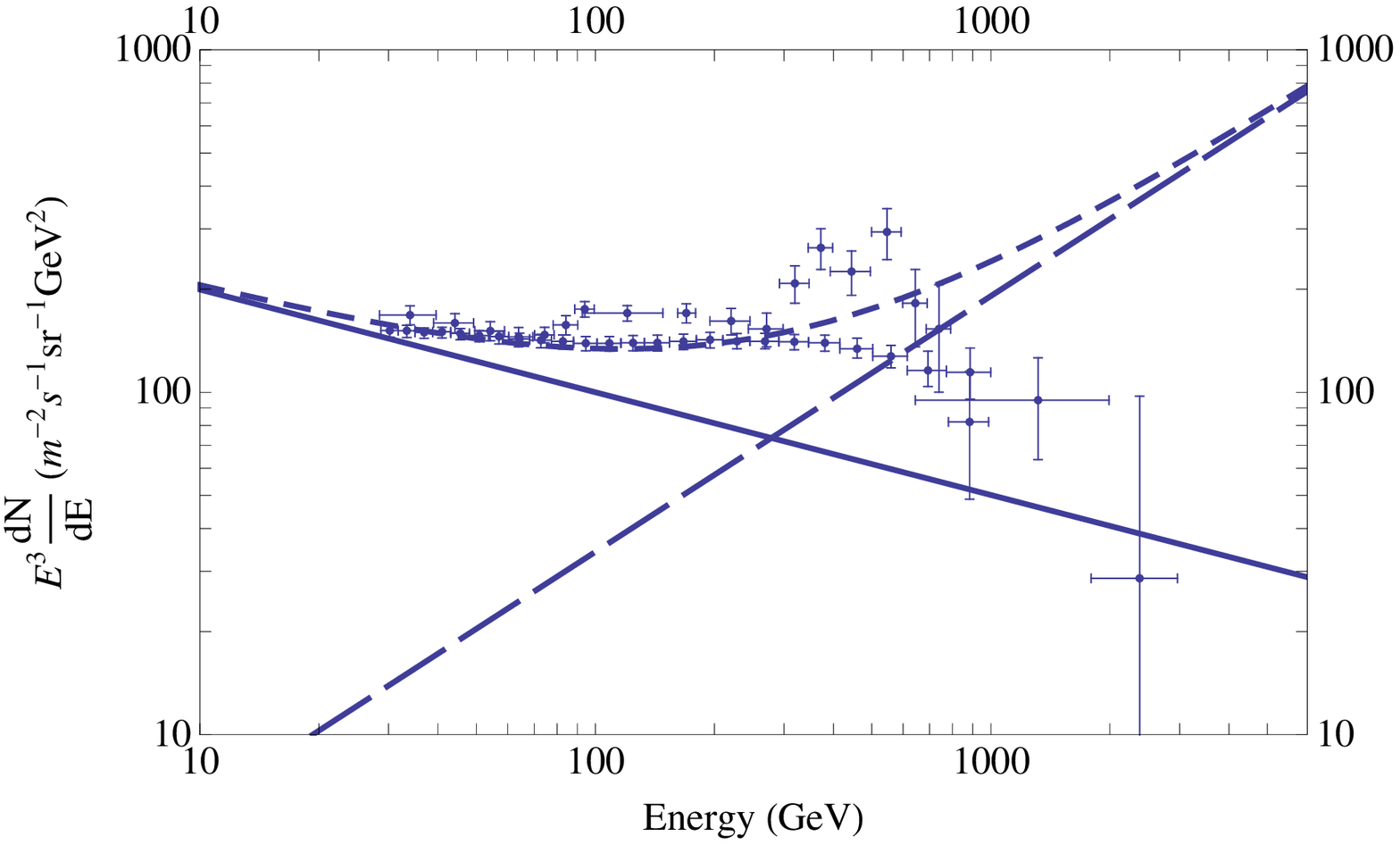}
\end{center}
\caption{(a) A new power-law component of $\epp$ fit to the PAMELA excess (\emph{long-dashed}) with expected background $e^+$ model (\emph{solid}) and total positron fraction (\emph{short-dashed}).  (b) Extrapolation of this new component to higher energies, assuming equal parts $e^+$ and $e^-$.  For reference, the new component positrons are about 12\% of the PAMELA fraction at 100 GeV, and therefore constitute about 1/4 of the Fermi flux $\phi(e^+)+\phi(e^-)$ at 100 GeV.  The fact that the high energy electron+positron and PAMELA excesses are connected by such a generic argument suggests that a single mechanism, producing equal numbers of $e^+$ and $e^-$, explains both.}
\label{fig:pamextrap}
\end{figure*}

Before proceeding, it is instructive to make a simple empirical
connection between PAMELA and Fermi, ATIC and PPB-BETS. For a casual observer of
the PAMELA data, it seems clear that for energies between 10 and 80
GeV, the rise is well described by a power law added to the expected
background spectrum.  If we assume that
this new component provides equal numbers of electrons and positrons,
it constitutes $\sim 25\%$ of the total electronic activity at 100 GeV
(i.e. positrons are $\sim 12\%$ of the total).  A natural exercise is
to fit this rise by assuming some new component of $\epp$ which
scales as a power of energy. One can extrapolate this power law to
higher energies and see where it makes an order unity contribution to
electronic activity (i.e., where the positron fraction is $\gsim 30\%$).
The extrapolation of PAMELA (Figure \ref{fig:pamextrap})
naturally yields an excess in the range of Fermi, ATIC and PPB-BETS similar
to the one observed. While this argument is heuristic, it nonetheless
makes a strong case that these signals arise from the same source,
rather than \emph{two} new, independent sources of high energy $\epp$
coincidentally within a factor of a few in energy of each other.

This apparent connection between the high energy electron and PAMELA excesses
is intriguing, but does not advance the argument that they arise from
DM annihilation; it merely suggests that they arise from the
\emph{same} source and that source produces equal numbers of $e^+$
and $e^-$.  In order to distinguish electrons and positrons coming from DM annihilations from those coming from conventional astrophysical
sources, such as pulsars, we must look in regions where DM annihilation would produce
strong signals relative to pulsars. 
The most obvious place for this is in the galactic center (GC) where
pulsars are expected to have a relatively uniform abundance
\cite{1990ApJ...348..485P,1996ApJ...461..872S,2001A&A...368.1063Z}
while the DM density is expected to rise significantly.  Since the
annihilation power goes as the square of the WIMP density, any
production of high energy $\epp$ in the GC would likely be
significantly enhanced, possibly by many orders of magnitude, for DM
but not for astrophysical sources.

Remarkably, there are already indications of excess high energy
electronic activity in the GC. Finkbeiner
\cite{Finkbeiner:2003im}, while studying the foreground contributions to the WMAP
1-year data, found an excess of microwave radiation termed the ``WMAP
Haze''.  Recently, Dobler \& Finkbeiner showed that the Haze persists
in the WMAP 3-year data and that it is spectrally consistent with hard
synchrotron emission \cite{Dobler:2007wv}.  It was first argued in
\cite{Finkbeiner:2004us} that this hard synchrotron radiation could
possibly be produced by high energy $\epp$ products from DM
annihilations.  Additional studies have confirmed that the Haze
\emph{can} in fact arise from DM annihilations
\cite{Hooper:2007kb,Cholis:2008vb}, and that it would be natural for
the same parameters that yield significant positron abundance in the
10-100 GeV range \cite{Cholis:2008vb,pierce}.  

Furthermore, previous measurements of \emph{gamma-rays} in the GC
\cite{egret} have indicated an excess above expected background levels
\cite{Strong:2004de} in the 10-100 GeV range.  While this could arise
from various sources, a very natural explanation would be from ICS of
starlight photons from the same high energy $\epp$ that generate the
Haze.

Building on the remarkable qualitative consistency of these data, we explore
possibilities for generating all of these signals from a single source of
DM annihilation.  The broad consistency of dark matter annihilation with all cosmic ray $e^+ e^-$ data has been studied in 
 \cite{Cirelli:2008pk}, where it was shown that the best fit points to the whole data set were in leptonic modes. Others have also noted the
connections between sources of hard positrons for PAMELA
\cite{Nelson:2008hj,Cholis:2008qq} and the higher energy
signals. More recently, the connection of ATIC and decaying dark matter was studied in \cite{Chen:2008qs}, while \cite{Hall:2008qu} considered the possibility of distinguishing DM and pulsar explanations with atmospheric Cherenkov telescopes. In terms of other experiments,  \cite{Cholis:2008vb} showed the consistency between positron excesses at HEAT (which is consistent with current PAMELA results) and the Haze.  In this paper we argue that
modes with dominant annihilation into leptons (or through a light
boson into leptons) naturally provide good fits to the high energy electron, PAMELA, and Haze
data, with similar cross sections for each signal, along with fits to the EGRET data that are within the EGRET uncertainties.
We shall see that a robust consequence of this class of annihilation modes is a large signal of ICS photons towards the GC.

\section{Dark Matter as a Source of High Energy Particles}

In this section we explore the possibility that the $\epp$,
$\gamma$-ray, and microwave signals mentioned in the previous section
arise from WIMP annihilation.  We examine the difficulty of obtaining
a high-enough cross section to leptons without violating anti-proton
and $\pi^0$ $\gamma$-ray constraints, and propose a model that
overcomes these difficulties.

Thermal dark matter (DM) naturally annihilates into high-energy cosmic
rays, providing a source of electrons both locally and in the GC. The
fact that WIMP DM naively predicts an excess of cosmic rays within a
few orders of magnitude of what is observed makes it a natural
candidate for understanding the various excesses, in spite of the challenges associated with achieving a signal as large as is seen.  Dark matter may annihilate through various channels, which can produce
$e^+e^-$ directly, or in decays of intermediate partners. The final particles may be detected
directly by satellites or balloon experiments near Earth, or
indirectly by their ICS and synchrotron signals (either locally, in
the GC, in DM subhalos, or as an extragalactic background).  These
astrophysical signals of DM can constrain the particle mass, density,
annihilation cross section, and annihilation channels.

Though this picture is appealing, attempts to explain the PAMELA excess through 
conventional annihilation channels (for instance to heavy flavors or gauge 
bosons) produce a softer positron spectrum than required
\cite{Cholis:2008hb,Cirelli:2008pk}. This problem is compounded by the absence of any 
anti-proton excess \cite{Adriani:2008zq}, since anti-protons would have been copiously produced from these annihilation modes. At the same time, direct annihilations to leptons are a model-building challenge as Majorana fermions have helicity-suppressed annihilations to these states, unless accompanied by a photon, which is $\alpha$-suppressed \cite{Bergstrom:2008gr}. This prompts consideration of vector \cite{Cheng:2002ej,Hooper:2004xn} or Dirac \cite{Harnik:2008uu} particles. Nonetheless, in terms of annihilations directly to standard model modes, it appears at this point that only $\chi \chi \rightarrow e^+e^-,\, \mu^+ \mu^-$, and to a lesser extent $\, \tau^+ \tau^-$, are in good agreement with the data.

An alternative mechanism was proposed by \cite{Cholis:2008vb} as the
means to produce hard leptons without anti-protons or excess $\pi^0$'s.
Motivated by the requirement of a light
$(\lsim \, \gev)$ boson $\phi$ in the XDM model of
\cite{Finkbeiner:2007kk} to explain a separate signal, they showed
that annihilations $\chi \chi \rightarrow \phi \phi$, followed by
$\phi \rightarrow e^+ e^-$ or $\phi \rightarrow \mu^+ \mu^-$ would
naturally provide a hard lepton spectrum, explaining the HEAT and
AMS-01 signals, and producing the Haze with similar parameters \footnote{See \cite{Pospelov:2007mp} for a subsequent discussion of what has been termed ``secluded'' WIMP models which includes XDM as a subset.}. These predictions provide a good fit to the PAMELA data
\cite{Cholis:2008qq}, and are a central element of a proposed unified
explanation of anomalies observed by PAMELA, ATIC, INTEGRAL, and EGRET
\cite{ArkaniHamed:2008qn,ArkaniHamed:2008qp}.  Such models also
naturally allow inelastic scattering, providing a possible explanation
for the DAMA annual modulation signal \cite{Chang:2008gd}.
The light mass scale of the $\phi$ forbids or 
suppresses production of anti-protons.    Furthermore, these 
light bosons can also help explain the amplitude of the signals by yielding the 
large cross section required to explain the cosmic-ray data via the Sommerfeld 
enhancement \cite{ArkaniHamed:2008qn} \footnote{The importance of this effect in 
the context of multi-TeV scale dark matter interacting via W and Z bosons was 
first discussed by \cite{Hisano:2003ec, Hisano:2004ds}, and more recently 
emphasized with regards to PAMELA in the context of ``minimal dark matter,'' 
with similar masses by \cite{Cirelli:2008id, cirelli}.} or capture into bound 
states of ``WIMPonioum'' \cite{Pospelov:2008jd}.

In this paper, we explore whether these annihilation channels provide a 
consistent picture that explains the entire set of data, including limits from 
the GC. We normalize our signal at the highest energies - using Fermi, ATIC and 
PPB-BETS - and proceed to lower energies, showing that similar parameters fit 
PAMELA and the Haze, while enforcing limits from EGRET. Rather than explore the model building questions, as in \cite{ArkaniHamed:2008qn}, we attempt to study the annihilation channels in a more model independent way. We extend the analysis of \cite{Cholis:2008vb, Cholis:2008qq} by considering other modes, considering the total electron spectrum and considering the associated gamma ray signatures as well. We see that 
annihilations to $e^+e^-$, $\mu^+\mu^-$, $\phi \phi$ with $\phi \rightarrow 
e^+e^-$ and $\phi \rightarrow \mu^+ \mu^-$ give good fits to the data. Modes with $\tau$'s generally give too broad a peak to explain ATIC, but are otherwise consistent with the electronic data.
We also show that a natural consequence of these models is a significant ICS signal towards the 
GC from the scattering of starlight photons by the $\epp$ arising from DM annihilation, 
which may be observable by Fermi/GLAST. In the case of annihilation directly to 
$e^+e^-$, the energy cutoff of this signal should give a clear indication of the 
WIMP mass.

\section{Annihilation modes and their cosmic ray signals}
In this section we compute the expected cosmic-ray and photon signals
for six primary annihilation modes. We consider $\chi \chi \rightarrow
e^+e^-$, $\mu^+ \mu^-$, and $\tau^+ \tau^-$, as well as $\chi \chi
\rightarrow \phi \phi$, followed by $\phi \rightarrow e^+e^-$, $\phi
\rightarrow \mu^+ \mu^-$, or $\phi \rightarrow \tau^+ \tau^-$, the final three being modes that can arise in XDM models.  For
each annihilation mode, we produce a four-panel figure showing the
local particle spectrum for high energy electrons/positrons and PAMELA, and the synchrotron and
gamma-ray signals for the galactic center (Figures
\ref{fig:electrons}-\ref{fig:xdmtaus}).  We then consider the limits
from diffuse gammas more generally, and estimate the astrophysical
uncertainties of our predictions.

\subsection{Analysis}
\label{sec:analysis}
Galactic cosmic ray propagation is a decades-old problem in
astrophysics.  Propagation parameters (diffusion and energy-loss
coefficients, as well as boundary conditions) are constrained by
isotope ratios of CR nuclei, measured as a function of energy.  For
example, the ratio of boron to carbon provides a constraint, since
most carbon CRs were accelerated, but most boron CRs were made by
spallation.  Likewise, $^9$Be and $^{10}$Be are both made by
spallation, but the short half-life ($1.5\times10^6$ yr) of $^{10}$Be
serves as a clock, measuring the residence time of CRs in the galaxy.
Using these and other constraints, it is possible to constrain a
simple cylindrical model of the Milky Way galaxy to fit existing CR
data, while predicting the behavior of a new leptonic component
introduced by dark matter annihilation.  To do this, we use a slightly
modified version of GALPROP \cite{Moskalenko:1999sb} version 50.1p
\footnote{GALPROP and the resource files we employed are available at
http://galprop.stanford.edu; our modifications are available upon
request.}.

GALPROP computes the steady-state solution to the usual propagation
equations including diffusion, re-acceleration, nuclear interactions,
and energy loss to ICS, bremsstrahlung and synchrotron for electron
CRs; and hadronic collisions with interstellar gas for protons and
heavier nuclei.  Inputs to GALPROP include 3D estimates of the
starlight, far IR, and CMB energy density, gas density, magnetic field
energy density, diffusion parameters, and the primary source function
from ``conventional'' processes (e.g. SN shocks).  Outputs include 3D
density grids for each CR species; full-sky maps of gamma-rays due to
ICS, bremsstrahlung, and $\pi^0$ gammas from ISM interactions; and a
synchrotron map.

We run the code in 2D mode, solving the propagation equations on an
$(r,z)$ grid.  Note that this azimuthal symmetry is imposed to obtain
the steady-state CR distribution, but then skymaps are generated using
the (best-guess) 3D distributions of Galactic gas.  These assumptions
vastly oversimplify the Galaxy, and it is important to bear in mind
that the quantitative results derived from such a model are suspect at
the factor of 2 level (at least).

A recent addition to GALPROP is a feature allowing a DM profile to be
specified, and the $e^+e^-$ injected by DM annihilation to be tracked
independently of other primary and secondary $e^+e^-$.  Because the
results in the inner galaxy are so important for ICS and synchrotron
predictions, it is crucial to treat the central spatial bin of the
GALPROP grid carefully.  This pixel represents a cylinder of radius
$dr=0.1$ and height $dz=0.05$.  At a distance of $8.5\kpc$, $0.1\kpc$
corresponds to $0.67\degree$.  Version 50.1p sub-samples each spatial bin
and averages the DM density.  However, the annihilation power goes as
the square of the density, so it is important to compute
$\langle\rho^2\rangle$, not $\langle\rho\rangle^2$.  Furthermore, the
averaging code incorrectly samples the DM distribution near $r=0$, and
sometimes introduces large errors for divergent profiles.

We have fixed these problems by summing $d\Omega a^2\rho^2 da$ over
each annular ($r,z$) pixel, where $a$ is the 3D radial coordinate, and
$d\Omega$ is the solid angle subtended by the intersection of the
pixel with a sphere of radius $a$.  By handling this averaging
carefully down to scales of $dr/100$, we are able to obtain good
convergence properties for the $dr$ and $dz$ given above, even for
profiles as cuspy as $\rho\propto r^{-1.2}$. 

In this paper, we use the Einasto profile; following Merritt et
al. \cite{Merritt:2005} we assume a DM halo density profile of the
form
\be
\label{eq:einasto}
\rho = \rho_0 \exp\left[-\frac{2}{\alpha}
  \left(\frac{r^\alpha - R_\odot^\alpha}{r_{-2}^\alpha}\right)\right]\,,
\ee
with $\rho_0 = 0.3 \GeV \cm^{-3}$ as the DM density at
$r=R_\odot=8.5\kpc$, and $r_{-2}=25\kpc$.

For reference, the benchmark parameters used in our runs are the following.  The
cylindrical diffusion zone has a radius of $20\kpc$ and height
$\pm4\kpc$.  The diffusion constant, $K(E)=5.8\times10^{28} (E/4
\GeV)^\alpha \cm^2 \s^{-1}$ where $\alpha=0.33$, has the energy
dependence expected for a turbulent magnetic field with the Kolmogorov
spectrum.  The rms magnetic field strength $\langle B^2\rangle^{1/2}$ is modeled with an exponential disk, 
\be
\langle B^2\rangle^{1/2} = B_0 \exp(-(r-R_{\odot})/r_B - |z|/z_B)
\ee
where $r_B=10\kpc$ is the radial scale, $z_B=2\kpc$ is the vertical
scale, and $B_0 = 5.0\, \microG$ is the value locally.

\subsection{Results}
\begin{figure*}
\begin{center}
\includegraphics[width=.45\textwidth]{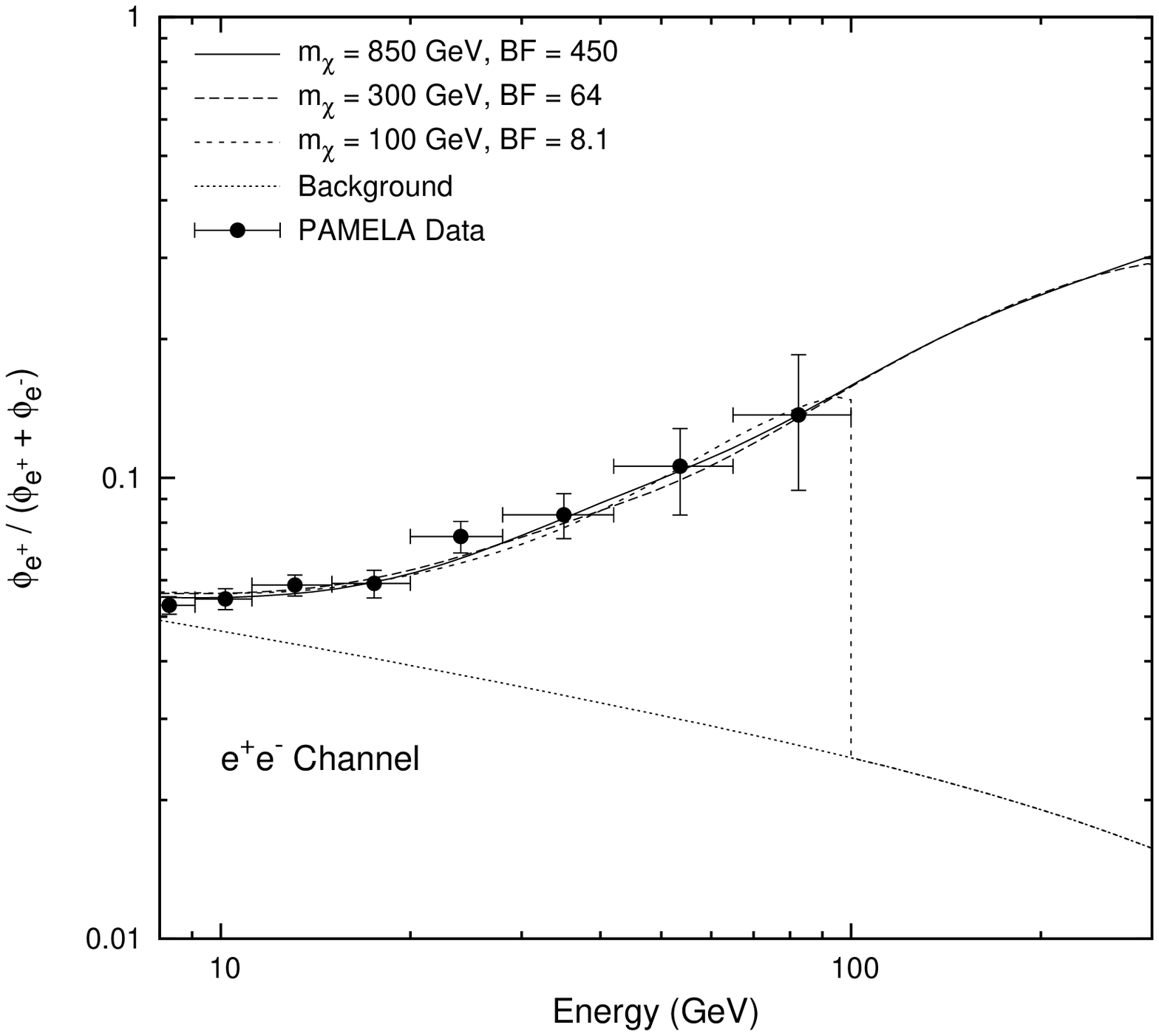}\hskip 0.2in
\includegraphics[width=.45\textwidth]{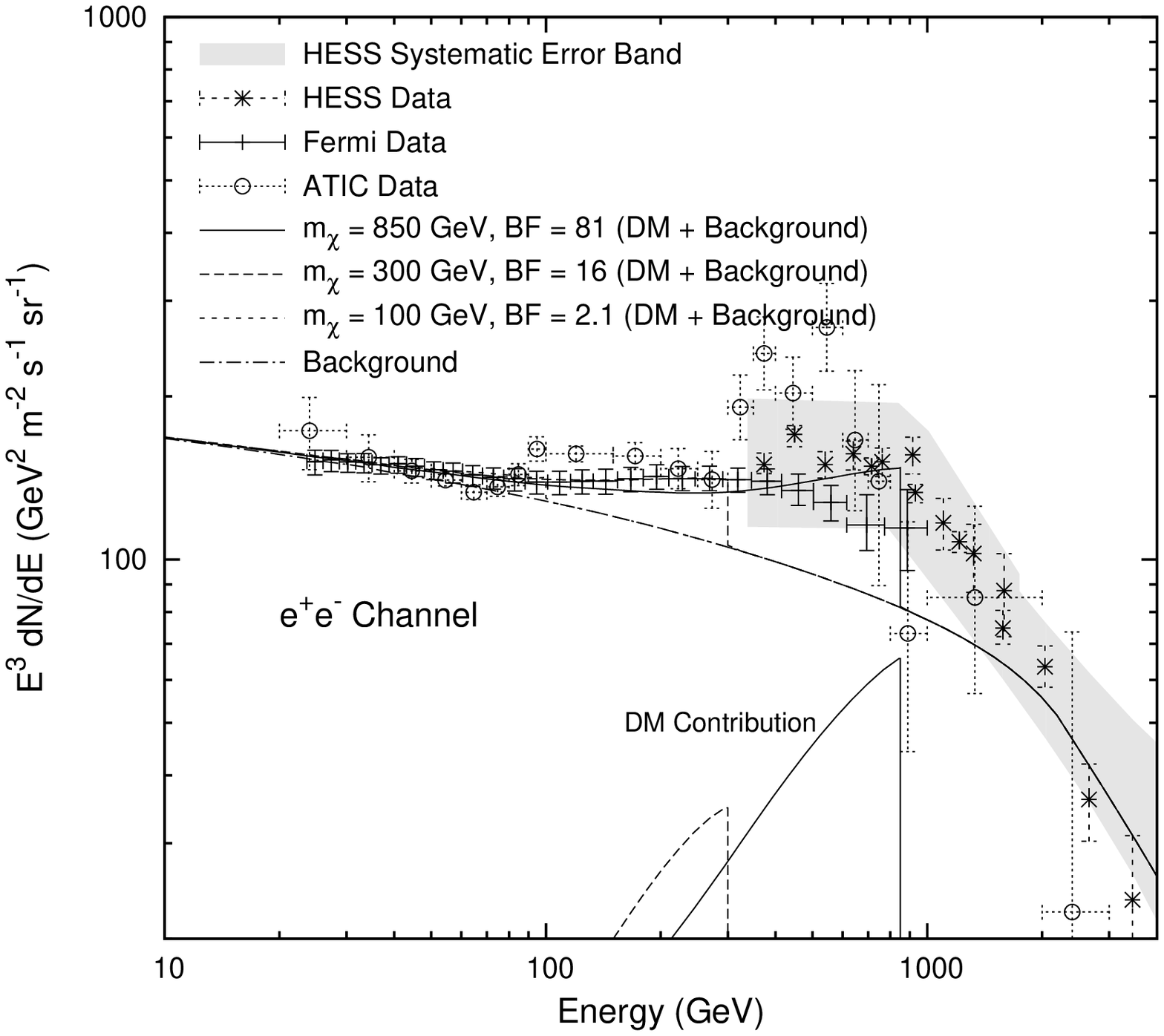}\\
\includegraphics[width=.45\textwidth]{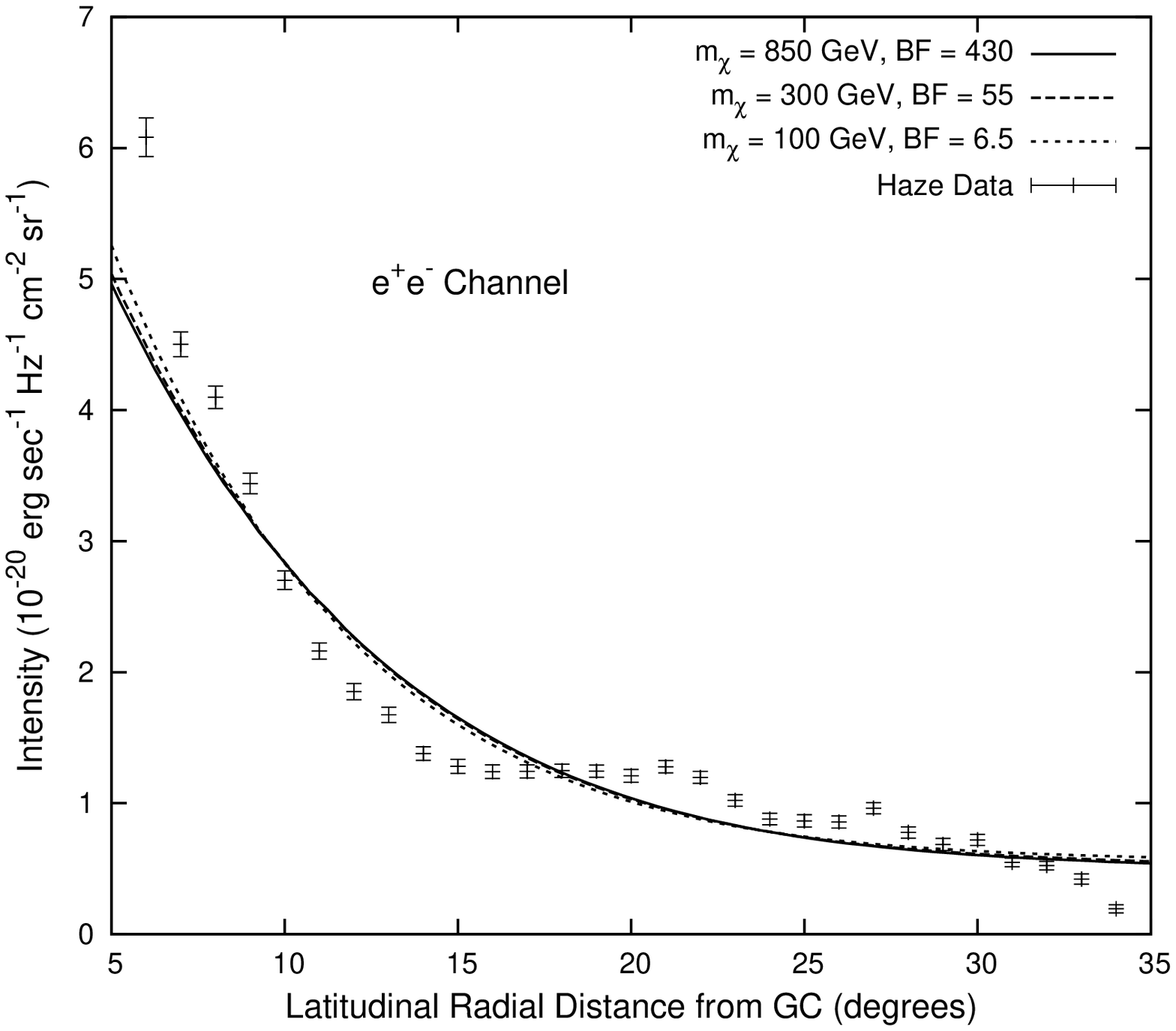}\hskip 0.2in
\includegraphics[width=.45\textwidth]{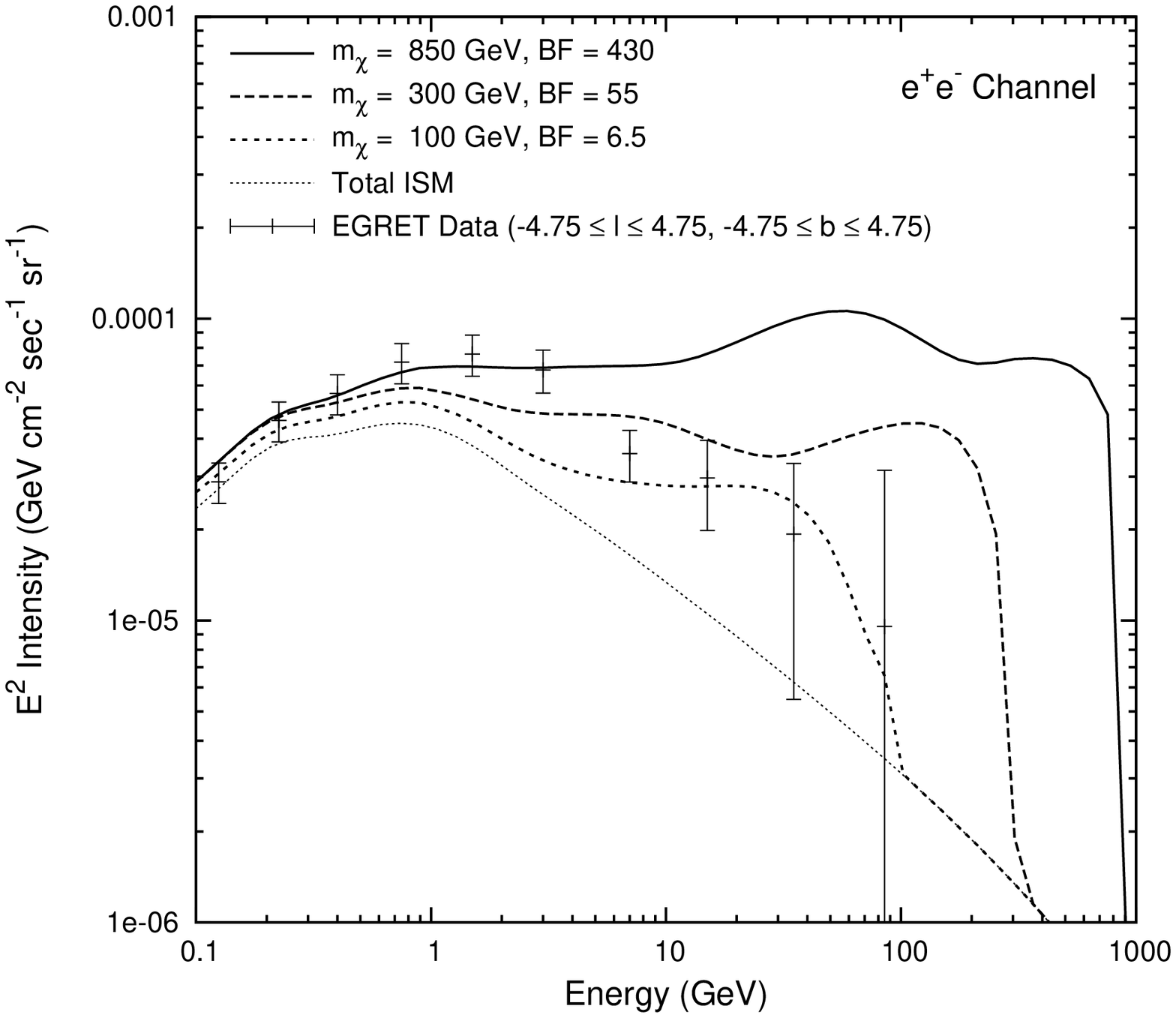}
\end{center}
\caption{The cosmic ray signals from dark matter annihilations $\chi \chi \rightarrow e^+ e^-$.
\emph{Upper left}: Predicted positron fraction vs.\ energy (\emph{solid and dashed
lines}), expected positron fraction vs.\ energy due to secondary production only (\emph{dotted}), and PAMELA
\cite{Adriani:2008zr} data points.  BF is the boost factor required
relative to $\sigmav = 3\times10^{-26}\cmcubps$ and the reference
local DM density of $\rho_0 = 0.3 \GeV \cm^{-3}$.
\emph{Upper right}: Spectrum of DM $e^+ e^-$ (\emph{solid and dashed}), background $e^+ e^-$ (\emph{dot-dashed}), and total (\emph{solid and dashed}) with data from Fermi \cite{Abdo:2009zk}, ATIC
\cite{ATIClatest} and PPB-BETS \cite{Torii:2008xu}.
\emph{Lower left}: Predicted WMAP Haze signal vs.\ galactic latitude at
23 GHz (\emph{solid and dashed}) and data points from WMAP \cite{Dobler:2007wv}.  Error bars are statistical only. 
\emph{Lower right}: Total diffuse gamma ray spectrum (\emph{solid and dashed}) and background diffuse gamma 
ray spectrum (\emph{dotted}) for the inner $5\degree$ of
the Milky Way, computed with GALPROP \cite{Strong:1999sv}.  Data points are from the Strong et
al. re-analysis of the EGRET data \cite{Strong:2005zx}, which found a
harder spectrum at $10-100$ GeV within a few degrees of the GC, using
improved sensitivity estimates from \cite{Thompson:2005}.  }
\label{fig:electrons}
\end{figure*}

\begin{figure*}
\begin{center}
\includegraphics[width=.45\textwidth]{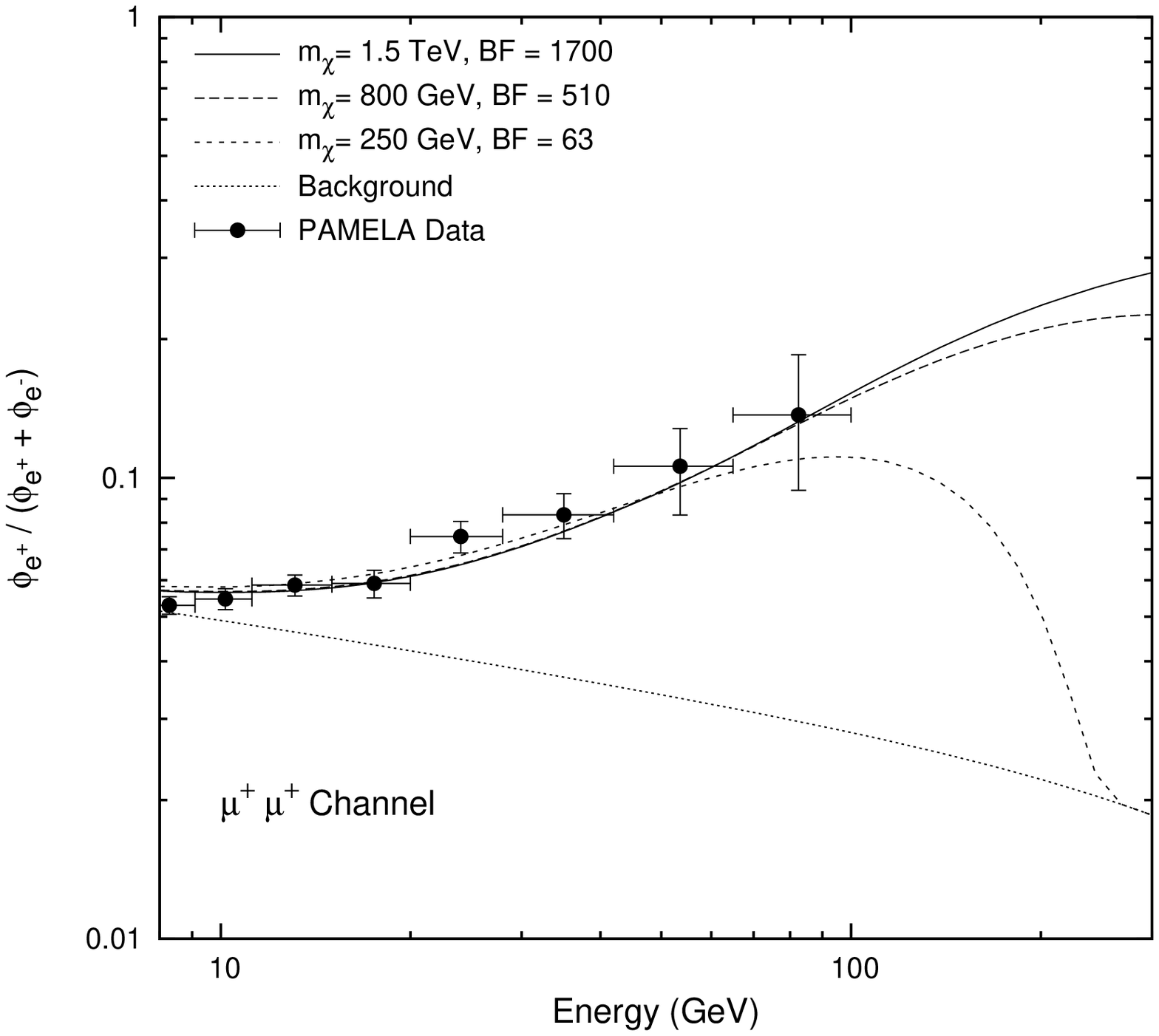}\hskip 0.2in
\includegraphics[width=.45\textwidth]{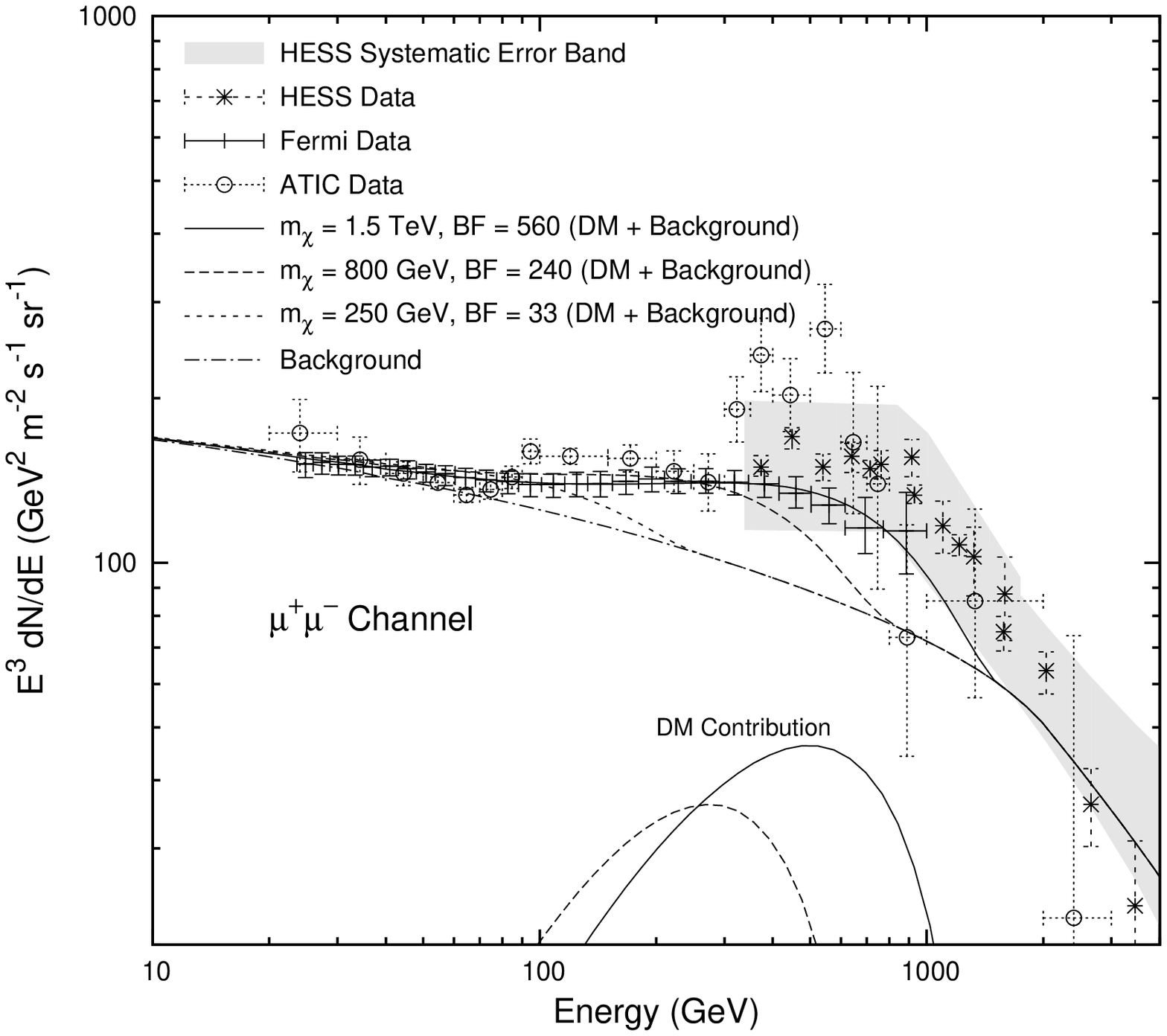}\\
\includegraphics[width=.45\textwidth]{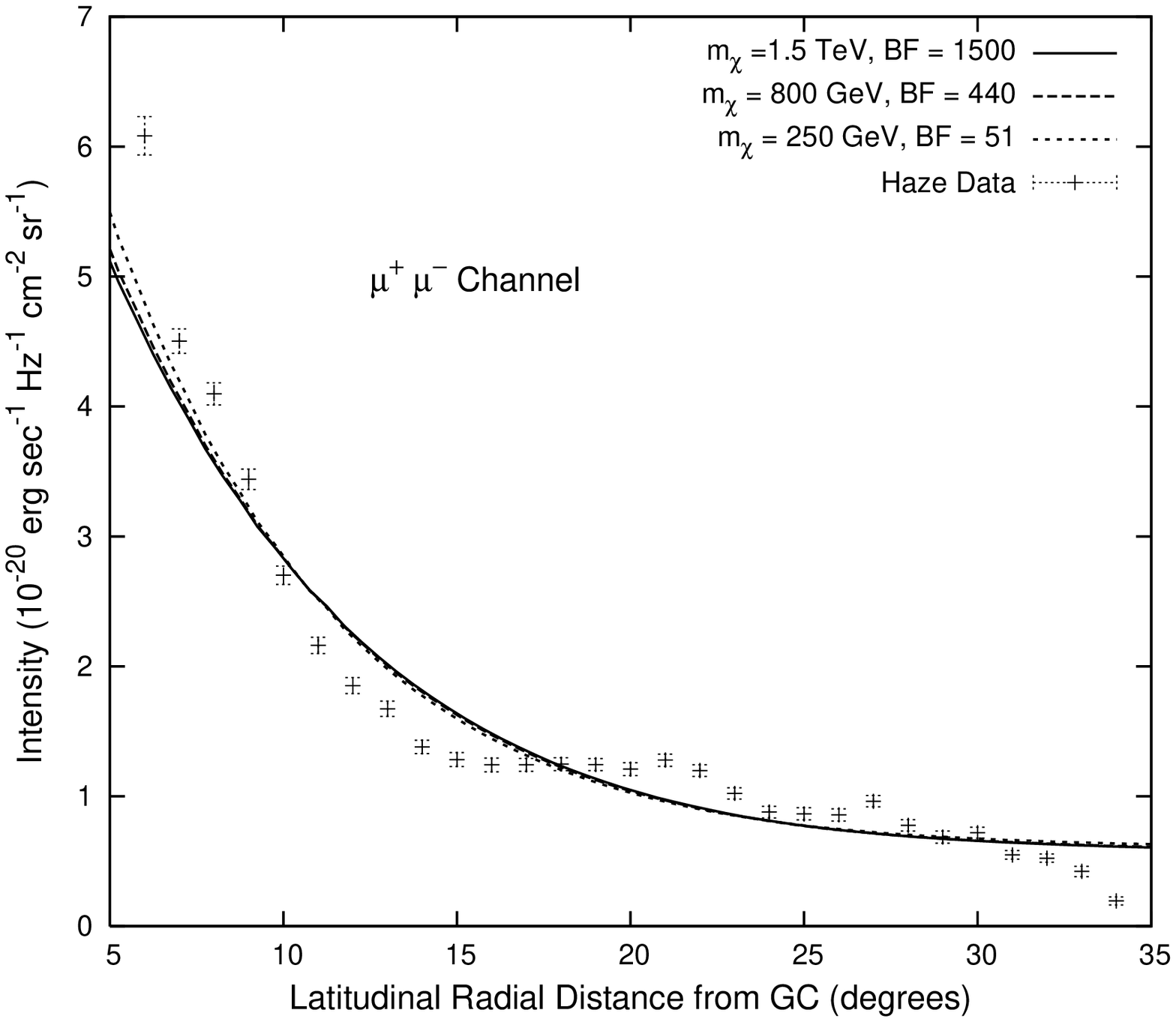}\hskip 0.2in
\includegraphics[width=.45\textwidth]{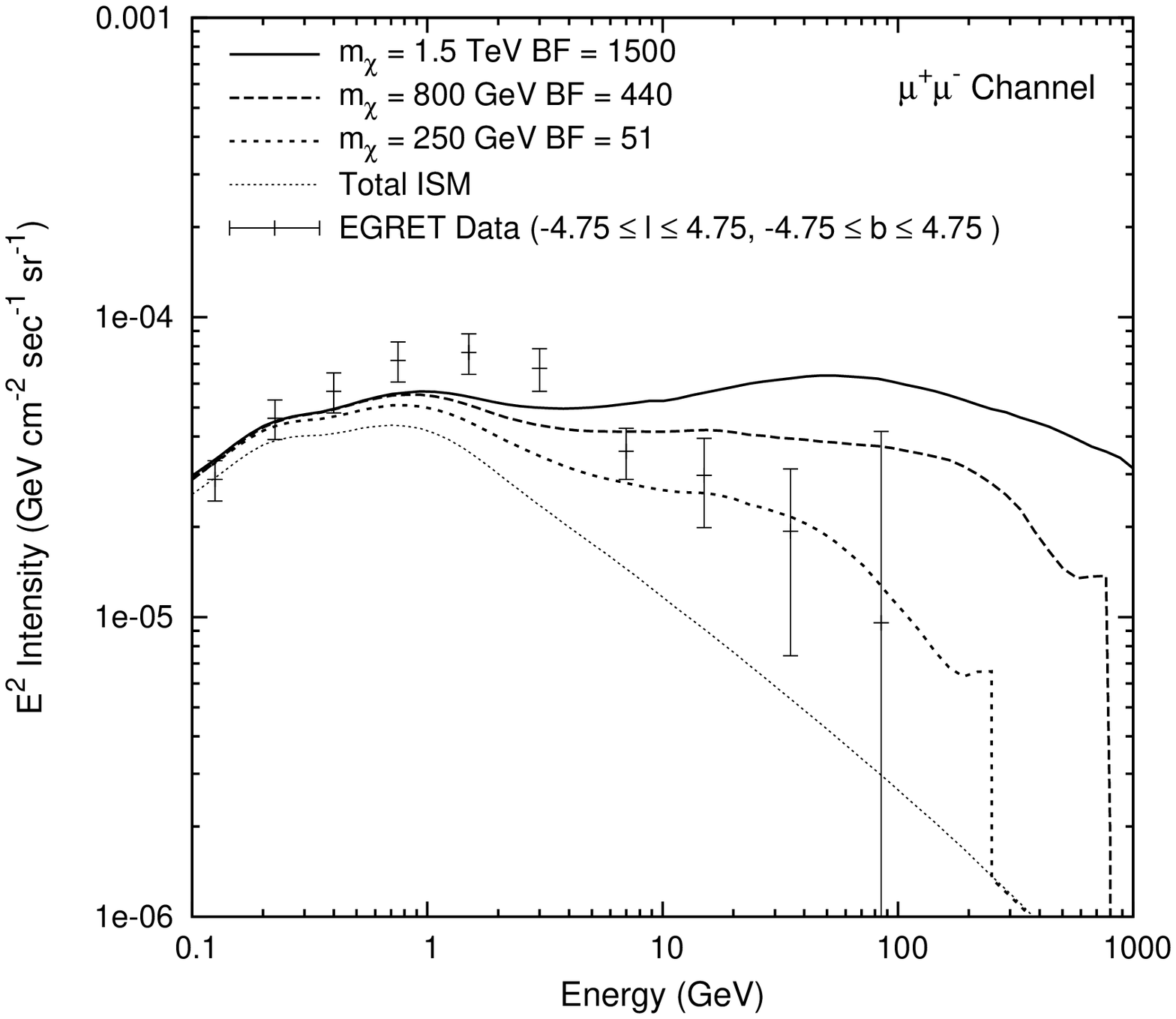}
\end{center}
\caption{The cosmic ray signals of dark matter annihilations as in
Figure \ref{fig:electrons}, but with $\chi \chi \rightarrow \mu^+
\mu^-$.}
\label{fig:muons}
\end{figure*}

\begin{figure*}
\begin{center}
\includegraphics[width=.45\textwidth]{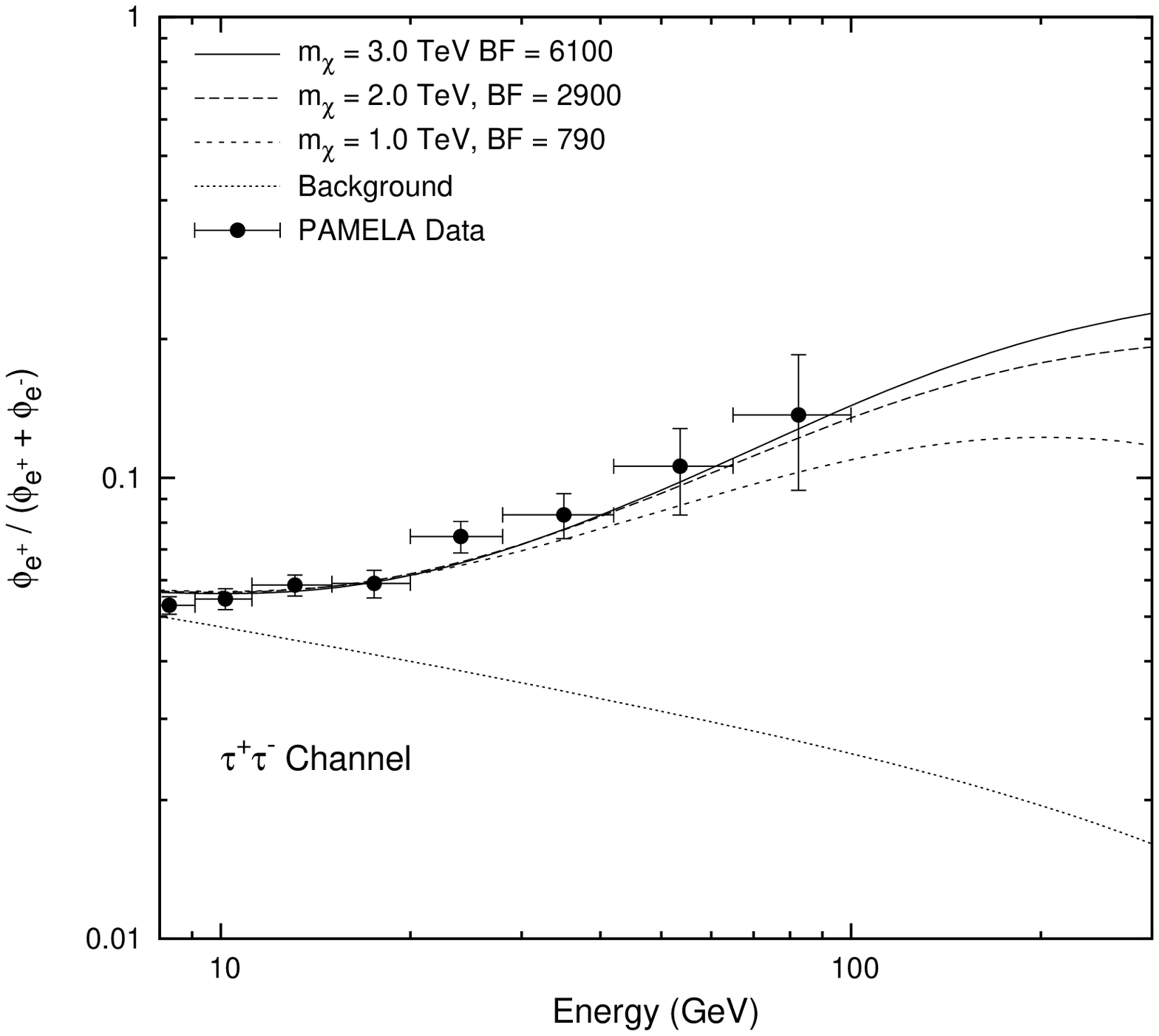}\hskip 0.2in
\includegraphics[width=.45\textwidth]{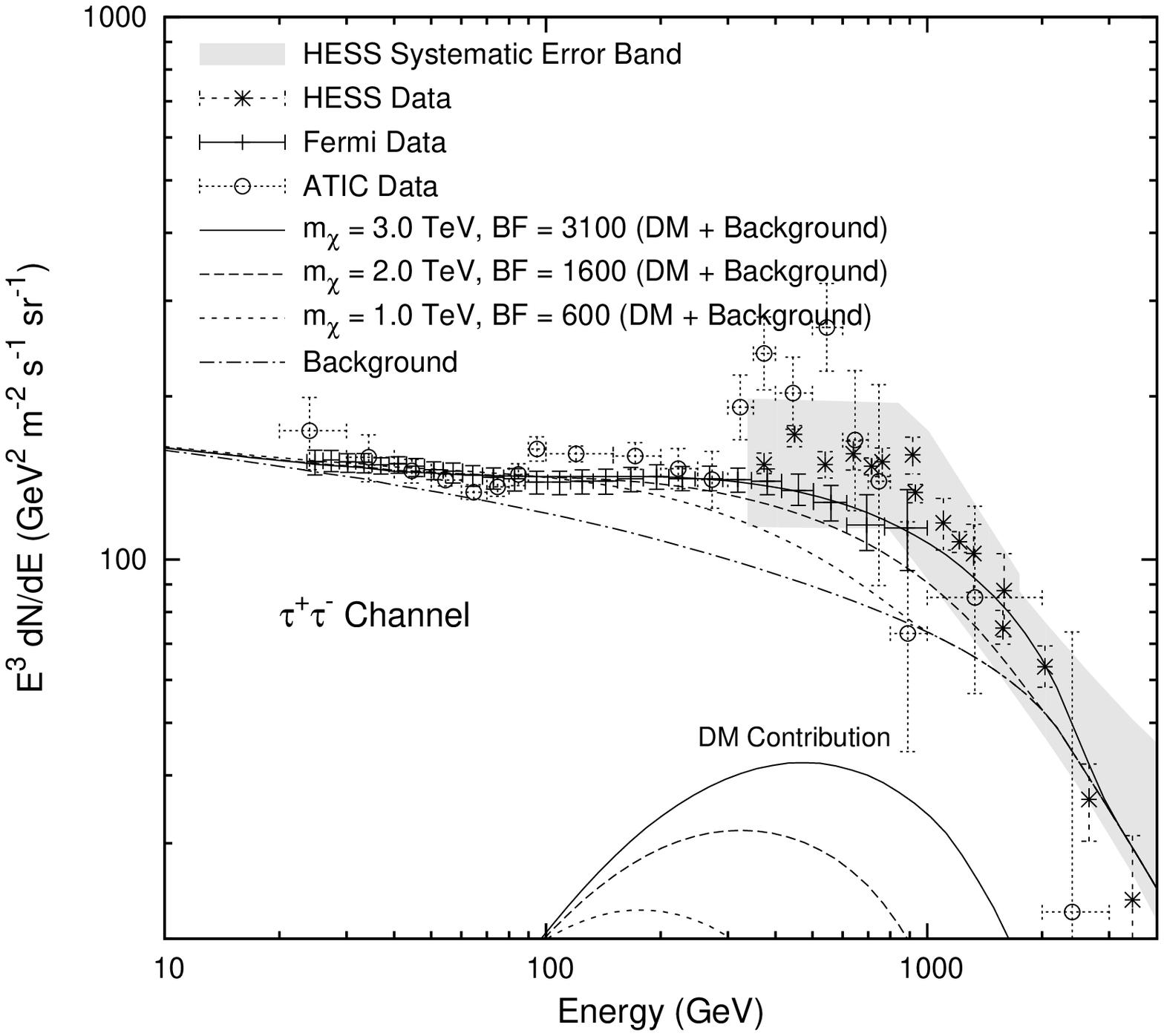}\\
\includegraphics[width=.45\textwidth]{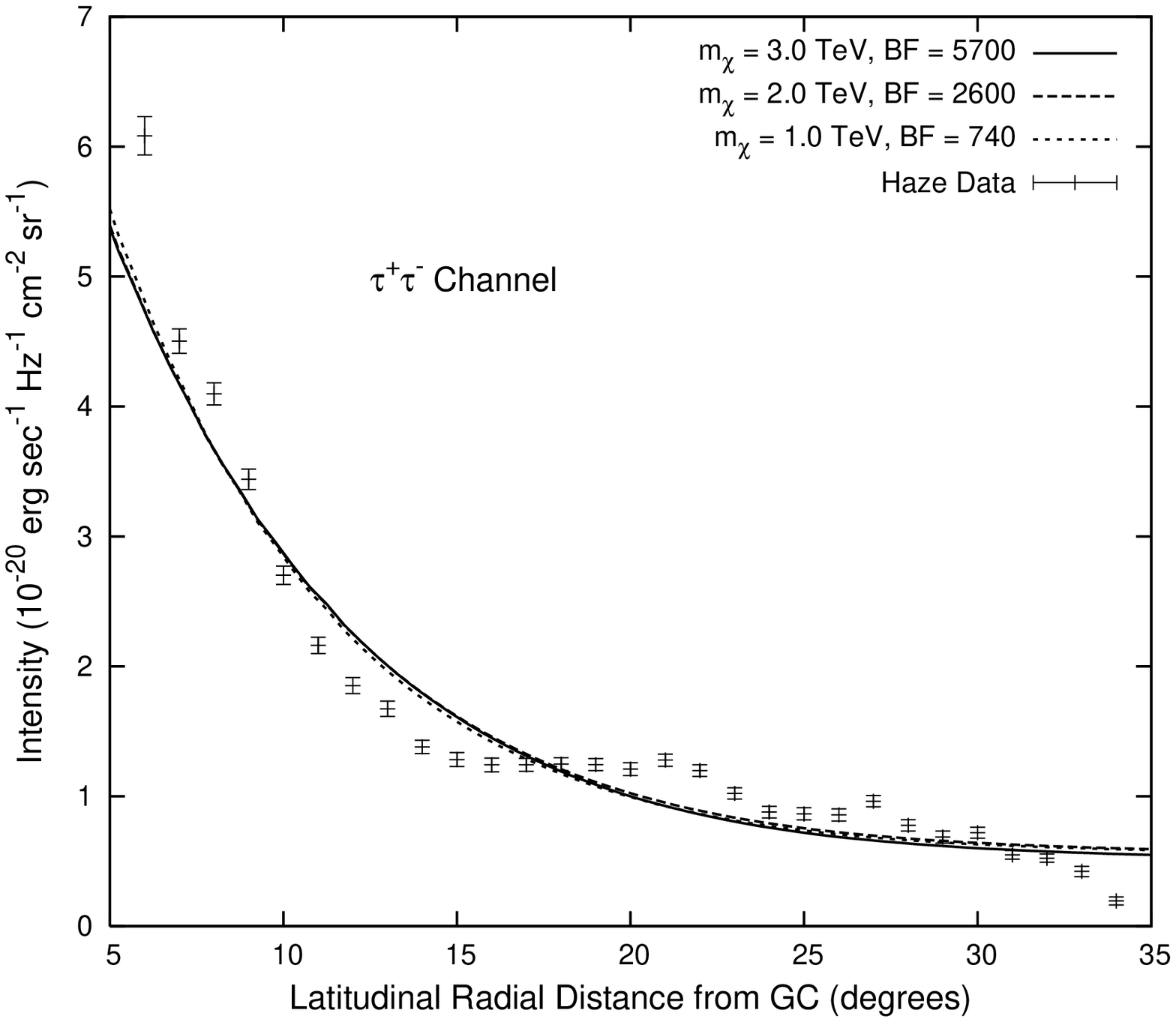}\hskip 0.2in
\includegraphics[width=.45\textwidth]{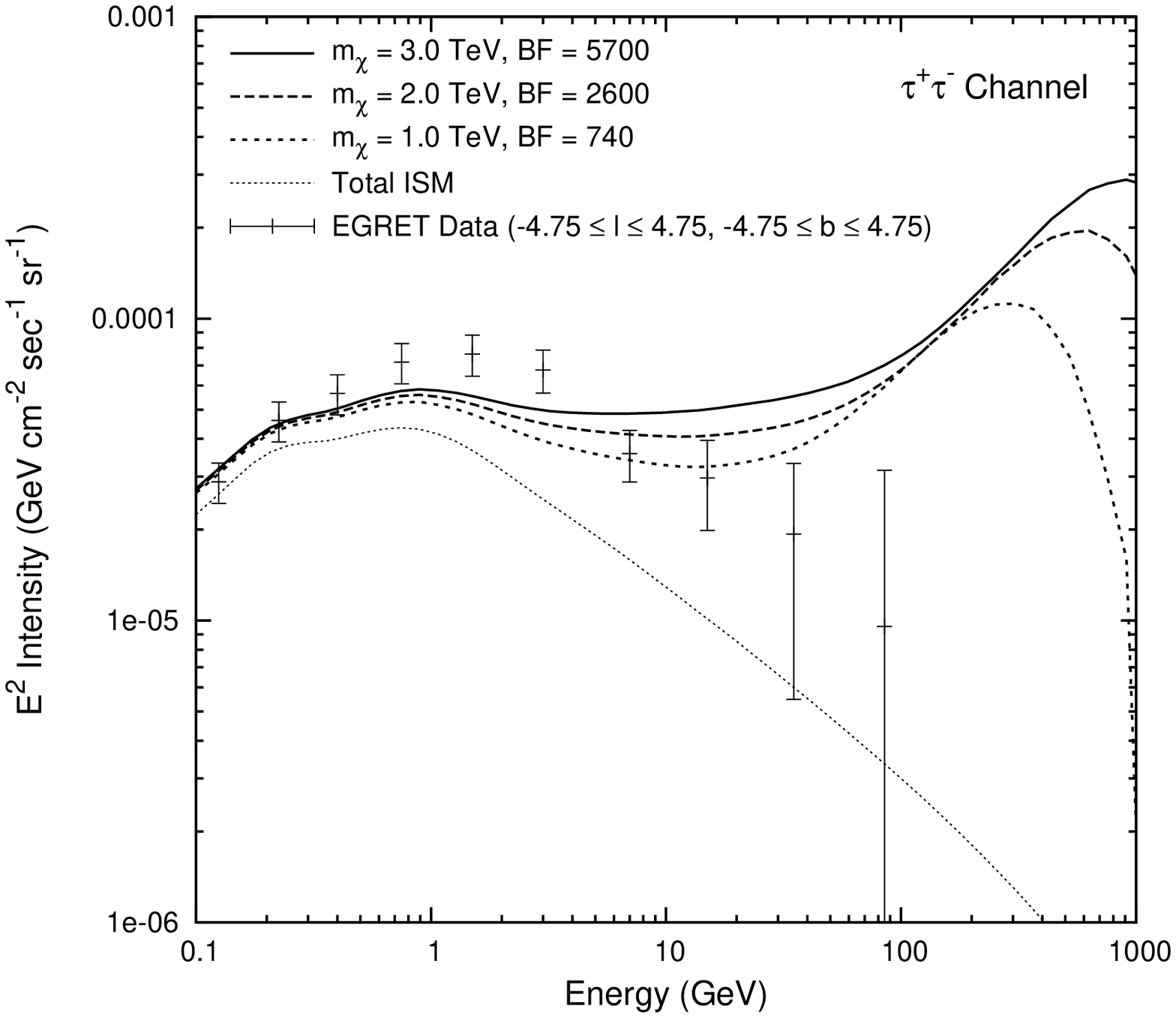}
\end{center}
\caption{The cosmic ray signals of dark matter annihilations $\chi \chi \rightarrow \tau^+ \tau^-$.}
\label{fig:taus}
\end{figure*}

\begin{figure*}
\begin{center}
\includegraphics[width=.45\textwidth]{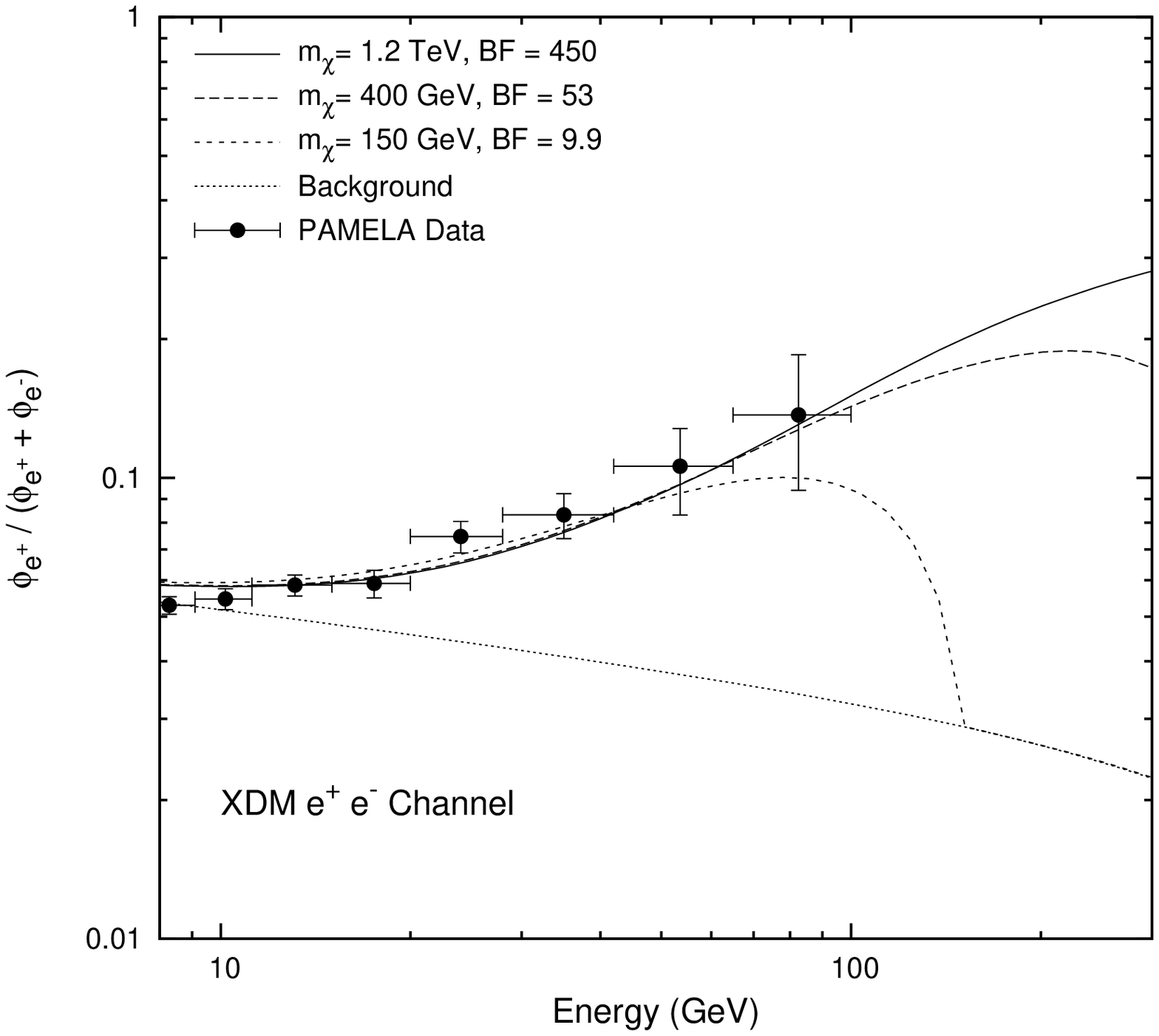}\hskip 0.2in
\includegraphics[width=.45\textwidth]{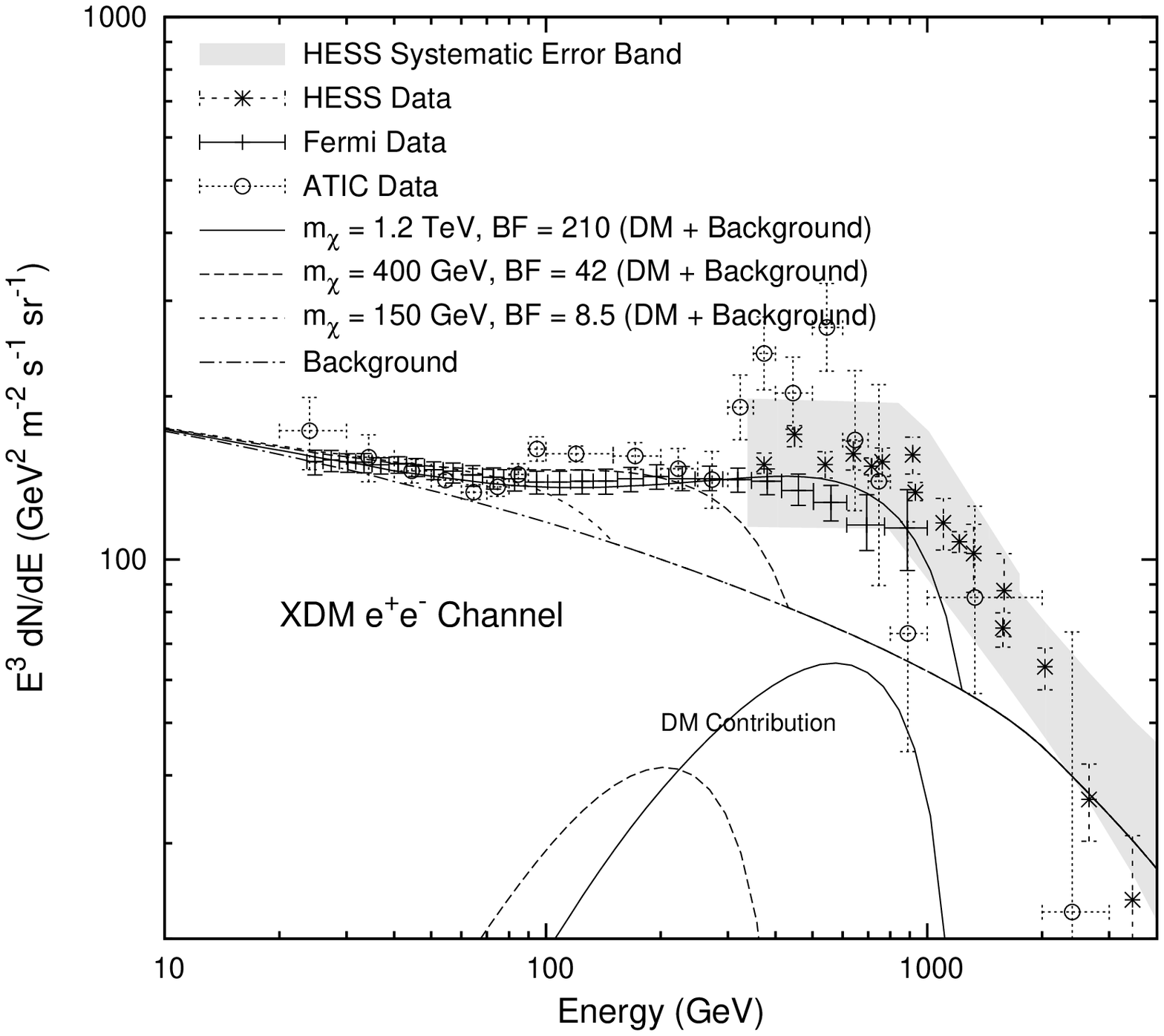}\\
\includegraphics[width=.45\textwidth]{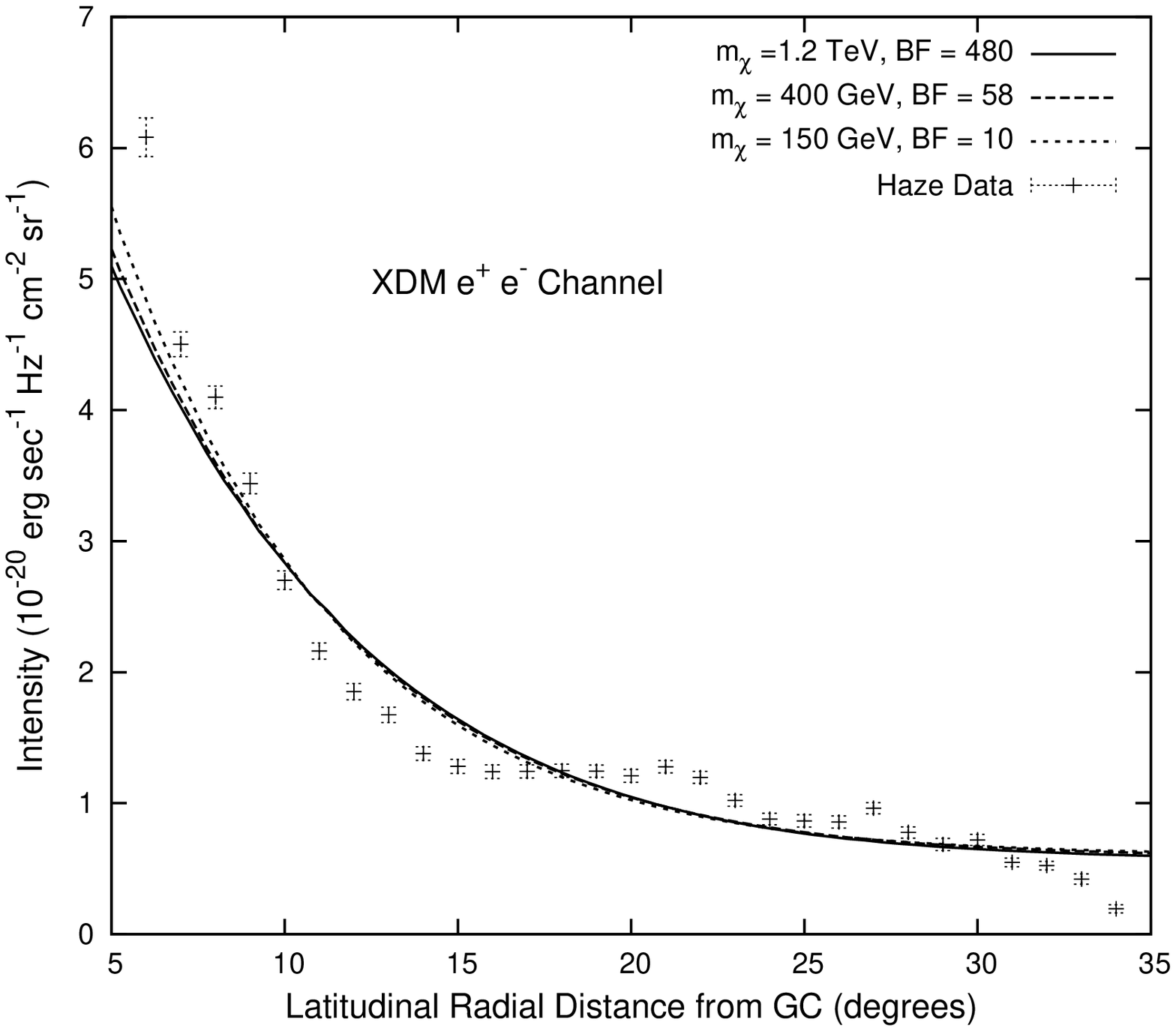}\hskip 0.2in
\includegraphics[width=.45\textwidth]{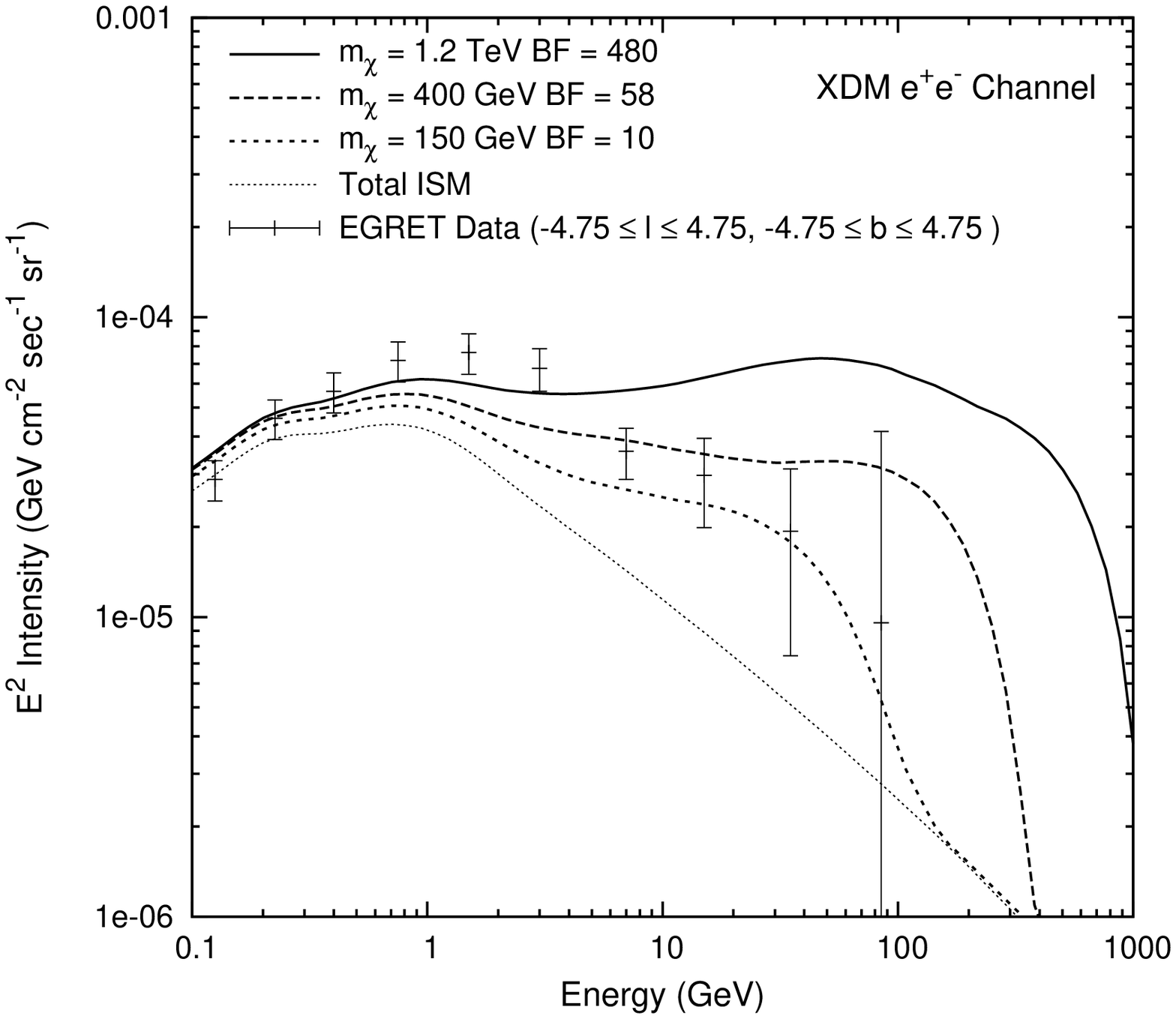}
\end{center}
\caption{The cosmic ray signals of dark matter annihilations $\chi \chi \rightarrow \phi \phi$, followed by $\phi \rightarrow e^+ e^-$.}
\label{fig:xdmelectrons}
\end{figure*}

\begin{figure*}
\begin{center}
\includegraphics[width=.45\textwidth]{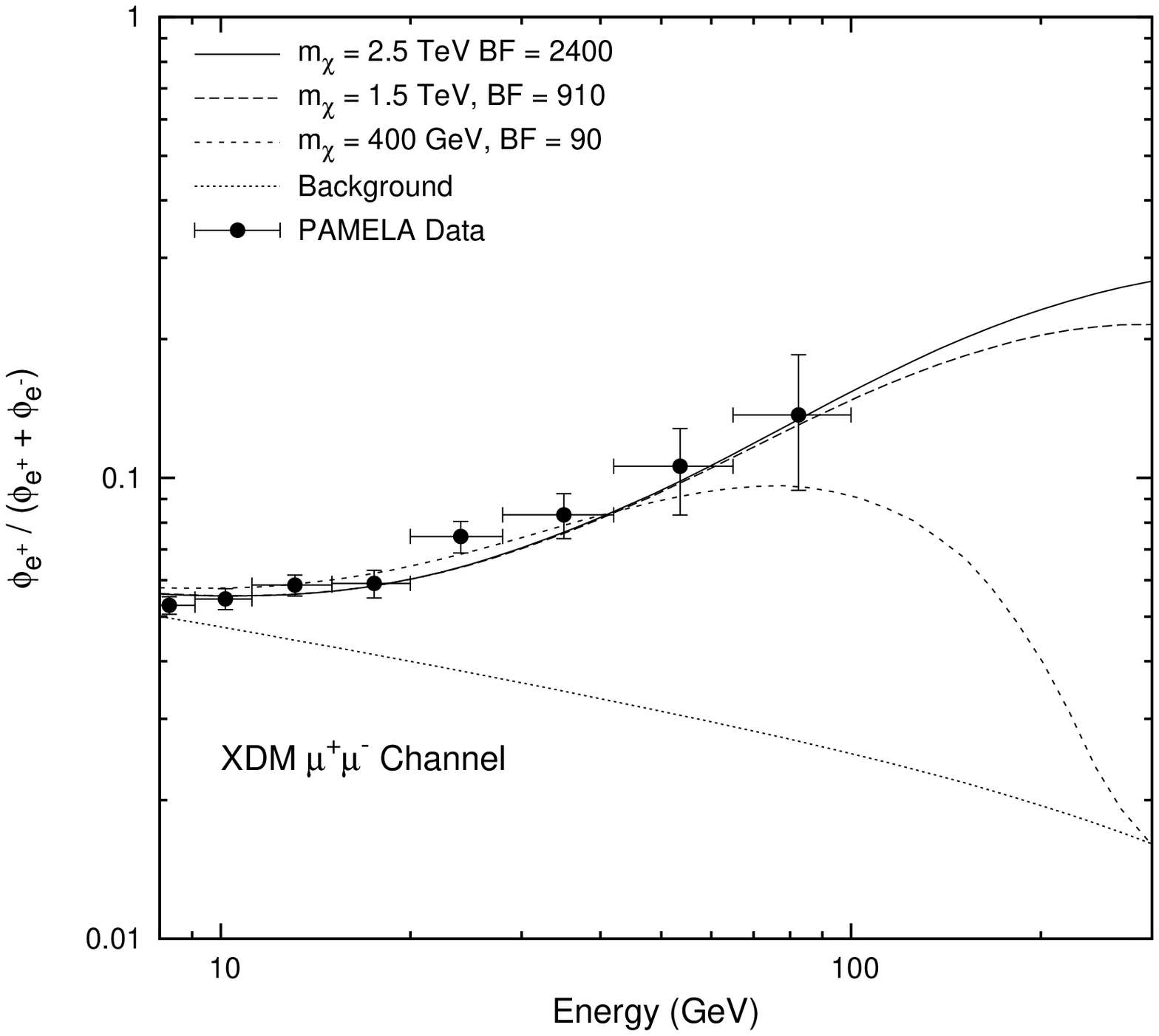}\hskip 0.2in
\includegraphics[width=.45\textwidth]{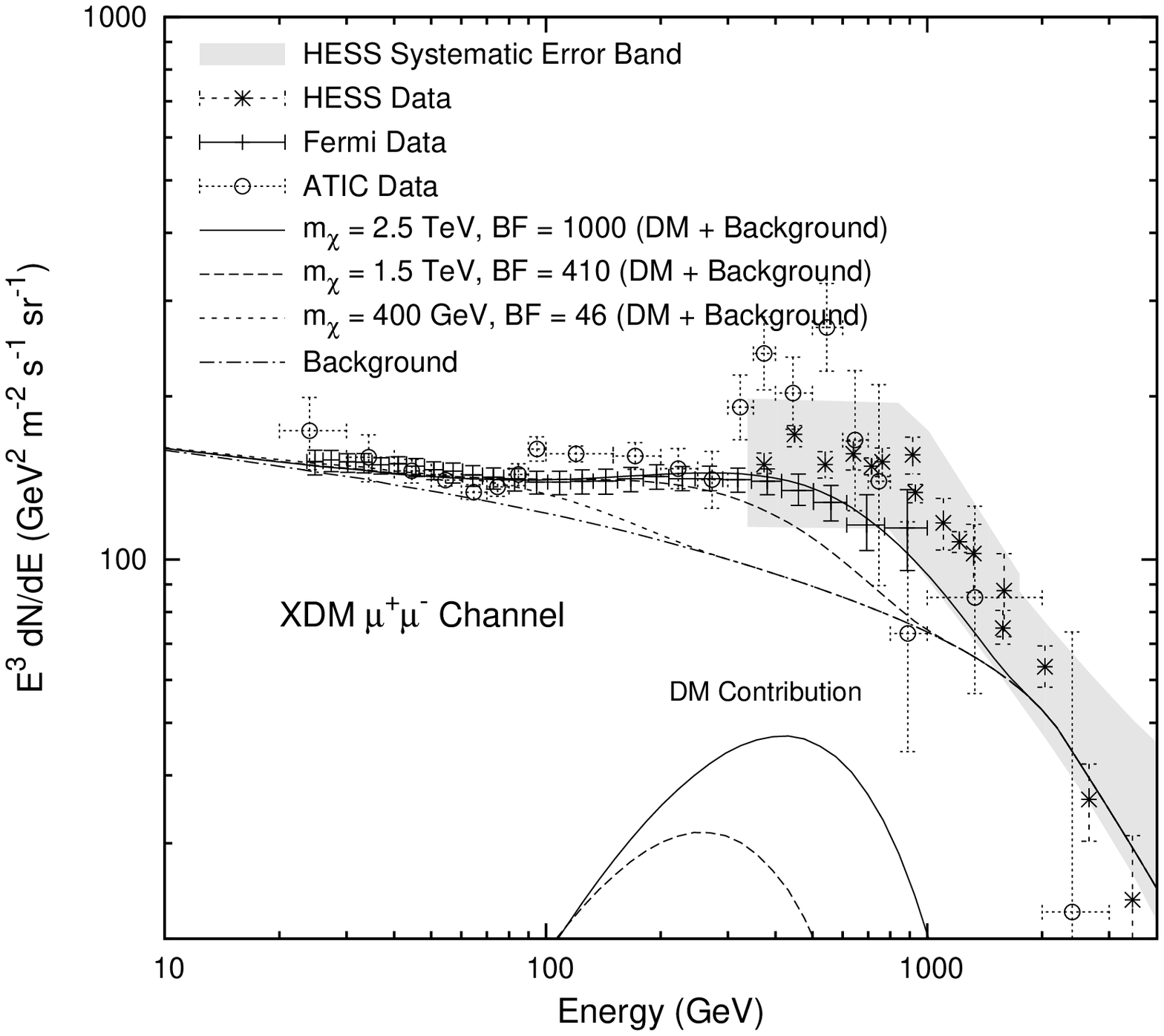}\\
\includegraphics[width=.45\textwidth]{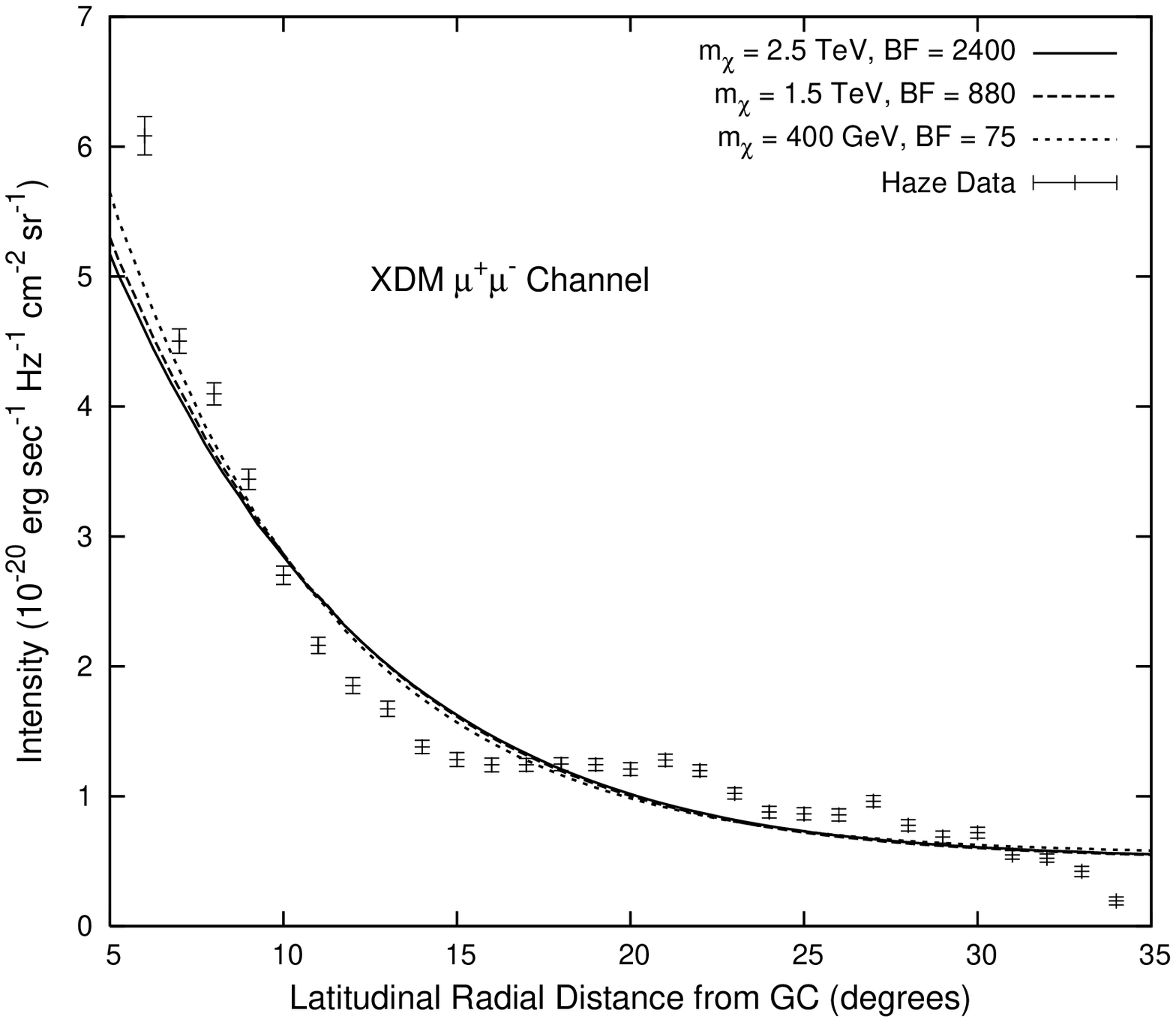}\hskip 0.2in
\includegraphics[width=.45\textwidth]{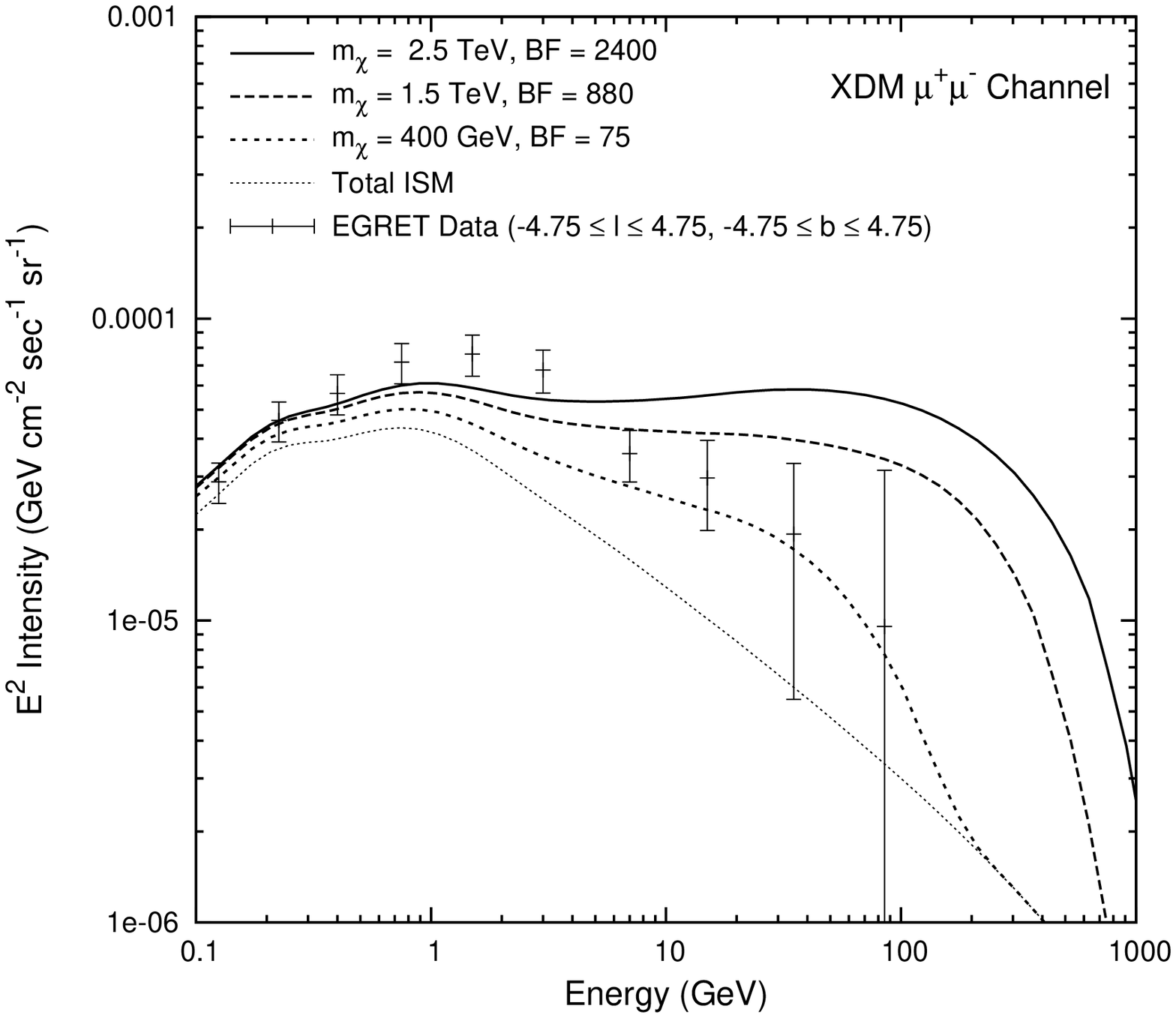}
\end{center}
\caption{The cosmic ray signals of dark matter annihilations $\chi \chi \rightarrow \phi \phi$, followed by $\phi \rightarrow \mu^+ \mu^-$.}
\label{fig:xdmmuons}
\end{figure*}

\begin{figure*}
\begin{center}
\includegraphics[width=.45\textwidth]{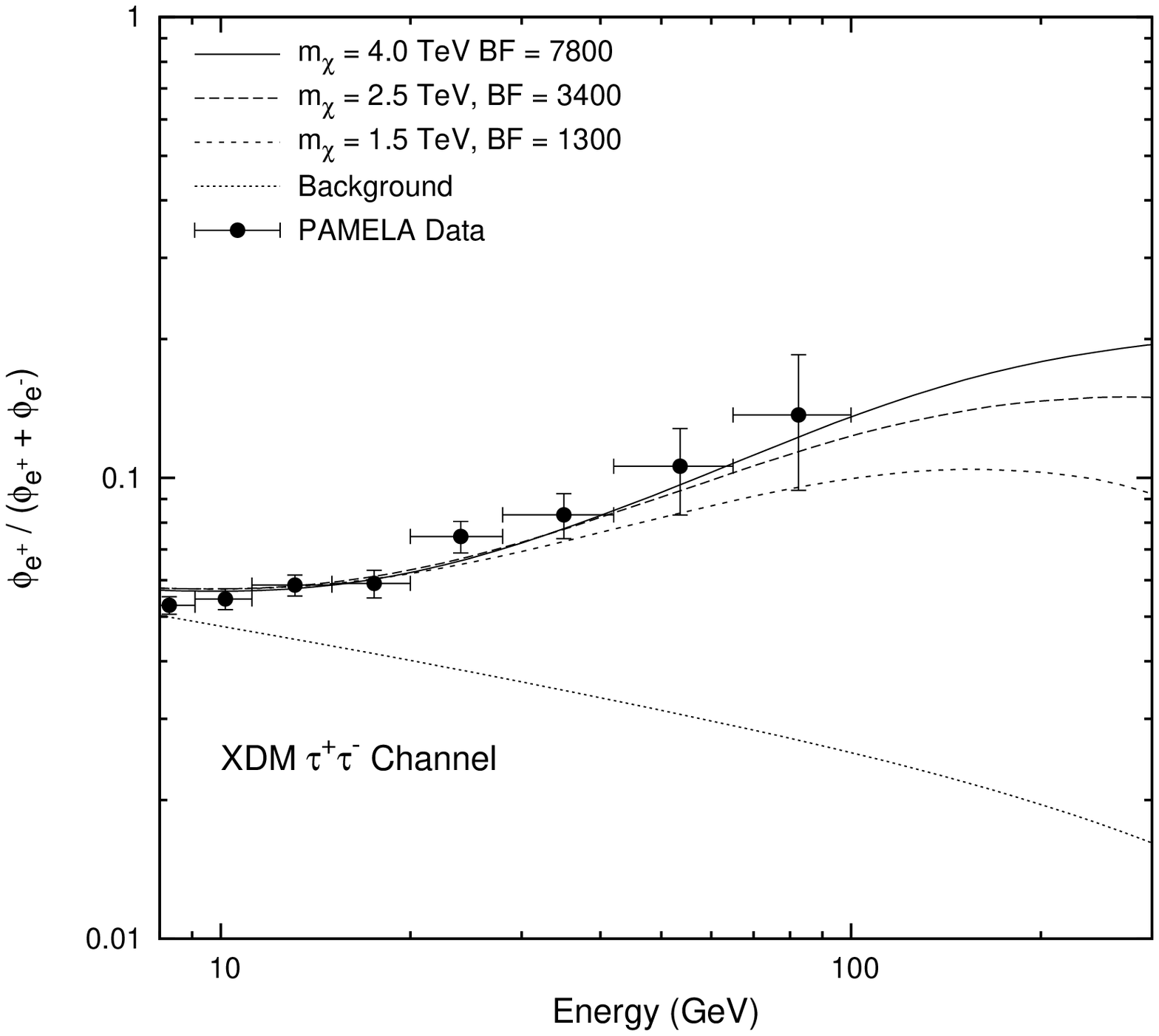}\hskip 0.2in
\includegraphics[width=.45\textwidth]{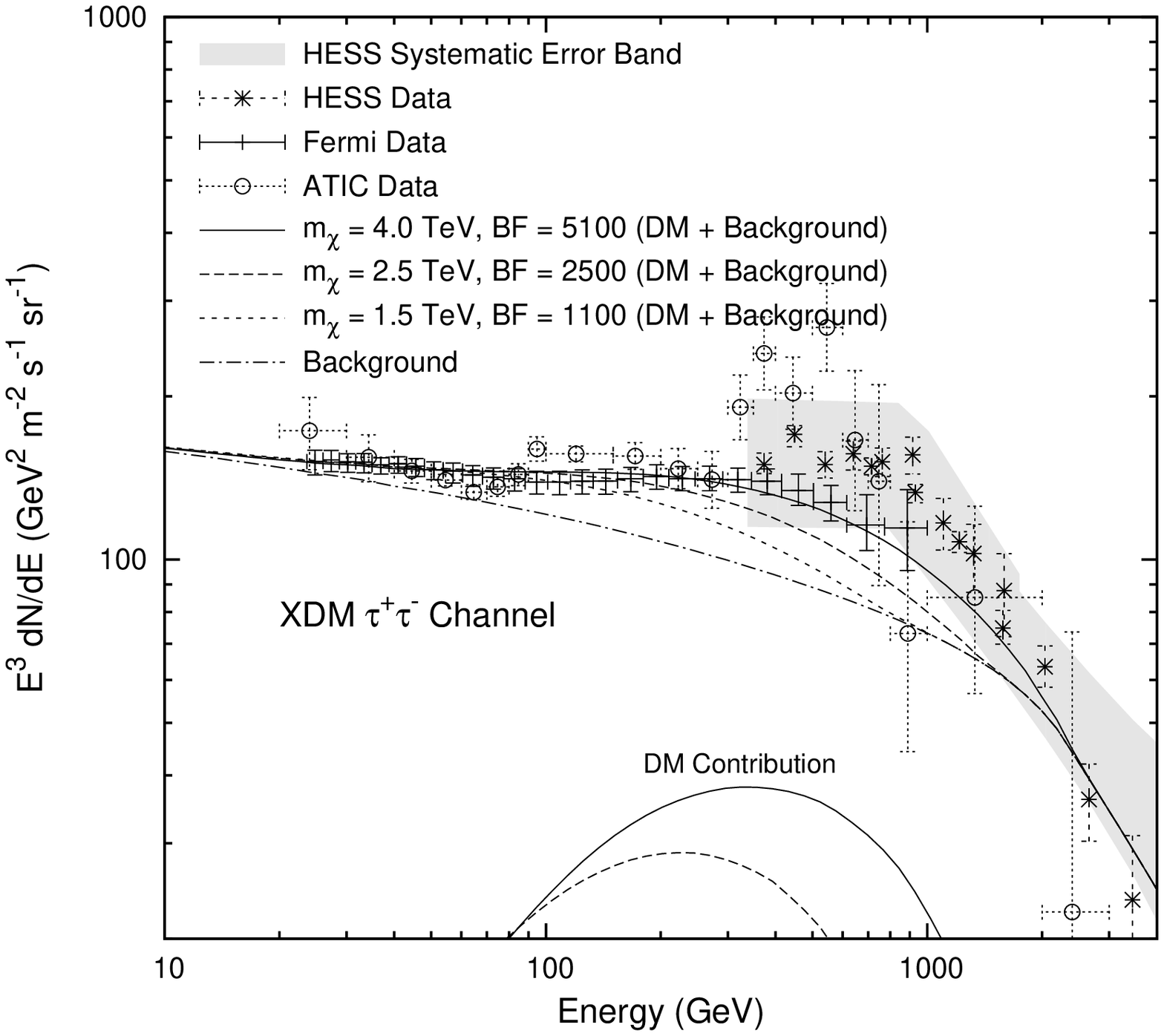} \\
\includegraphics[width=.45\textwidth]{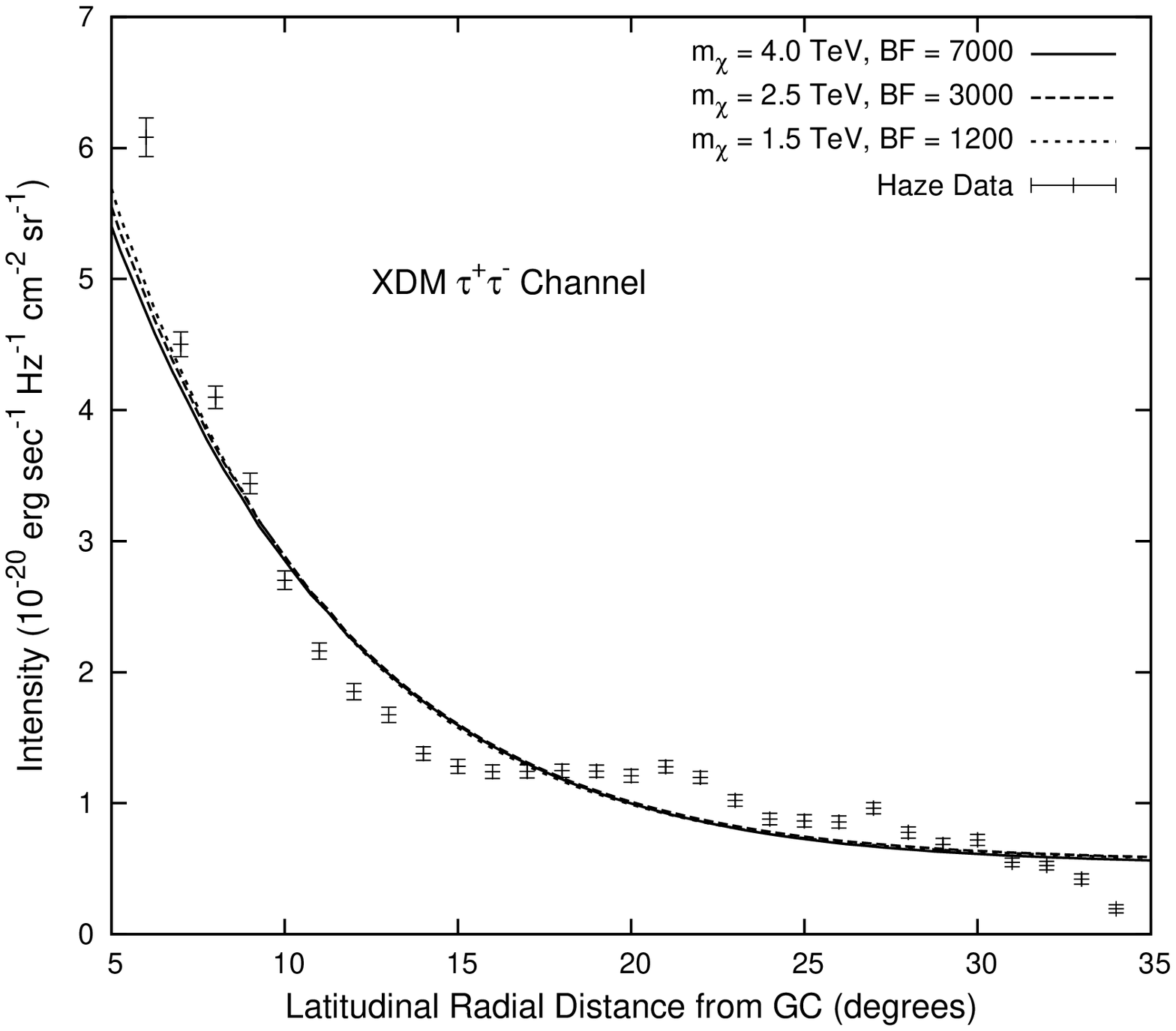}\hskip 0.2in
\includegraphics[width=.45\textwidth]{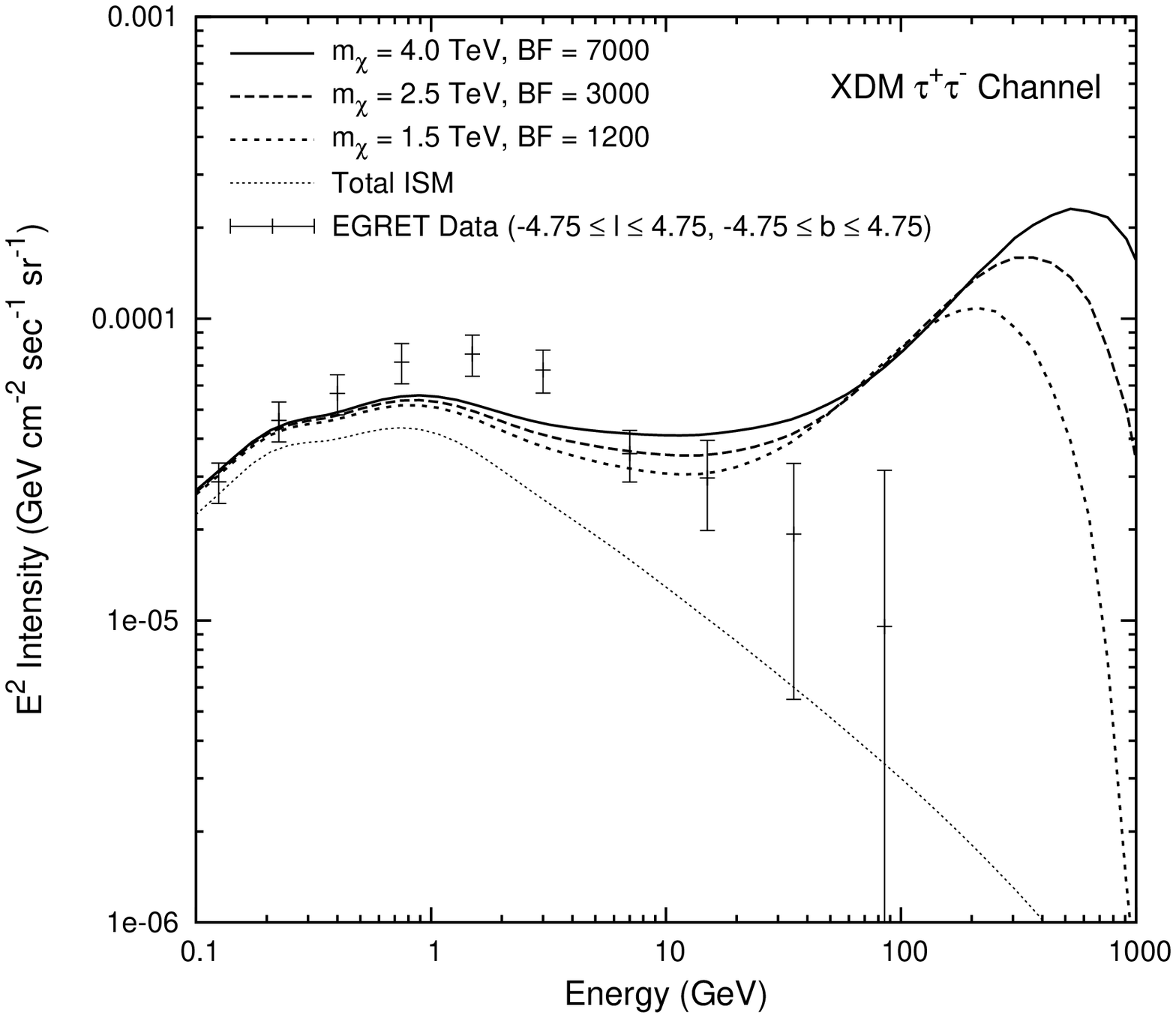}
\end{center}
\caption{The cosmic ray signals of dark matter annihilations $\chi \chi \rightarrow \phi \phi$, followed by $\phi \rightarrow \tau^+ \tau^-$.}
\label{fig:xdmtaus}
\end{figure*}

The results of our analyses are presented in Figures
\ref{fig:electrons}-\ref{fig:xdmtaus} and \ref{fig:ATICfits}.  Qualitatively, we
see that all modes considered give acceptable fits to the data.  A few
comments are in order, however.

The modes with annihilations directly to electrons or $\phi$-mediated
electrons give the best fit to the ATIC/PPB-BETS excess, with only direct annihilations to electrons yielding a true peak (see Figure \ref{fig:ATICfits}).  This is
simply a reflection of the hardness of the injection spectra
(monochromatic and flat, respectively). These modes also require the
lowest boost factors.  Again, this is because none of the energy is
partitioned into invisible modes (as with muons or taus) or into very
soft electrons and positrons (as with pions from tau decays).
Direct annihilation to muons also gives an acceptable fit to the ATIC
excess, while $\phi$-mediated muons are slightly too soft. Annihilations to $\tau$'s  and to 
$\phi$-mediated $\tau$'s are significantly too soft to achieve
the ``peaky'' nature of the excess.   In summary, modes with broad cutoffs (such as
taus) give poor fits to ATIC, while modes with sharp cutoffs
(electrons and $\phi$-mediated electrons) give better fits, as well as
smaller boost factors. 

In contrast, essentially all the annihilation modes {\em except} direct to electrons can fit the Fermi data well. The smooth cutoff (coupled with the larger error bars at higher energies) make essentially all of them compatible, although the $\tau^+\tau^-$ and XDM $\tau$  modes require somewhat uncomfortably large boost factors, if this is to be a thermal relic.

A careful observer will note that
there appear to be two components to the ATIC excess: a ``plateau''
from approximately 100-300 GeV, and a ``peak'' from 300-700 GeV. None
of the modes we have considered independently give both of these
features, but it is possible that multiple annihilation channels could
exist, for instance some linear combination of $e^+e^-$, $\mu^+ \mu^-$
and $\tau^+ \tau^-$. This could arise if $\phi$ could
decay into multiple states, or multiple types of $\phi$ particles were
produced in the annihilation as would be expected for a non-Abelian
version of $\phi$-mediated DM annihilation
\cite{ArkaniHamed:2008qn,ArkaniHamed:2008qp}. However, the absence of such a signal at Fermi makes such elements probably unnecessary.

All modes give a good fit for the PAMELA positron fraction. It is not
unexpected, given Figure \ref{fig:pamextrap}, that a reasonable, smooth fit to Fermi and ATIC also fits PAMELA
with similar boost factors.  Quite remarkably, the boost factors required for the Haze are within a
factor of two of those needed for Fermi, ATIC and PAMELA. This
relationship was noted in \cite{Cholis:2008vb}, where $\phi$-mediated
electron and muon modes were studied as explanations for the excesses
observed by previous positron cosmic ray experiments (HEAT,
AMS-01). There it was found that the Haze signal naturally required a
boost factor very similar to that needed for the excess that has since been
seen at PAMELA.

The overall boosts needed for Fermi and PAMELA differ approximately by a factor of two. Too much should not be made of this. To begin with, the fits are done separately, and in that light, the closeness of the required boosts is actually very encouraging. Secondly, there are uncertainties in the instruments; in particular, the Fermi acceptances are not precisely known. Additionally, since PAMELA actually measures a {\em ratio}, a similar signal can be found with a lower boost by employing a softer primary electron spectrum. This would change the Fermi fit somewhat and would require a larger boost to fit the Fermi data. Finally, we have attempted to hit the central value of the highest PAMELA data points. Should these come down, so, too, will the boost. Thus, absent a measurement of the electron spectrum only (i.e., not $\epp$) from PAMELA, it is difficult to say whether they are in tension, or if there are simply small systematics that will allow agreement. Future data should clarify this.

We should note that we find larger boost factors needed for the Haze
than calculated in \cite{Hooper:2007kb} (and noted in
\cite{ATIClatest}). The principle reasons are the very cuspy profile
used in \cite{Hooper:2007kb} and the relatively low levels of inverse
Compton scattering assumed by them.  In \cite{Hooper:2007kb} the cuspy profile $\rho \propto r^{-1.2}$
was used, leading to more power in DM annihilations in the center of the galaxy, resulting
in lower boost factors to fit the Haze. In contrast, we use the Einasto profile
(Eq. \ref{eq:einasto}) with fits established by Merritt et al.
\cite{Merritt:2005}.  As was shown in \cite{Cholis:2008vb}, the Einasto profile
yields similar boost factors as the NFW profile with $\rho \propto
r^{-1}$. For electrons with $E\gsim 5 \GeV$, the dominant energy losses are due to
ICS and synchrotron radiation. \cite{Hooper:2007kb} considered a homogeneous magnetic field of 10 $\microG$
and homogeneous energy loss rates of $5\times10^{-16}(E/\GeV)^{2} \GeV/$s and
$2.5\times10^{-16}(E/\GeV)^{2} \GeV/$s.  As the magnetic field that we
consider here has a peak value of 11.7 $\microG$ at the center of the
galaxy (corresponding to $3.0\eV\cm^{-3}$), our energy losses to synchrotron radiation in the GC are very similar to those in 
\cite{Hooper:2007kb}. However, our energy losses to inverse Compton
scattering increase with decreasing distance from the center of the galaxy 
due to the inhomogeneity of the interstellar radiation field.  The radiation field model used in GALPROP is large in the center of the galaxy,
with energy densities of $18.8\eV\cm^{-3}$ at $r=0$ and $14.3\eV\cm^{-3}$ at $r=0.5\kpc$.
As a result, the larger ICS losses compete
against the synchrotron losses, and less power is emitted in
synchrotron radiation. Thus, less energy is radiated into the
$23\GHz$ band, resulting in higher boost factors. We note that
\cite{Hooper:2007kb} did not include bremsstrahlung energy
losses, but this has a minor effect on the overall boost. Nonetheless, one should conclude that there is at least a factor of a few uncertainty in the overall normalization of the cross section needed to explain the Haze. And, while we take it as a positive that the conventional parameter choice leads to very similar boosts for PAMELA, Fermi and the Haze, too much should not be made of this in light of this uncertainty.

\vskip 0.1in
{\noindent \bf Signals at EGRET and Fermi/GLAST}

With the successful operation of Fermi/GLAST, there is naturally great interest in gamma ray signatures of dark matter models. It is especially important within the context of the present PAMELA, Fermi electron, ATIC and PPB-BETS results, because, as we have emphasized, the galactic center may offer the ideal testing ground of the dark matter explanation of these data. It is then noteworthy
that a generic prediction of scenarios involving WIMP
annihilation is a significant production of ICS gamma rays in the
galactic center. We believe this to be independent of the particular annihilation mode which explains the electron and positron signals, and only that the significant electronic production here is reflected in the galactic center.

\begin{figure*}
\begin{center}
\includegraphics[width=0.45\textwidth]{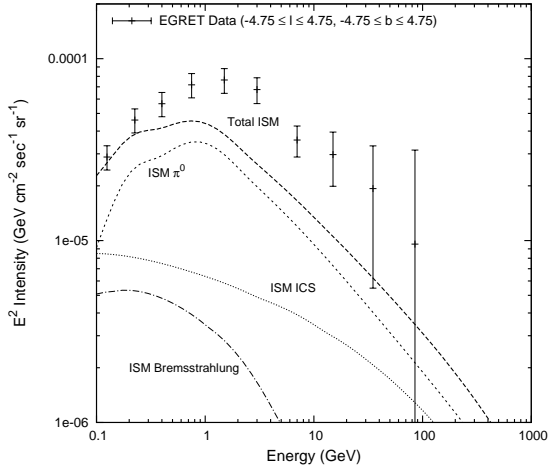}
\end{center}
\caption{The dominant contributions to gamma rays in the galactic center (inner $5\degree$) from conventional astrophysical CR interactions. Shown are the photons from $\pi^0$'s, electron ICS signals and bremsstrahlung.  In this and following figures, ``ISM'' denotes gammas produced by CR interactions with the interstellar radiation field, as well as the interstellar medium (gas and dust).  The $\pi^0$ line refers to $\pi^0$ $\gamma$-rays produced by the hadronic CRs interacting with the interstellar medium.  The slope of the $\pi^0$ curve at high energies is tied to the primary $p$ CR spectrum, and is better constrained than the amplitude. }
\label{fig:conventionalsources}
\end{figure*}

Background gamma rays in the galactic center are expected to come from a variety of sources, specifically, $\pi^0$'s from protons interacting with the ISM, inverse Compton scattering (ICS) from cosmic ray electrons, as well as a small component of bremsstrahlung. (See Figure \ref{fig:conventionalsources}.) Note that the spectrum due to conventional sources is expected to fall dramatically as one moves to higher energies.

In the context of our present discussion, the behavior of the gamma ray spectrum at the higher energies detectable by Fermi/GLAST should be dramatically changed. If dark matter is responsible for an O(1) correction to the local flux of cosmic ray electrons in the 400-800 GeV range, then it would be expected to be the dominant contributor to the electron flux in this energy range in the galactic center. Such electrons are responsible for upscattering photons to energies $E_{upscatter}\approx 2E_{initial} \gamma^2$. For (initially) optical photons, this will result in a major contribution to photons in the 10 - 1000 GeV range. Note that these photons are independent of other sources, such as the final-state radiation (FSR) gammas from charged particle production or $\pi^0$'s produced in e.g., $\tau$ decay. Rather, these photons are intimately linked to the production of high-energy electrons and positrons, the same population of particles which we believe should be yielding the Haze.

Indeed, this basic intuition is borne out in Figures \ref{fig:electrons}-\ref{fig:xdmtaus}. One sees very significant gamma ray signatures at high energies in all channels. One should not read too much into the fact that many modes seem to exceed the EGRET values - as we shall discuss shortly, there is a factor of a few uncertainty arising from astrophysics. While modes with $\tau$'s can have gamma ray fluxes dominated by $\pi^0$ production (often exceeding present EGRET limits quite significantly), the fluxes for the electron and muon modes are dominated by ICS gammas. 
FSR signals tend to be significant only at the highest energies, which are out of the range of Fermi/GLAST for the dark matter masses suggested by the break at $\sim$ TeV seen at Fermi, HESS and ATIC.  (Fermi/GLAST has an upper energy reach of $\sim 300$ GeV for photons.)  The collinear approximation we have used for the FSR gammas is not appropriate at the highest energies, where FSR becomes important, and should probably be taken as an upper bound on the FSR signal in that energy range.  Additionally, the size and shape of the FSR signal is model dependent; it depends on the parameters involved (notably the $\phi$ mass in the $\phi$-mediated modes). 
That said, it is possible that a lower value for the ISRF energy density could lead to a lower ICS signal, as we shall discuss shortly, in which case FSR signals could be quite relevant in the Fermi range, even for the large masses considered.  In contrast, the ICS signal should be present for any population of electrons with the same spectrum which explains the high energy electron signal and PAMELA, and, more to the point, almost any population of DM-produced electrons which explains the Haze. This strong signal should be observed by Fermi/GLAST and would be a smoking gun of this massive high energy electronic production in the galactic center.

 We note that none of our current models seem to give a good fit to
 the EGRET data points in the $2-5$ GeV range.  The difficulty these EGRET 
 observations present for standard $\gamma$-ray production mechanisms was pointed out previously by
 \cite{Strong:2004de}, who proposed an ``optimized'' model, with
 modified electron and proton CR spectra.  Because this excess
 appears everywhere on the sky, not just the galactic center, it is
 unlikely to be related to ICS from dark matter.  A modified proton
 spectrum is also not likely in our model, given that no significant
 anti-proton excess has been observed.  However, the potential for DM
 annihilation to affect the average electron spectrum is obvious.

\begin{figure*}
\begin{center}
\includegraphics[width=.45\textwidth]{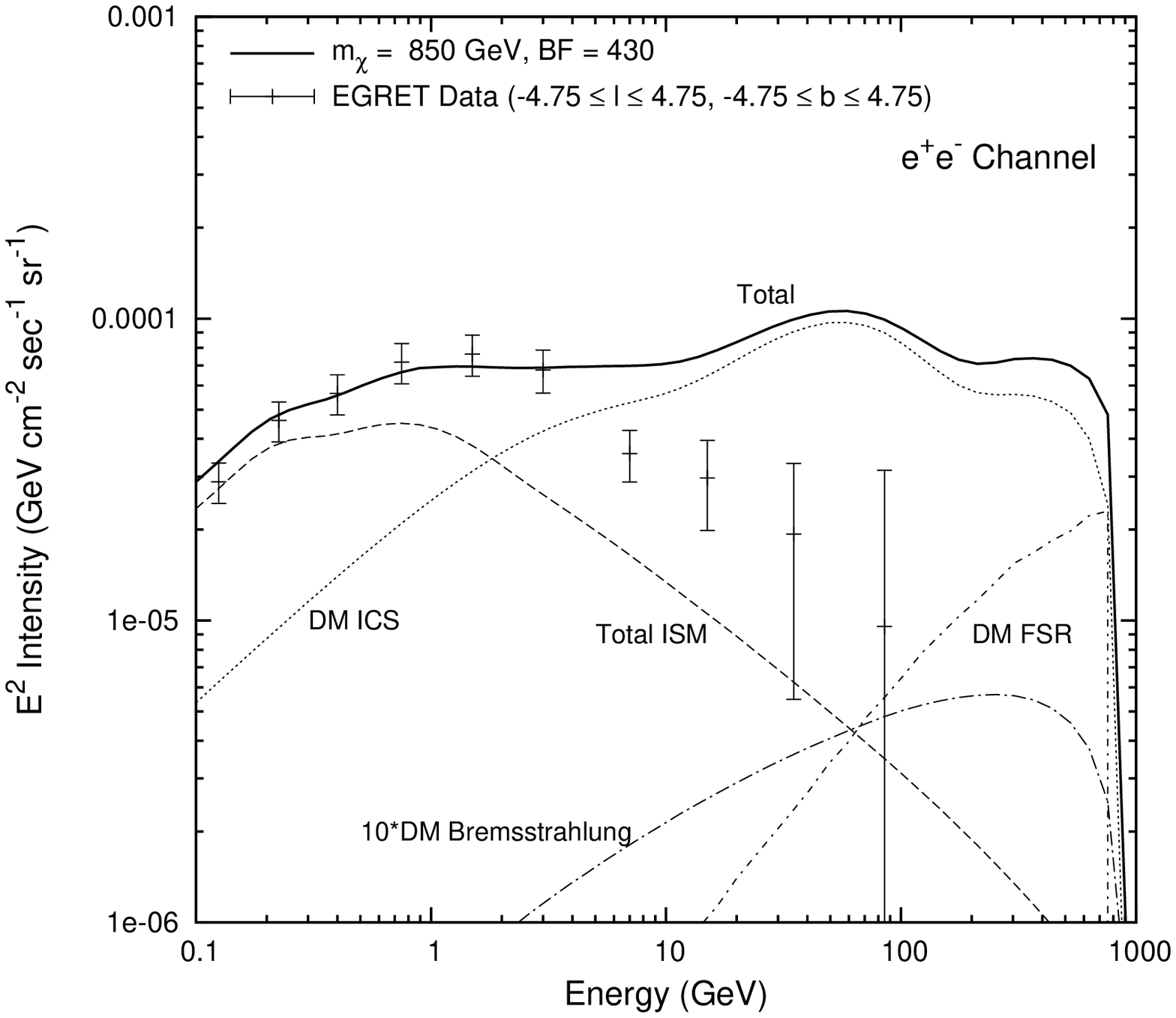}\hskip 0.2in
\includegraphics[width=.45\textwidth]{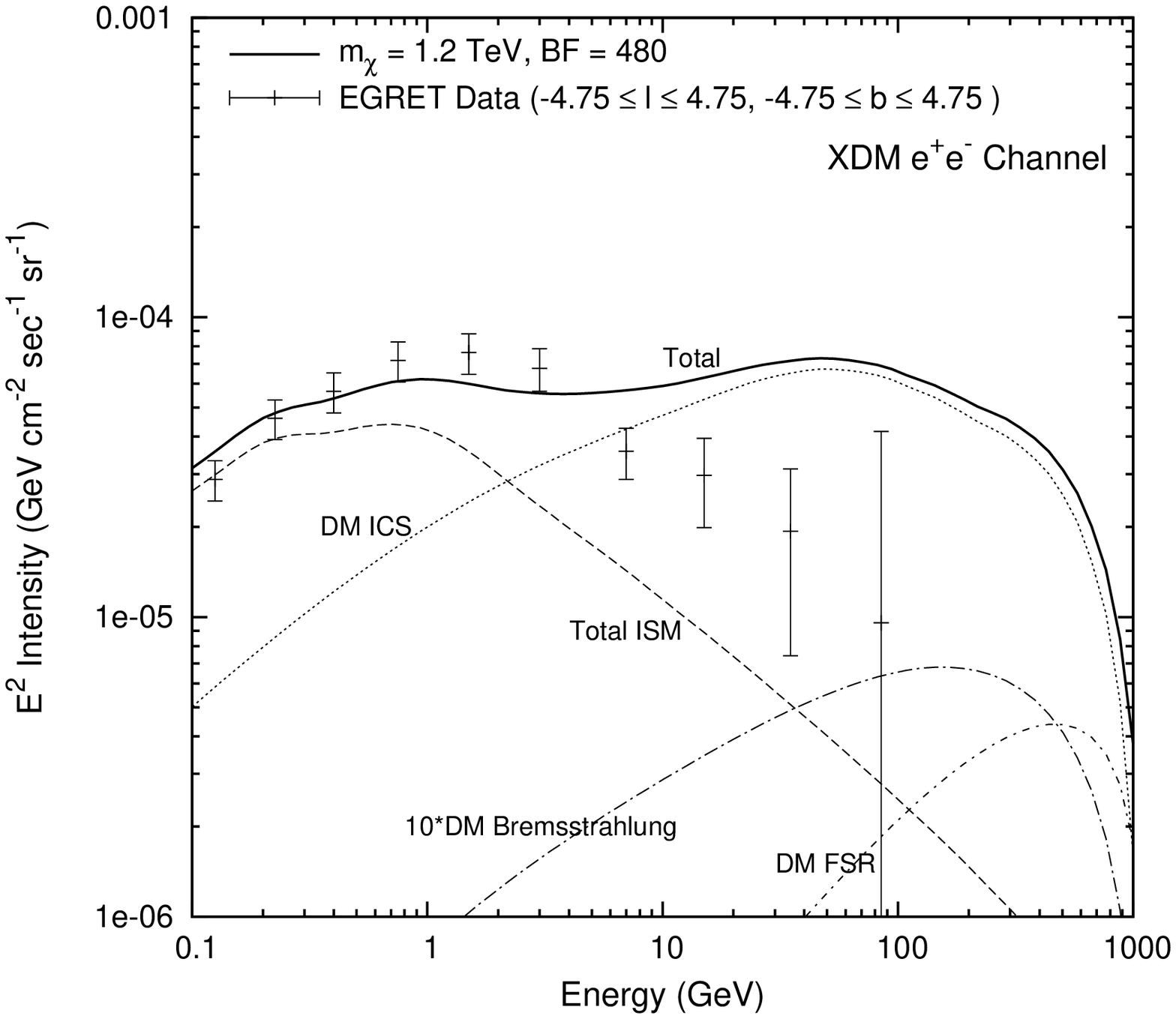}\\
\includegraphics[width=.45\textwidth]{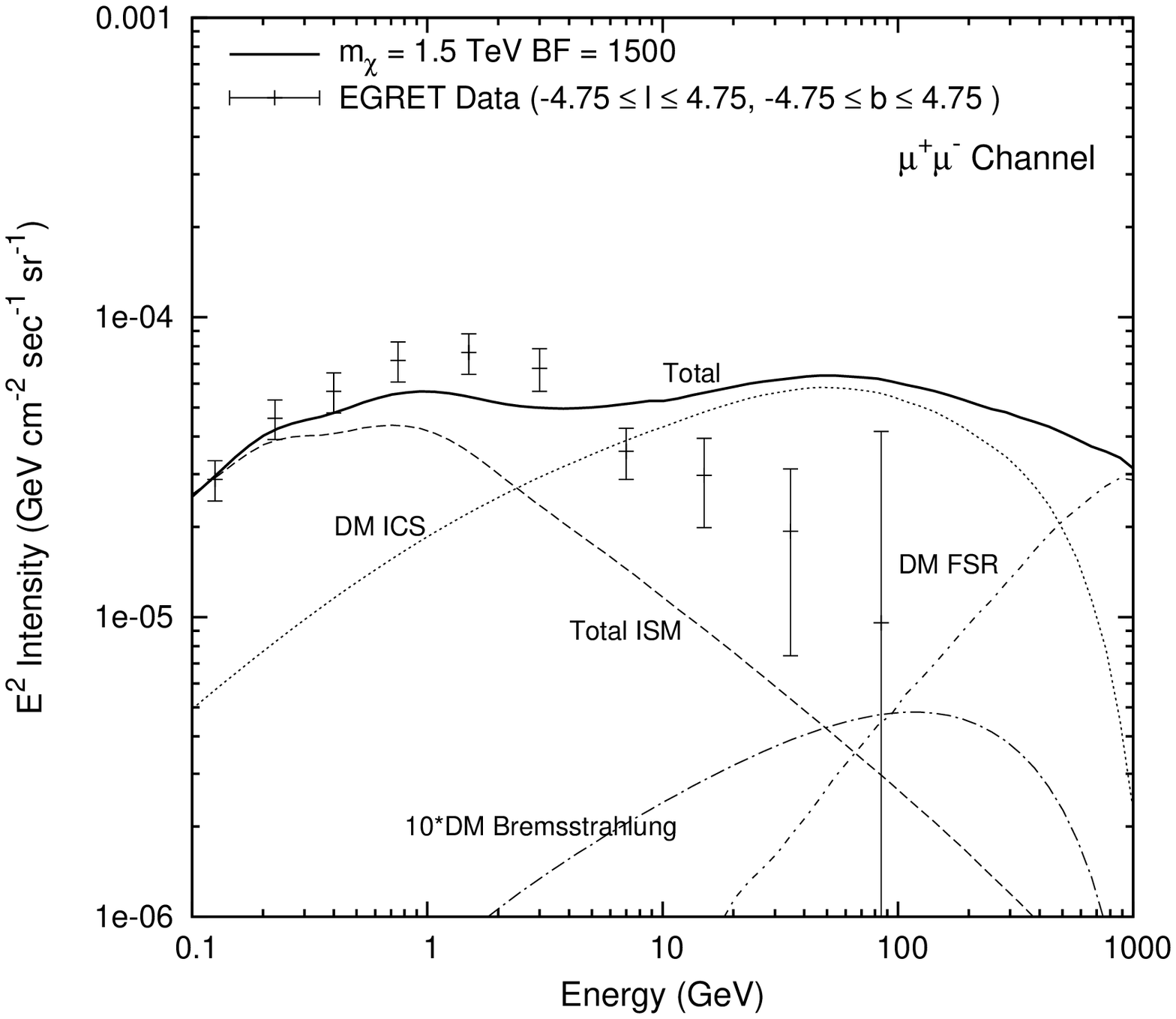}\hskip 0.2in
\includegraphics[width=.45\textwidth]{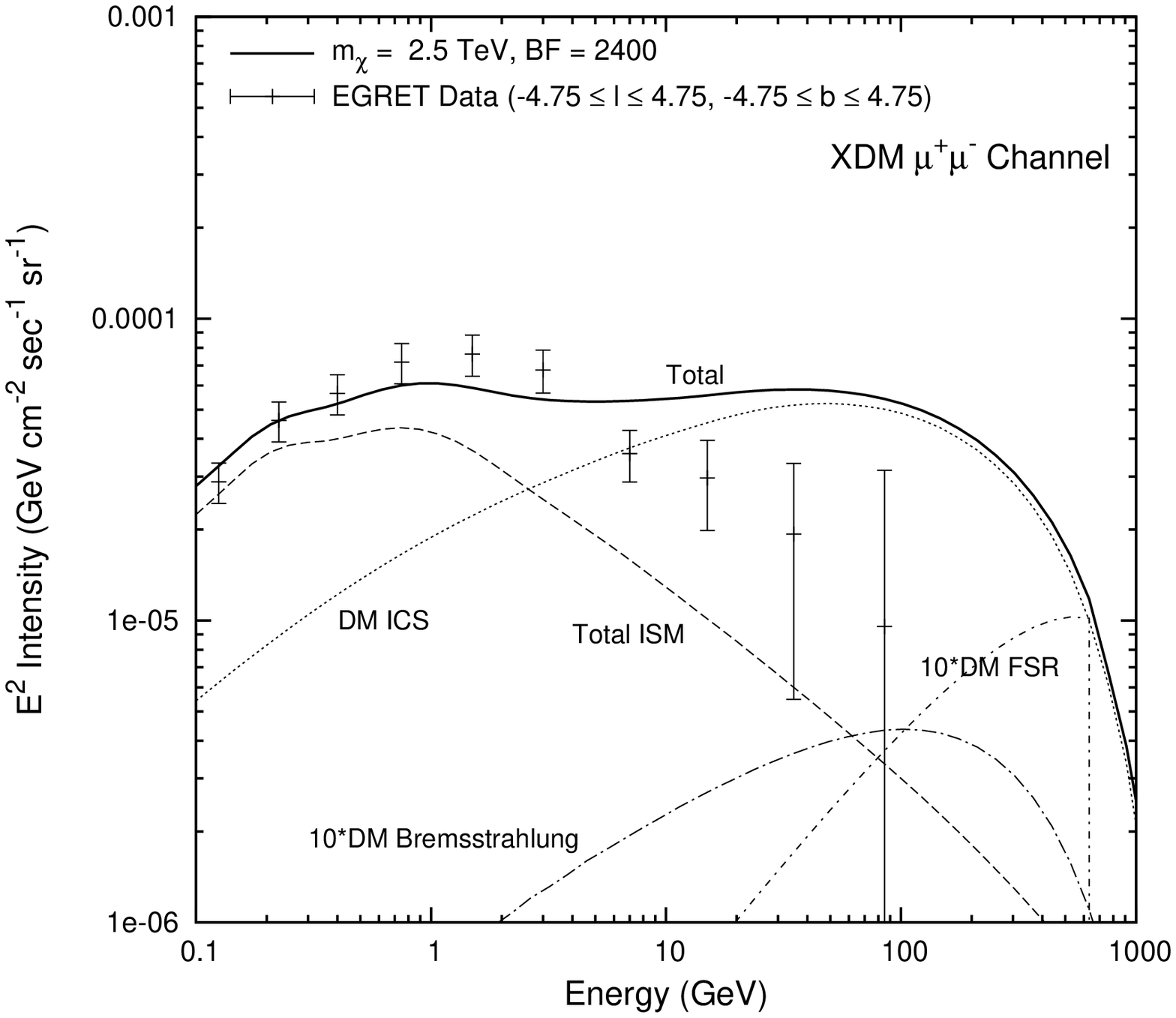}\\
\includegraphics[width=.45\textwidth]{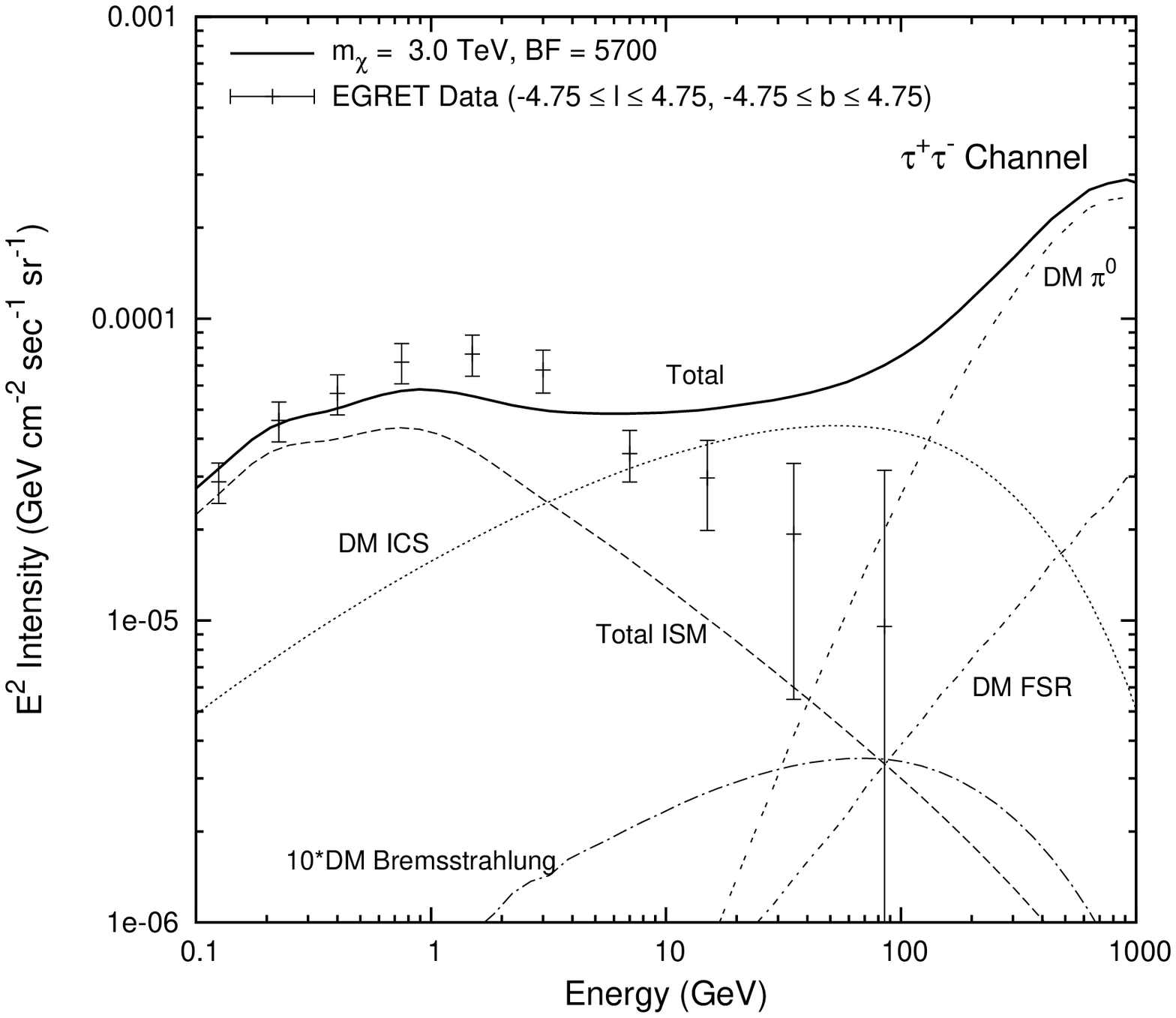}\hskip 0.2in
\includegraphics[width=.45\textwidth]{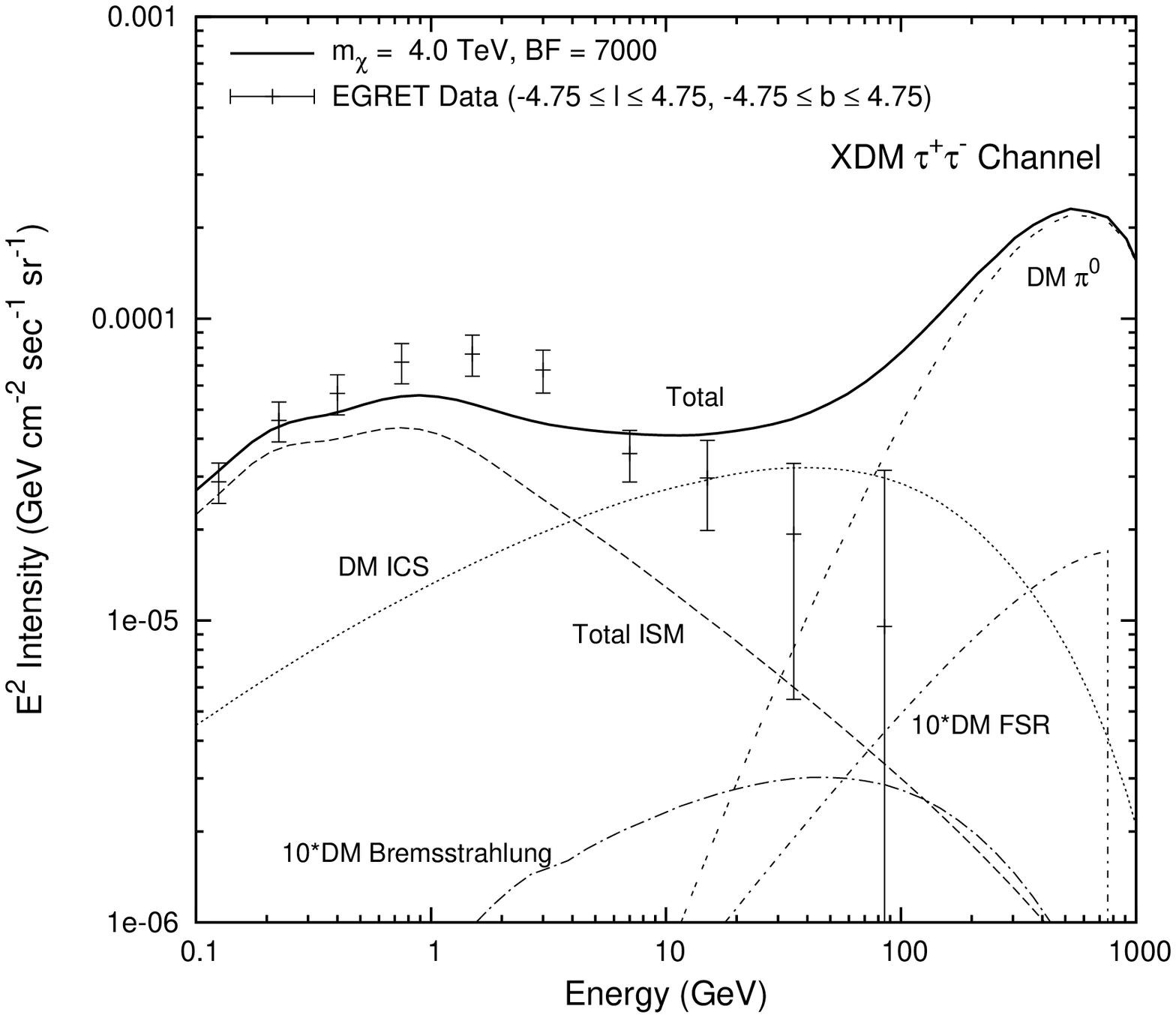}
\end{center}
\caption{Different contributions to gamma rays in the galactic center for different dark matter annihilation modes.}
\label{fig:EGRETsources}
\end{figure*}

\subsection{Limits from diffuse gamma rays}
One of the most important consequences of this scenario is the
presence of significant ICS gamma ray production in the inner few
degrees of the galaxy.  This is a natural byproduct of these new
sources of $\epp$.  These signals are naturally dependent on the
positrons and electrons produced, and hence on the cosmic ray
propagation models. In contrast, predictions of photons from $\pi^0$'s
or final state radiation \cite{Beacom:2004pe,Bergstrom:2004cy,Birkedal:2005ep,Mack:2008wu,Bergstrom:2008gr,Bertone:2008xr,Bergstrom:2008ag,Meade:2009rb,Mardon:2009rc,Meade:2009iu} depend only on the DM halo profile and
annihilation modes. It is worth considering the constraints these
prompt photons place on DM annihilation, using both galactic center
and diffuse extra-galactic limits.

\begin{figure*}[htpb]
\begin{center}
a)\includegraphics[width=.45\textwidth]{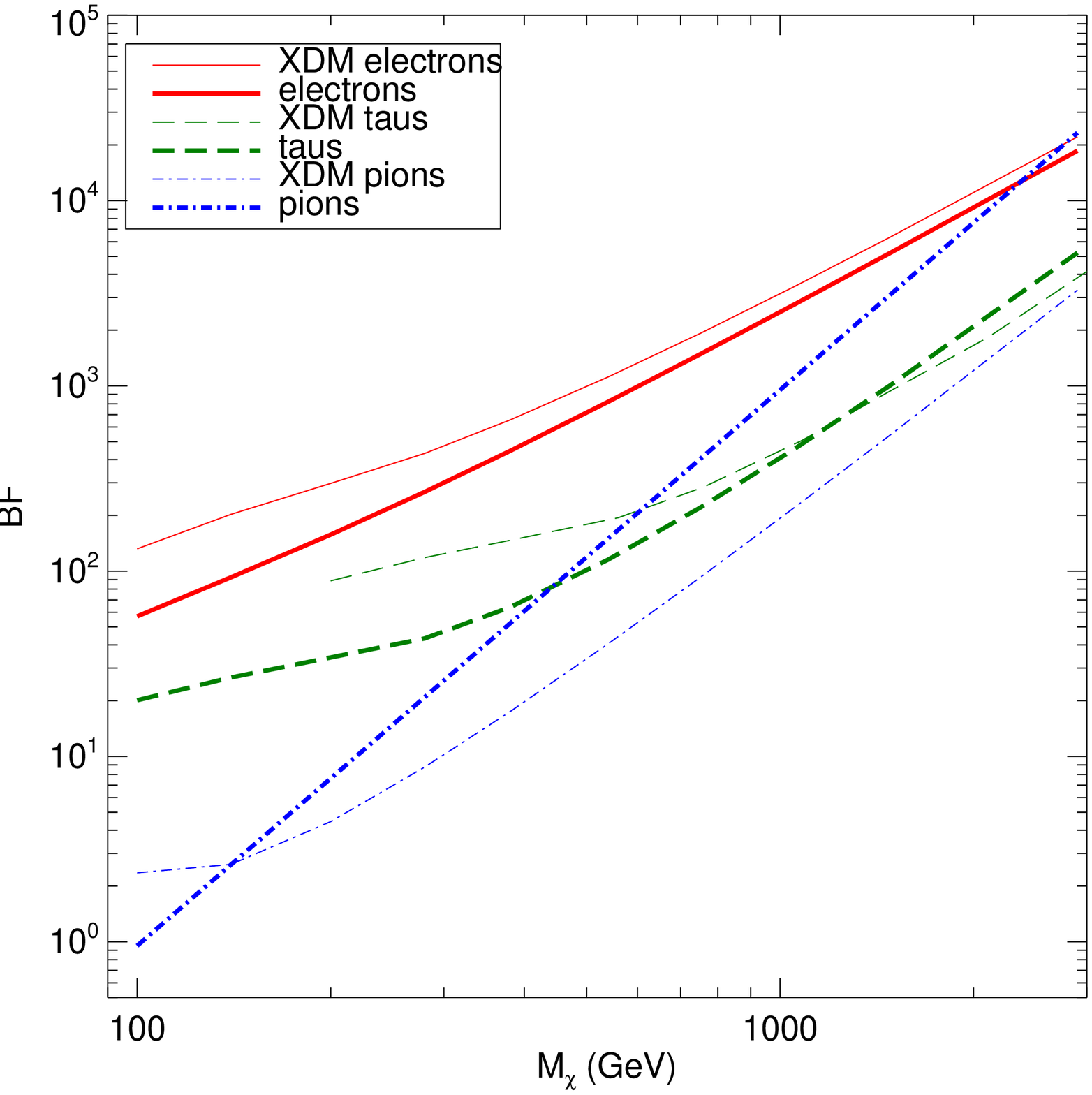}\hskip 0.2in
b)\includegraphics[width=.45\textwidth]{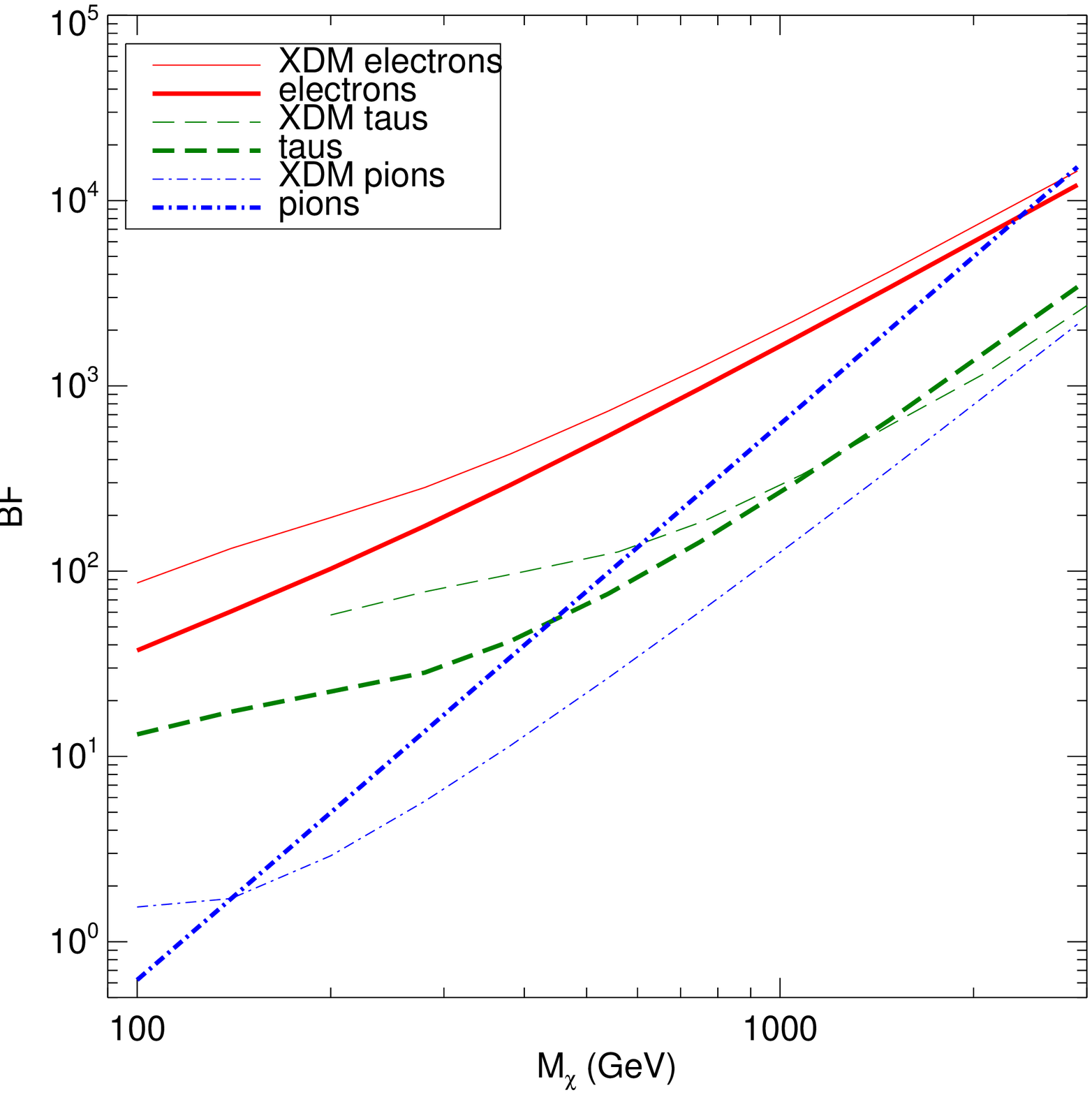} \\
\end{center}
\caption{Boost factor limits (compared to $\sigma v = 3 \times 10^{-26}\rm\, cm^{3}s^{-1}$ and $\rho_0 = 0.3 {\rm GeV\, cm^{-3}}$ for a) NFW profile, b) Einasto profile.
In each panel, lines are shown for XDM electrons ($\chi \chi \rightarrow \phi \phi$, $\phi \rightarrow e^+ e^-$), direct electrons ($\chi \chi \rightarrow e^+ e^-$), XDM taus ($\chi \chi \rightarrow \phi \phi$, $\phi \rightarrow \tau^+ \tau^-$), direct taus ($\chi \chi \rightarrow \tau^+ \tau^-$), XDM pions ($\chi \chi \rightarrow \phi \phi$, $\phi \rightarrow \pi^0 \pi^0$), and direct pions ($\chi \chi \rightarrow \pi^0 \pi^0$).  The limits come from prompt FSR and $\pi^0$ gammas only. Limits are determined by requiring that no EGRET data point is exceeded by more than $2\sigma$.}
\label{fig:totgammalim}
\end{figure*}

We show in Figure \ref{fig:totgammalim} the limits on boost factors (BF) for different $\gamma$-ray production modes, both for NFW and Einasto profiles. To generate these plots, we calculate the total flux of prompt photons, either from FSR or $\pi^0$ decays, from the different sky regions as defined in \cite{Strong:2004de}. We consider annihilations to $\tau^+ \tau^-$, $\pi^0 \pi^0$, and $\epp$, as well as the mediator decay modes $\chi \chi \rightarrow \phi \phi$, with $\phi \rightarrow \tau^+ \tau^-$, $\phi \rightarrow \pi^0 \pi^0$ and $\phi \rightarrow \epp$. To generate these limits, we use the EGRET bounds from \cite{Strong:2005zx}, and take the 2-sigma upper bounds to be conservative (no background subtraction is included). We show limits for both NFW (dashed) and Einasto (solid) profiles. Note that the inclusion of backgrounds can make these limits significantly stronger - we present them in this form to be clear what an incontestable upper bound is.

Other limits can be found from HESS data \cite{Bell:2008vx} or from the inner $1\degree$ of the Milky Way. These limits are highly dependent on the details of the profile, and we limit ourselves to the more conservative ones from the EGRET data. \footnote{Subsequent to the initial version of this paper, specific limits from HESS have appeared for many of the scenarios considered here \cite{Bertone:2008xr,Bergstrom:2008ag,Meade:2009rb,Mardon:2009rc,Meade:2009iu} and we refer the interested reader to those works. These works find, as we do, that modes with $\tau$'s are highly constrained, as are modes annihilating directly to leptons. Annihilation into hard electrons or muons are constrained except in very soft halos, while XDM annihilation modes are consistent for Einasto profiles.}

\subsection{Other modes}
\begin{figure*}
\begin{center}
\includegraphics[width=.45\textwidth]{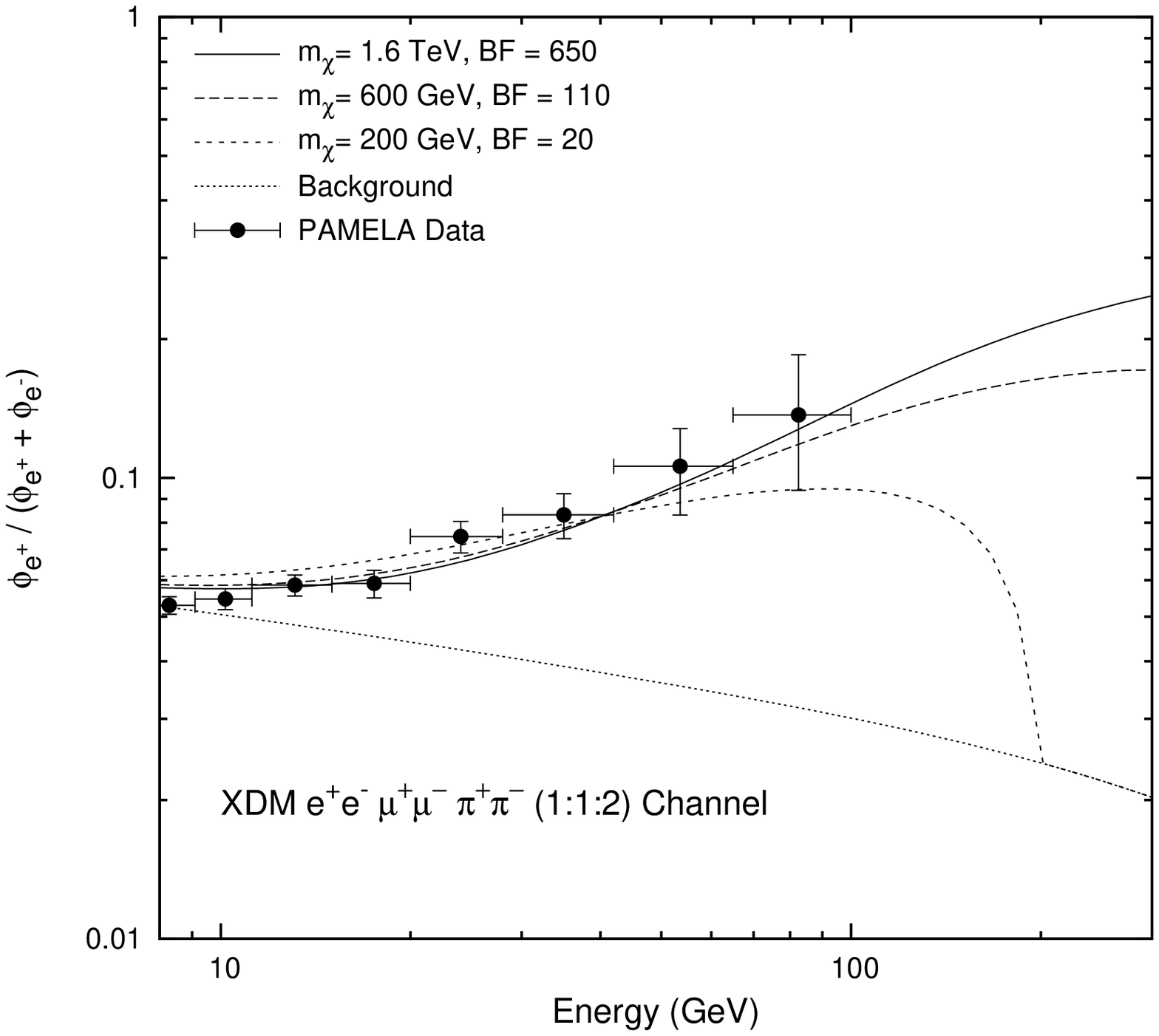}\hskip 0.2in
\includegraphics[width=.45\textwidth]{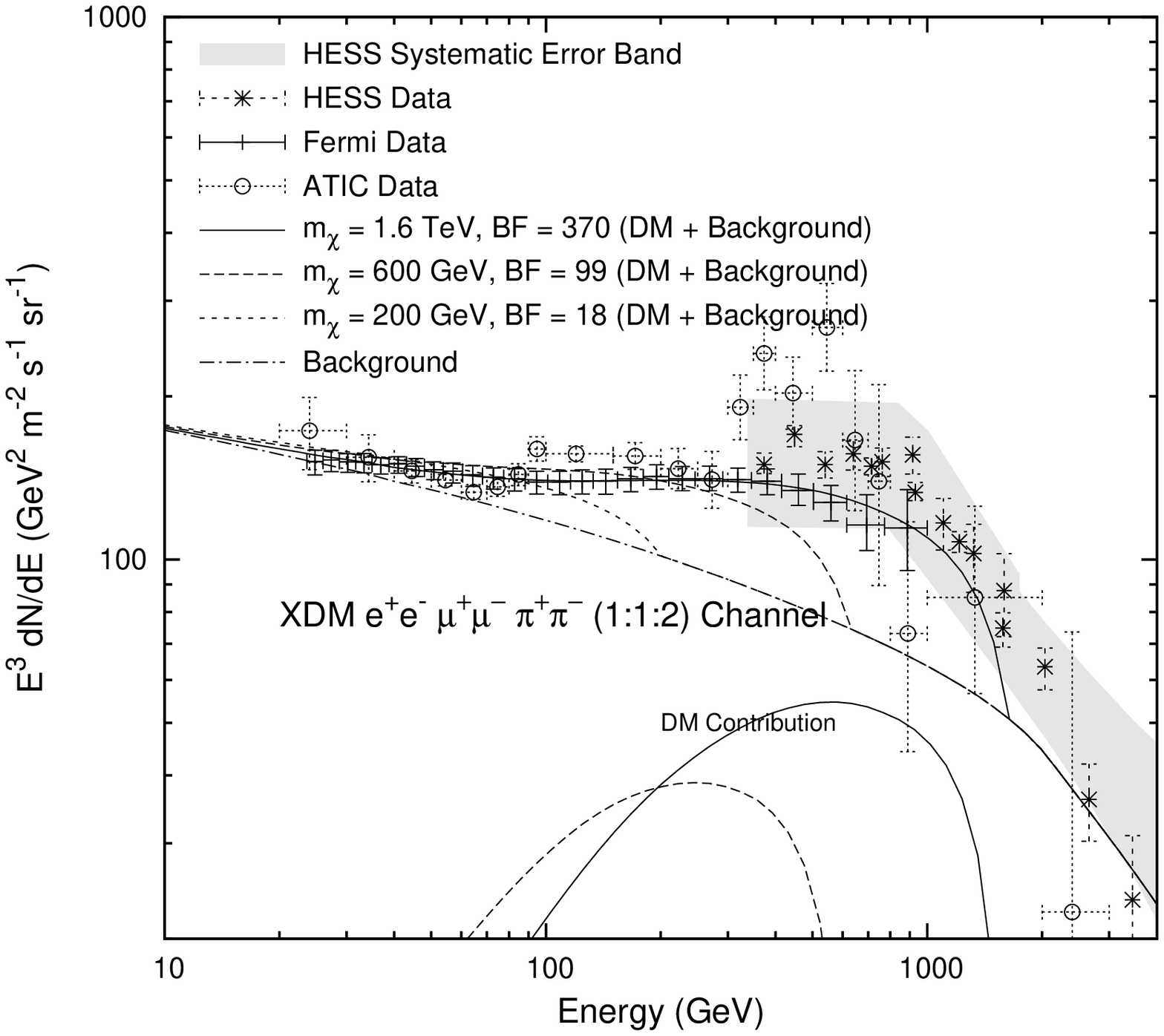}\\
\includegraphics[width=.45\textwidth]{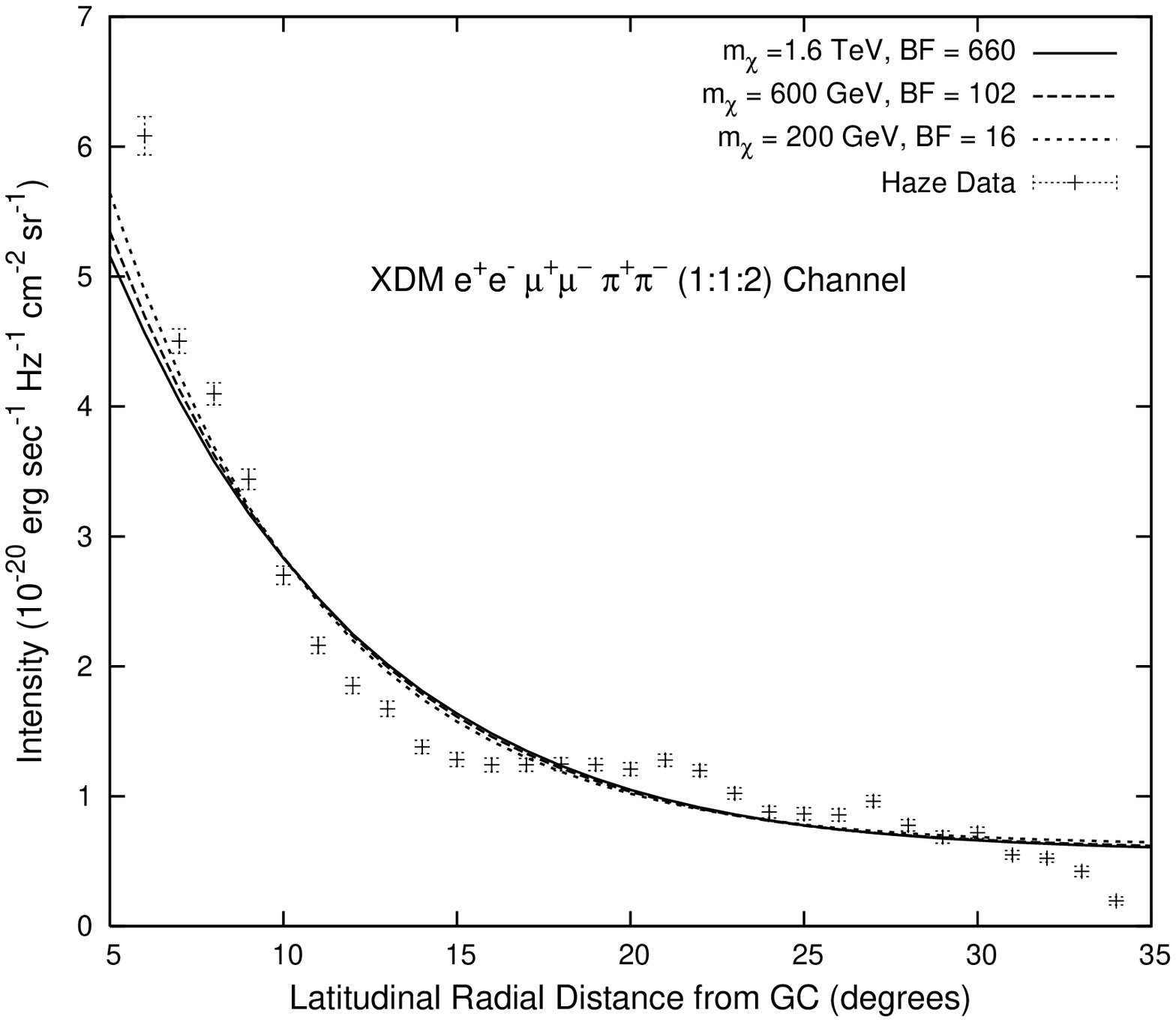}\hskip 0.2in
\includegraphics[width=.45\textwidth]{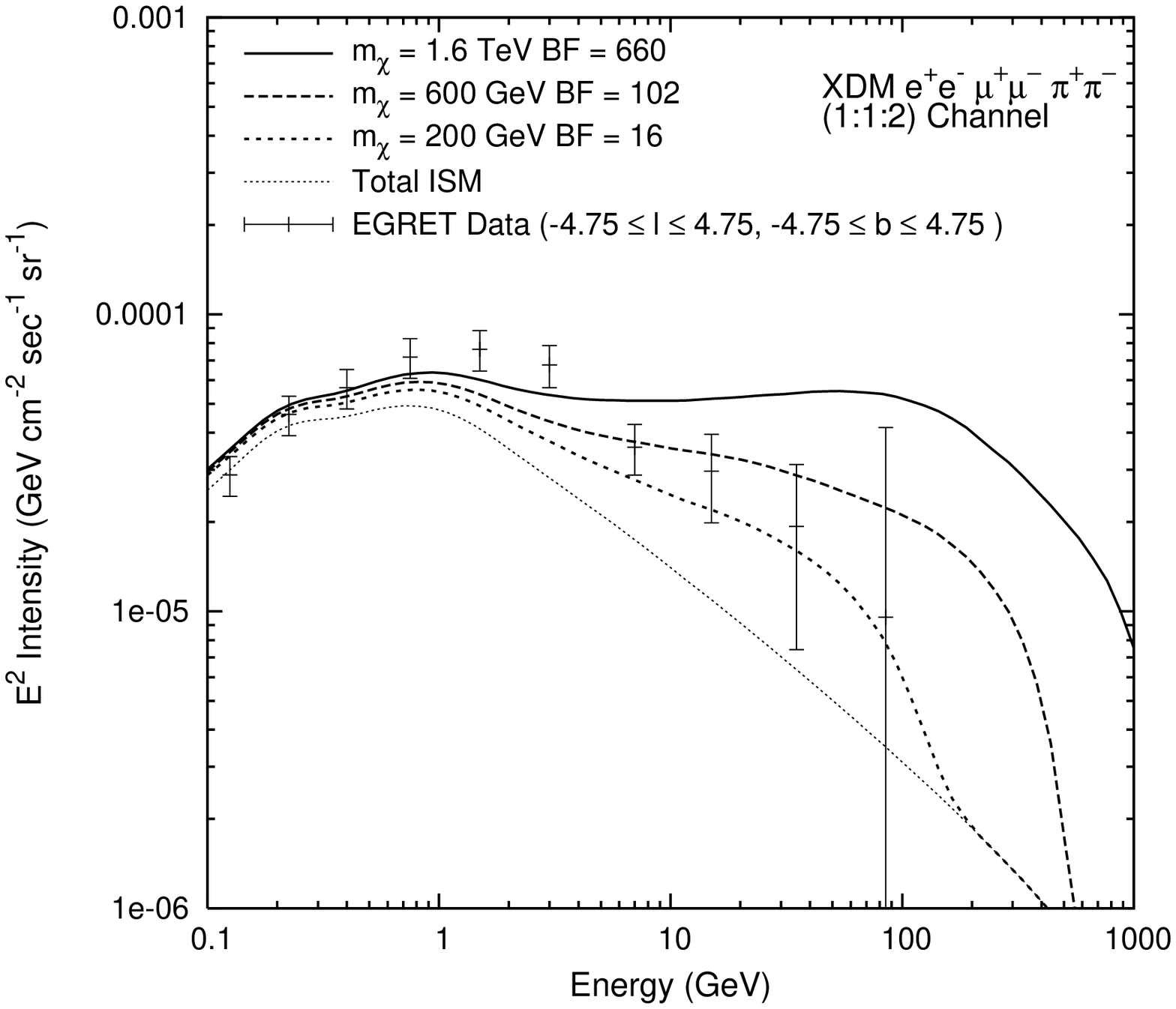}
\end{center}
\caption{The cosmic ray signals of dark matter annihilations as in Figure \ref{fig:electrons}, 
but with $\chi \chi \rightarrow \phi \phi$, followed by $\phi \rightarrow \epp$:$\phi \rightarrow \mu^+ \mu^-$:$\phi \rightarrow \pi^+ \pi^-$ in a $1:1:2$ ratio.}
\label{fig:221}
\end{figure*}

\begin{figure}[htpb]
\begin{center}
\includegraphics[scale=.45]{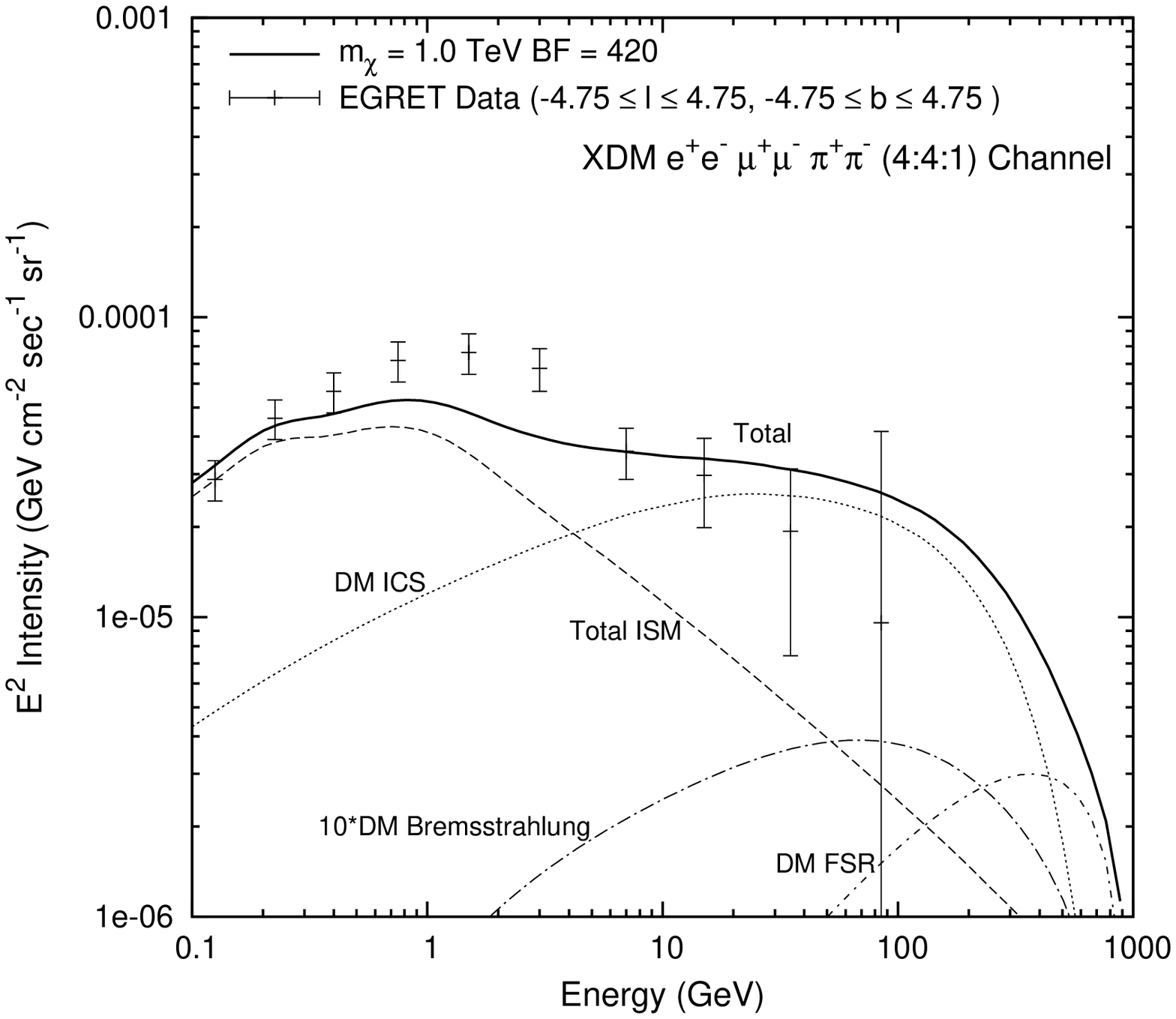}
\end{center}
\caption{The gamma ray signals from a model that annihilates into electrons, muons, and pions in a 4:4:1 ratio. The gamma ray spectrum is noticeably softer than that for annihilations into just electrons with the same mass.}
\label{fig:icswithsoft}
\end{figure}

While we have focused on the six simple modes, there are many alternatives. In particular, if $\phi$ is a vector, it can go to combinations of various modes \cite{ArkaniHamed:2008qn,Cholis:2008qq,Meade:2009iu}. This tends to have little effect on the electronic spectra of Fermi and ATIC \cite{Cholis:2008qq}, which are dominated by the hardest significant mode, and only moderate effect on PAMELA. Similarly, the Haze, which is dictated by the same energy range as PAMELA, is not extremely sensitive to such changes in the annihilation channel, but only the total power in it (this is in part due to the fact that the overall spectrum of the Haze is still not well known).  However, the presence of additional soft electrons and positrons can change the spectrum of ICS photons. As an example we show the cosmic ray signatures for a mode with an additional soft component in Figure \ref{fig:221}, where $\chi \chi \rightarrow \phi \phi$ where $\phi$ decays to $e^+e^-$, $\mu^+\mu^-$ and $\pi^+ \pi^-$ in a 1:1:2 ratio.  For ICS photons in a more intermediate situation, we show in Figure \ref{fig:icswithsoft} the gamma ray spectrum for the case of annihilations $\chi \chi \rightarrow \phi \phi$, where $\phi$ subsequently decays to $e^+e^-$, $\mu^+\mu^-$ and $\pi^+ \pi^-$ in a 4:4:1 ratio.

\begin{figure*}[htpb]
\begin{center}
\includegraphics[scale=.4]{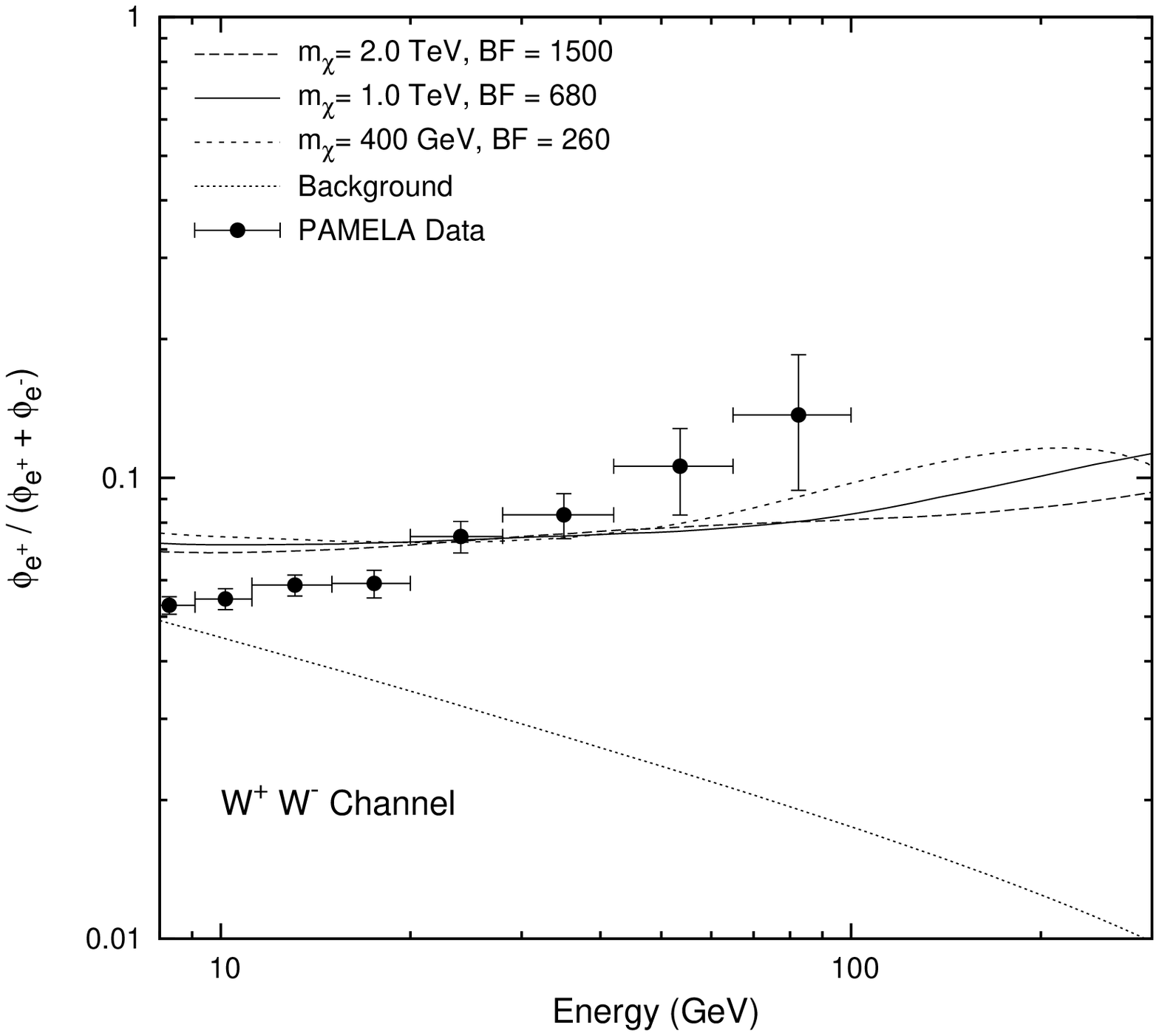}\hskip 0.2in
\includegraphics[scale=.4]{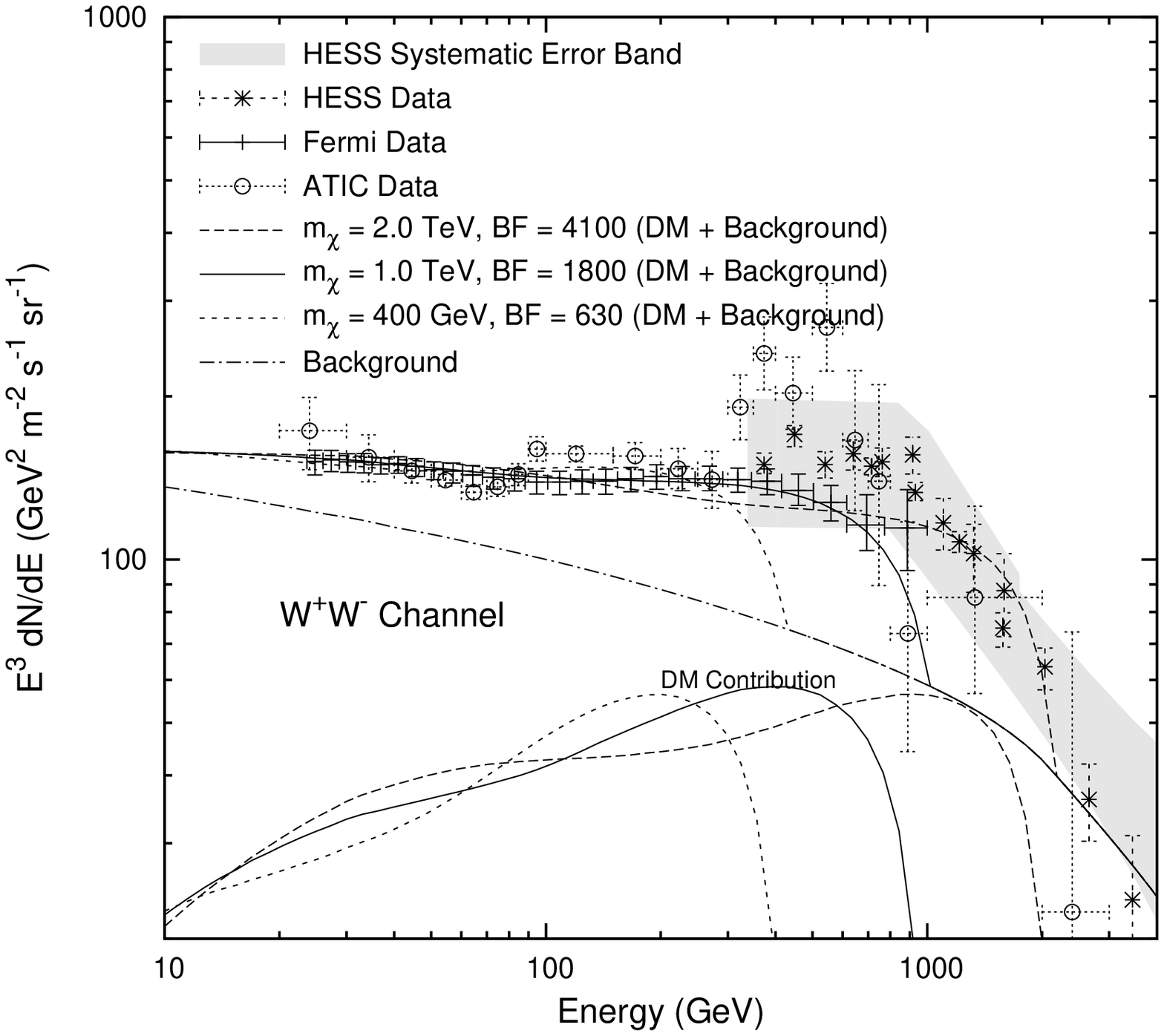}
\end{center}
\caption{PAMELA and high energy electron signals of dark matter annihilations $\chi \chi \rightarrow W^+W^-$.}
\label{fig:wcosmics}
\end{figure*}

Similarly, we consider how well an annihilation mode with hadronic
elements fits the data. We show in Figure \ref{fig:wcosmics} the electronic signals of $\chi \chi\rightarrow
W^+ W^-$. Such a mode is strongly excluded by anti-protons
\cite{Cirelli:2008id,Donato:2008jk}, but it is still instructive to see
how modes with significant soft components fit the data. An approximately 1 TeV WIMP gives the best fit to the high energy break observed by Fermi, ATIC and HESS (dominantly through its hard leptonic decays), but it does a poor job of reproducing the spectrum. Going to higher masses does not help, as in the PAMELA energy range, dark matter would significantly overproduce positrons and provide a poor fit, even with smaller boost factors  (note that the approximate fits to the PAMELA data have boosts only one third as large as those for Fermi). Extragalactic limits on diffuse gammas are stronger than this, as discussed by \cite{glast}, where boosts of $\sim 100$ were found to be borderline with EGRET within the context of an NFW profile.  With the strong $\bar p$ limits, any explanation of the high energy electron excess and PAMELA with W bosons would be expected to be due to significant local production.  Consequently, neither a Haze nor an ICS signal would be expected (as the similar boost would not be expected in the center of the galaxy).

\begin{figure*}[htpb]
\begin{center}
\includegraphics[scale=.45]{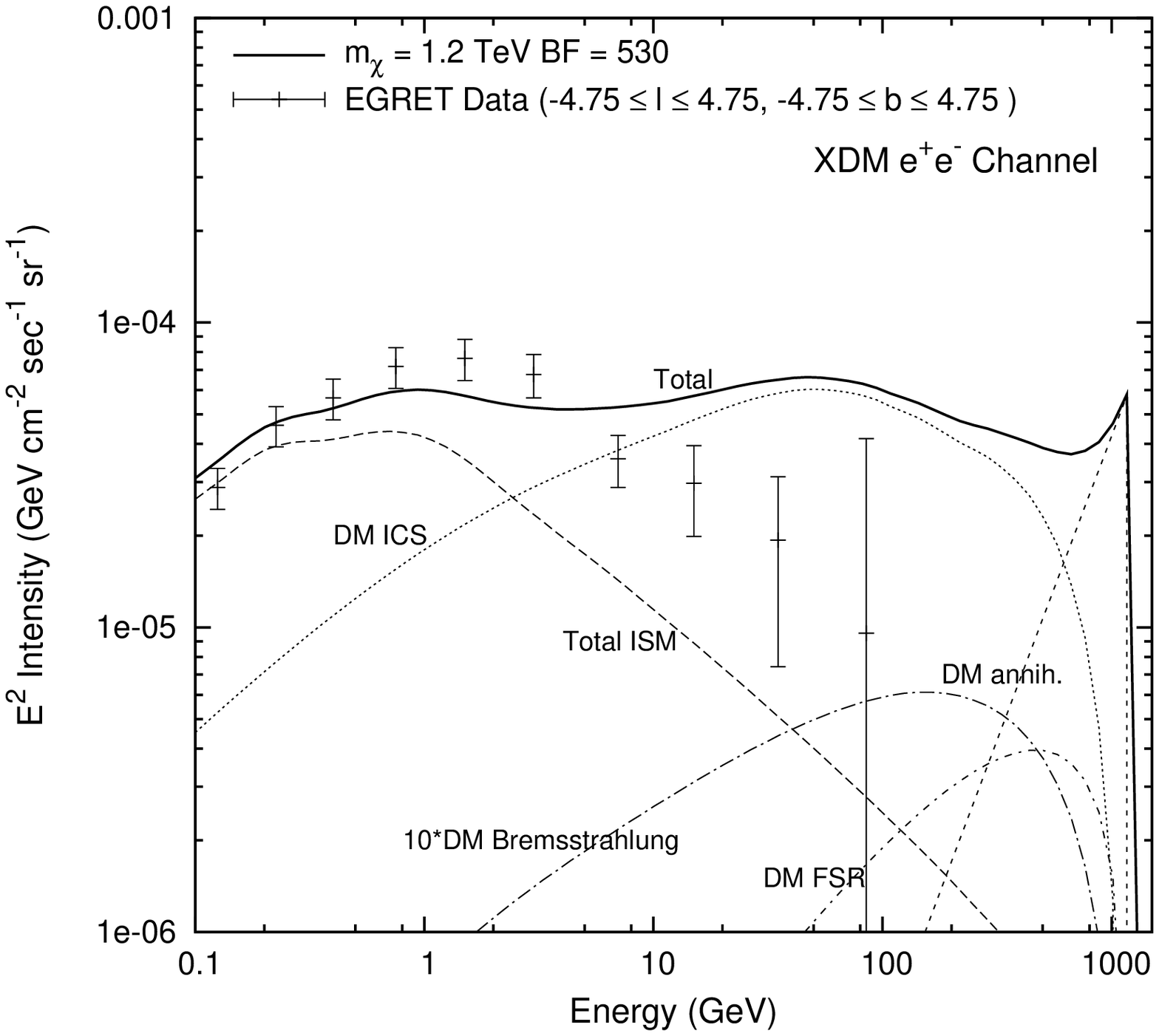}
\end{center}
\caption{Gamma ray signals including a $10\%$ BR of $\phi \rightarrow \gamma \gamma$.}
\label{fig:FERMIxdmelectronwgamma}
\end{figure*}

Finally, if $\phi$ is a scalar, then one can expect a branching ratio to $\gamma \gamma$ up to $\sim 10\%$ \cite{Cholis:2008vb}. We show how such a contribution to the Fermi/GLAST signal would appear in Figure \ref{fig:FERMIxdmelectronwgamma}. One can see that in the Fermi range, the gamma ray spectrum is swamped by ICS contributions. Only at the highest energies are the gammas from DM annihilation visible. Should the high energy ($\gsim 300 \gev$) electrons not arise from dark matter, and if the $\phi$ is sufficiently light ($\lsim 300$ MeV), this contribution may appear as a small peak at the end of the $E^2$ gamma distribution (as the spectrum of these photons is flat in energy).

\subsection{Astrophysical Uncertainties}
One of the most robust predictions of the dark matter scenarios
considered above is the ICS signal in the center of the galaxy. One
might naturally wonder how robust this is.

The greatest source of uncertainty is the extrapolation of the local
annihilation signal, as observed by PAMELA, Fermi, and ATIC, to the galactic
center. This extrapolation is uncertain for many reasons, not least
because of the assumption of a smooth dark matter distribution in the
galactic halo, in spite of the fact that simulations show clumpy
substructure.  Dark matter in subhalos, which might ordinarily have 
annihilation power comparable to that of the smooth halo even in
conventional models, can dominate by a large factor in the case of
Sommerfeld enhanced annihilation \cite{ArkaniHamed:2008qn}, because
the velocity dispersion of subhalos is small (see \cite{Lattanzi:2008qa,MarchRussell:2008tu,Robertson:2009bh,Essig:2009jx,Bovy:2009zs,Kuhlen:2009is,Yuan:2009bb,Kuhlen:2009jv} for further discussion).  In this case the
annihilation power in the galactic center has more to do with the mass
distribution and survival rate of clumps in the galactic center than the
properties of the smooth halo.  Additionally, dark matter can
form a dark disk, leading to high annihilation rates
\cite{Read:2008fh}.  On the other hand, the halo could be less cuspy
than we've assumed here, and substructure could be destroyed in the
center, lowering the annihilation signal relative to our
expectations.  Taking into account the uncertainties in both the dark
matter distribution and the particle physics, our simple extrapolation
from local annihilation rate to the galactic center is probably
uncertain by one or two orders of magnitude. 

However, if the central density is too low, we would be unable to
explain the Haze with dark matter annihilations, or the 511 keV signal from
INTEGRAL through the exciting dark matter mechanism.  But it happens
that the boosts needed to explain the Haze are very similar to those
needed for Fermi and PAMELA, and so we constrain most of these
uncertainties using the normalization of the Haze.  This still is not
perfect; there is always a factor of a few uncertainty in the
boosts associated with the Haze, simply arising from uncertainties in
the strength of the magnetic field in the galactic center, as well as its spatial distribution.

Putting aside this uncertainty, we examine the range of signals
possible for Fermi/GLAST. In doing this, it is important to distinguish
the different components contributing to the Fermi DM signal, namely,
ICS, FSR and $\pi^0$ gammas.  The $\pi^0$ gammas arise only from the
modes with $\tau$'s, but all the modes we consider have an FSR signal,
whether the annihilation goes directly into charged particles or via
intermediate $\phi$ states (see appendix).  Given a WIMP model, these
photons (FSR and $\pi^0$) are uncertain only at the level of the
overall dark matter annihilation in the galactic center, which is
naturally the appropriate size to produce the Haze.

\begin{figure}[htpb]
\begin{center}
\includegraphics[width=.45\textwidth]{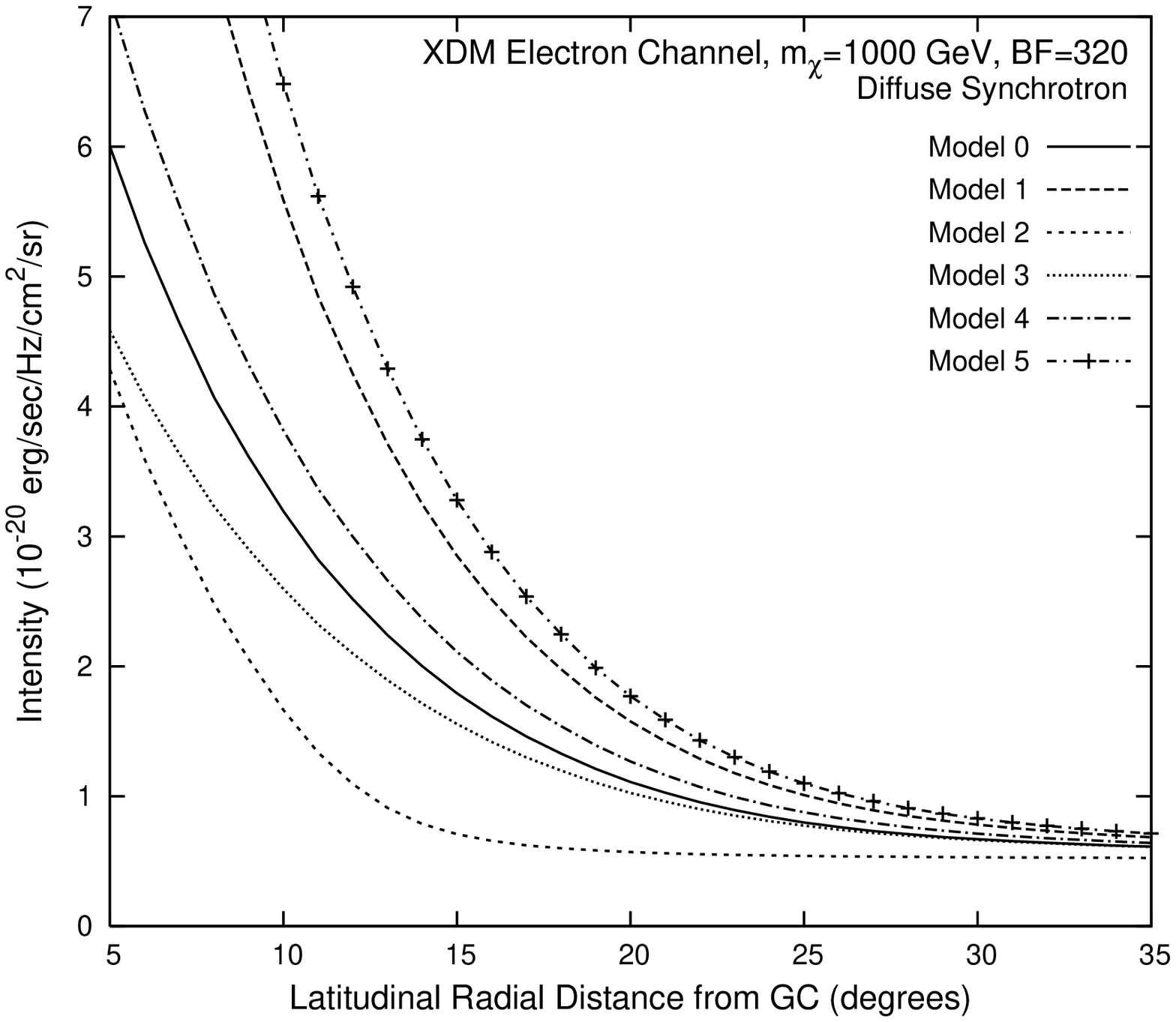}\hskip 0.2in
\includegraphics[width=.45\textwidth]{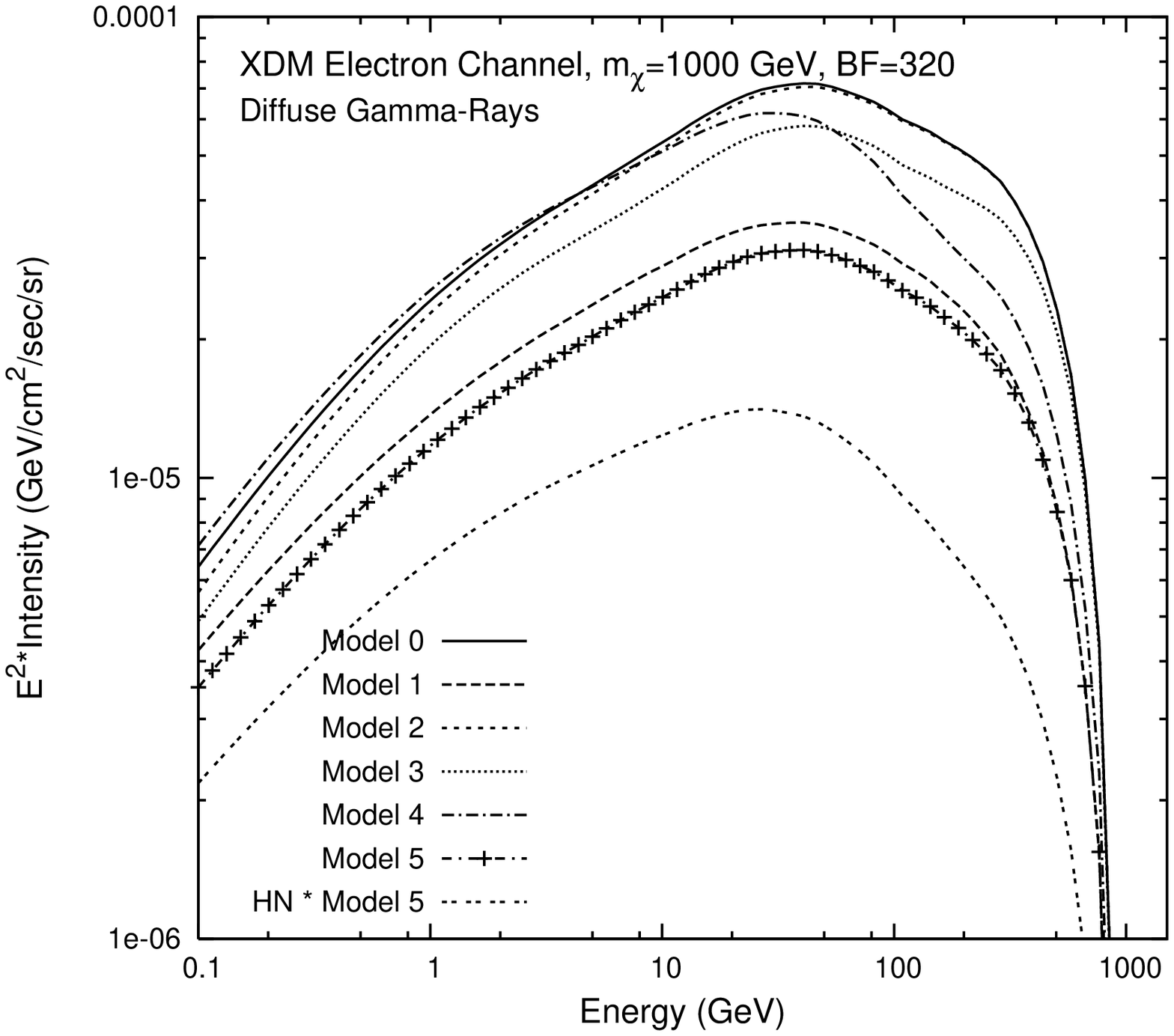}
\end{center}
\caption{Synchrotron (as a function of radial distance from the GC, \emph{left}) and ICS
(\emph{right}) towards the GC for five different diffusion models.  Model 0: the
``benchmark'' model given in \S\ref{sec:analysis}.  Model 1: $B_0$ doubled leading to more energy losses from synchrotron compared to ICS.
Model 2: Diffusion zone scale height decreased from 4 kpc to 2 kpc.  Model 3:
Diffusion index, $\alpha$, changed from 0.33 (Kolmogorov spectrum) to 0.5.
Model 4: Optical intensity in the interstellar radiation field (ISRF) 
reduced by half leading to less ICS of optical photons (decreased ICS at
energies above 10 GeV) and increased synchrotron.  Model 5: $B_0$
doubled \emph{and} optical intensity in the ISRF reduced by half.  Since Model 5
yields the largest synchrotron output, if we normalize the boost factor to fit
the Haze data with this model, the implied ICS (``HN * Model 5'') emission is
roughly a factor of 4 below the ``benchmark'' Model 0.
}
\label{fig:astrophys}
\end{figure}

The ICS photons are also produced by all the modes under
consideration, and are, in general, the most model-independent
prediction of these scenarios, arising directly from the interactions
of electrons and positrons with starlight. This, however, can vary
significantly, because the electrons share their energy between ICS
and synchrotron, depending on the relative energy densities of starlight
and the magnetic field. If the magnetic field in the galactic center
is larger than the $11.7\,\microG$ we assume ($=3 \eV\cm^{-3}$), then
the signal in WMAP is increased, at the expense of the ICS signal,
assuming constant power.  If the annihilation power is normalized to
the Haze, then the inferred ICS power drops further.  Additionally, 
a change in the diffusion constant can result in fewer electrons trapped in the galactic center region,
allowing more signal to propagate away from the inner $5\degree$. 
Finally, if the energy density of starlight is 
lower than in the ISRF maps employed in GALPROP \citep[for a discussion 
of the ISRF used in GALPROP and the associated uncertainties, see][]{porter08}, the ICS signal 
can be decreased by a corresponding factor of up to roughly two at higher energies. We
attempt to quantify these uncertainties in Figure
\ref{fig:astrophys} for the ICS signal from $\chi \chi \rightarrow
\phi \phi$, with $\phi \rightarrow e^+e^-$. We see that the signal can
be suppressed by a factor of 3-6, depending on energy.

The dominant gamma ray contribution for most modes is from ICS, as we show in Figure \ref{fig:EGRETsources}. Aside from this contribution and its uncertainty, we see that for the tau modes there is significant production of gamma ray radiation from $\pi^0$ gammas. Normalizing to the Haze, we expect a possibility of variation in this contribution at roughly a level of a factor of 2, that is, just in proportion to the overall changes in DM annihilation needed for the Haze. A similar level of uncertainty can be made for the FSR gammas, although they are less important. Thus, explaining the existing data, Haze+High Energy Electrons+PAMELA, with dominant tau modes seems in significant tension with existing EGRET data. This should be easily clarified by Fermi/GLAST, however.

\section{Conclusions}
The hard electron spectrum in the energy range $\sim 400 - 800$ GeV, observed previously by earlier experiments of ATIC and PPB-BETS, has now been confirmed by the recent run of ATIC and the even more recent Fermi results (though we should add that Fermi did not observe the pronounced peak in the spectrum). This gives a new piece of evidence for a new primary source of electronic production in the halo. Recent results by the PAMELA experiment point in a similar direction. Simple extrapolations of PAMELA seem to connect the experiments, calling out for consideration of a unified explanation.

In this paper, we have studied the possibility that dark matter annihilations could simultaneously explain the excesses seen by Fermi, ATIC, PPB-BETS, PAMELA and WMAP. We find remarkable agreement, with very similar boost factors to explain each one of them. The presence of the Haze in the center of the galaxy gives some weight to the dark matter interpretation of the signal, as astrophysical sources would not be expected to give significant positron fluxes in the galactic center, much less at latitudes of $5\degree$ to $10\degree$ (700 pc to 1400 pc) off the galactic disk. The Fermi/GLAST observations of ICS and other photons in the galactic center should give clear evidence for this scenario. Should these gamma rays extend to the reach of the Fermi/LAT experiment, the ubiquity and scale of excess electronic production would make dark matter the definitive explanation.

\vskip 0.25in

{\bf \noindent Acknowledgements}
\vskip 0.1in
\noindent We would like to acknowledge helpful discussions with Tracy Slatyer.  DPF and GD are partially supported by NASA LTSA grant NAG5-12972.
This work was partially supported by the Director, Office of Science,
of the U.S.  Department of Energy under Contract
No. DE-AC02-05CH11231.  NW is supported by NSF CAREER grant
PHY-0449818. LG, IC and NW are supported by DOE OJI grant
\#DE-FG02-06E R41417. NW acknowledges the hospitality and support of the Kavli Institute for Theoretical Physics China, CAS, Beijing 100190, China, where some of this work was undertaken.

\appendix
\section{Fits to ATIC}
In light of the current Fermi data, it is natural to focus on those data for normalizing our electron spectra at high energies. However, prior to Fermi, ATIC gave the first indication of a mass scale, and the presence of an excess at higher energies. In a previous version of the paper, prior to the Fermi data release, fits to ATIC were presented as well. While the updated Fermi data may cast significant doubt on the need for a sharp "peak" in the spectrum, it is still worthwhile to see how "peaky" a signal can be achieved with these modes. Thus, we show these fits here.

\begin{figure*}
\begin{center}
\includegraphics[width=.45\textwidth]{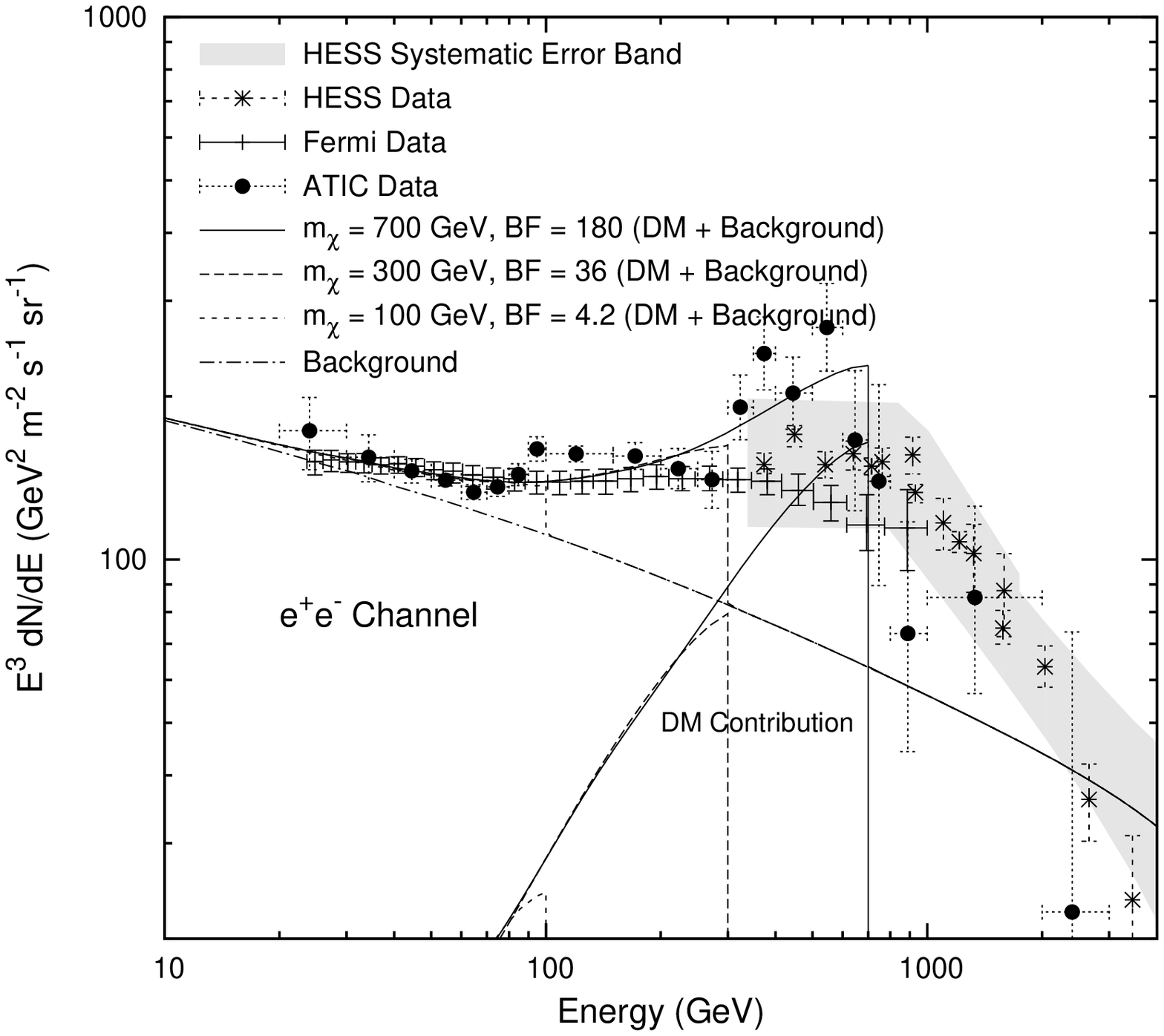}\hskip 0.2in
\includegraphics[width=.45\textwidth]{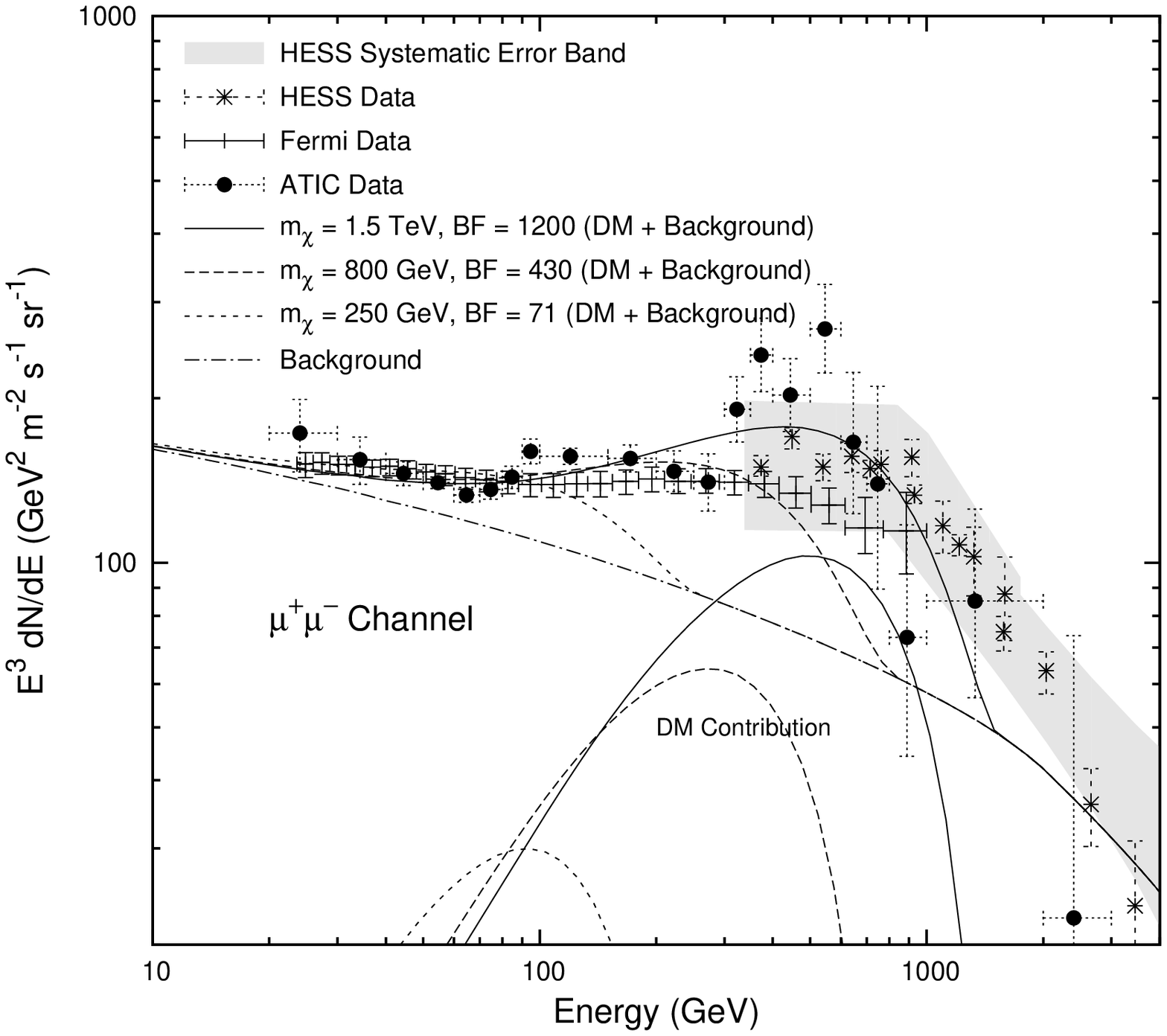}\\
\includegraphics[width=.45\textwidth]{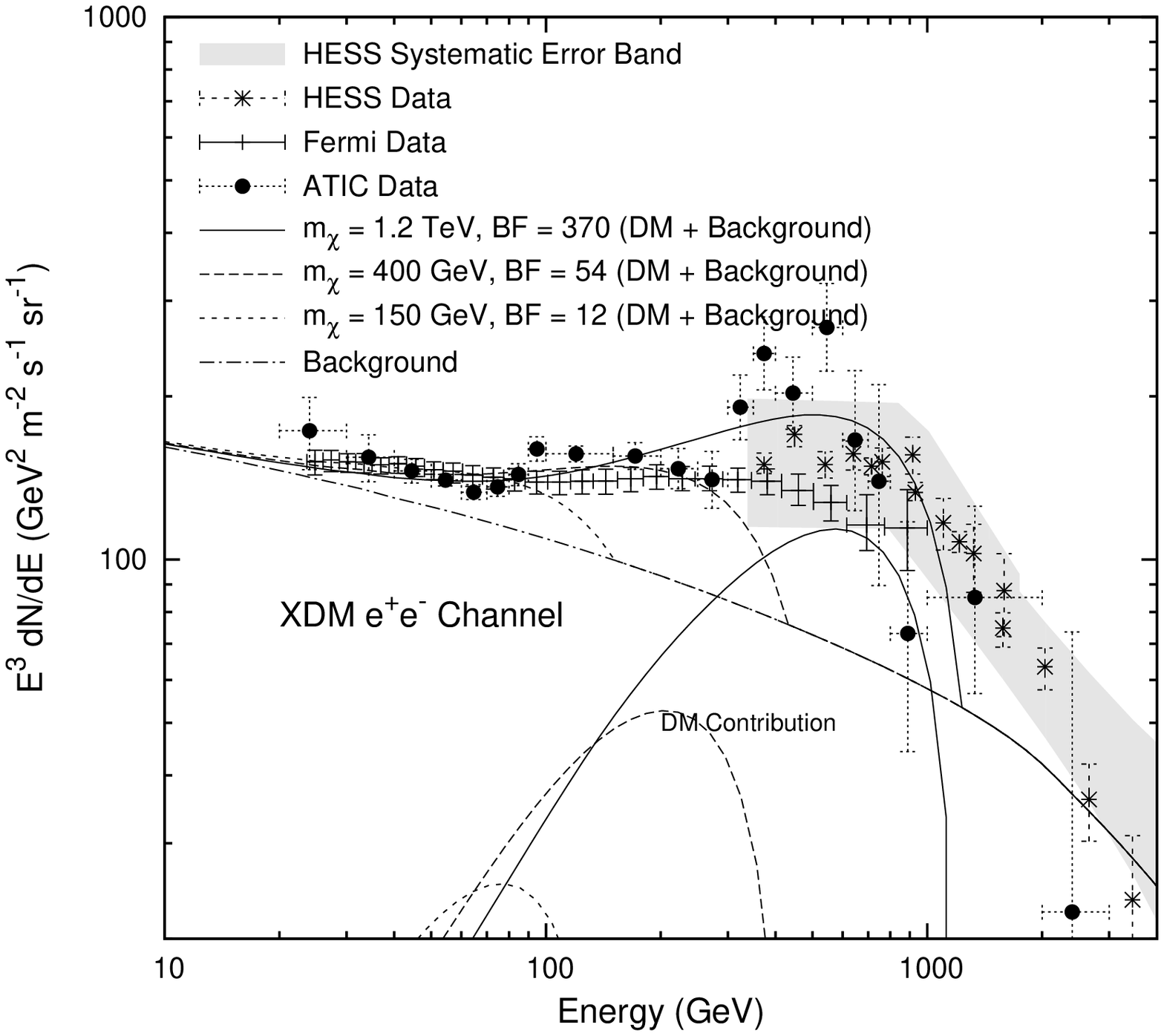}
\includegraphics[width=.45\textwidth]{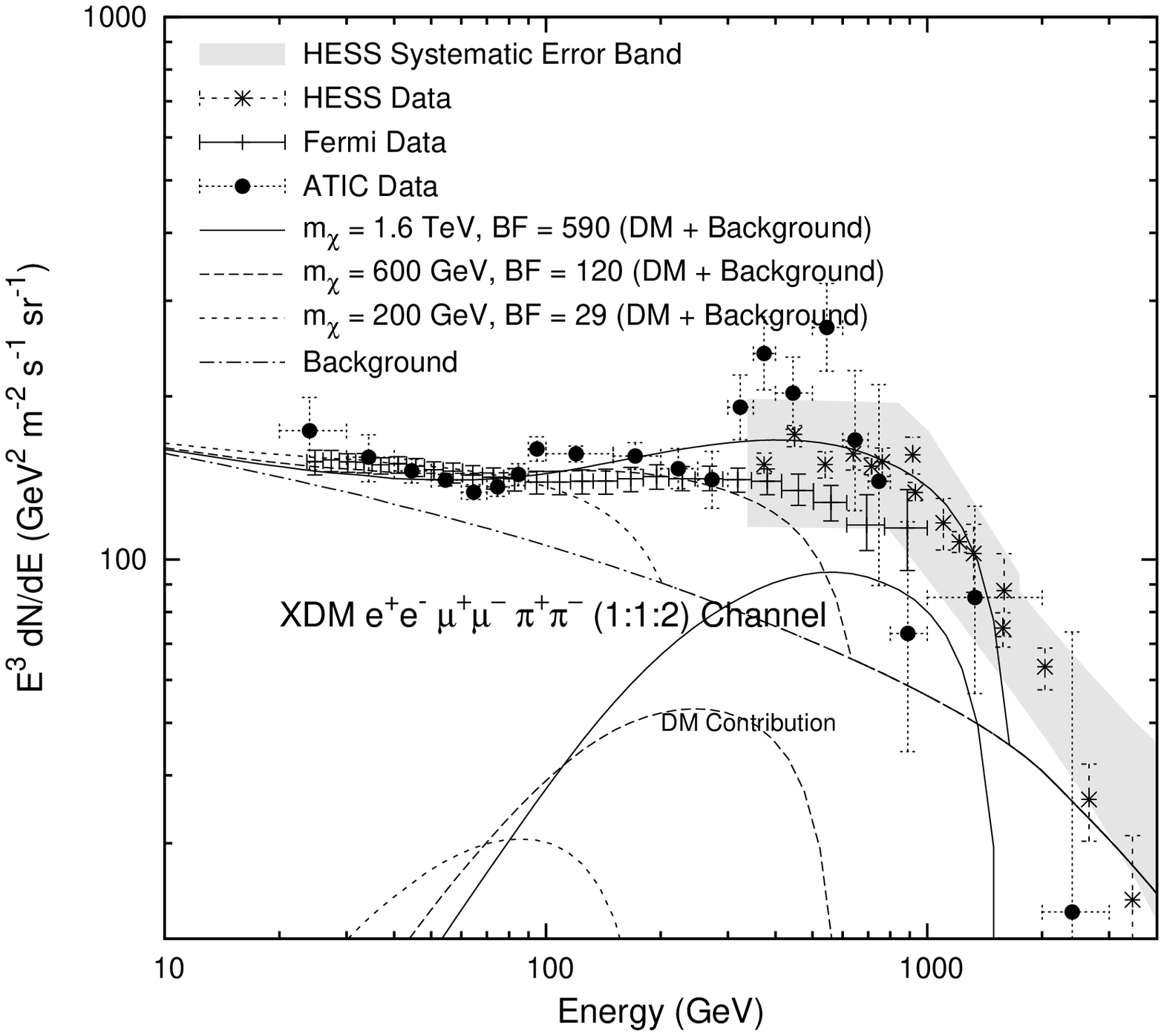}\hskip 0.2in
\end{center}
\caption{Electron spectra fits to ATIC from various annihilation modes.}
\label{fig:ATICfits}
\end{figure*}

\section{Final State Radiation}
Final state radiation is very important for annihilations into charged states \cite{Beacom:2004pe,Bergstrom:2004cy,Birkedal:2005ep,Mack:2008wu,Bell:2008vx}. An exhaustive study of FSR is well beyond our purposes here, but we include it in the spectra presented. The FSR for the dark matter direct annihilation to fermions is \cite{Bergstrom:2004cy,Birkedal:2005ep}
\be
\frac{d \sigma(\chi \chi \rightarrow f \bar f \gamma)}{d x} \approx \frac{\alpha Q_f^2}{\pi} {\cal F}(x) \log\left( \frac{s (1-x)}{m_f^2} \right) \sigma( \chi \chi\rightarrow f \bar f)
\label{eq:fsr1}
\ee
where $s = 4 m_\chi^2$, $x= 2 E_{\gamma}/\sqrt{s}$, and ${\cal F}(x) = ((1 - x)^2 + 1)/x$.

For the decays of $\phi \rightarrow f \bar f$, we need a slightly different formula.

Working from eq. \ref{eq:fsr1}, we can instead consider the case of $\phi$ decay, which has the same essential formula, but with $s= m_\phi^2$. If $\phi$ is boosted by an amount $\gamma = m_\chi/m_\phi$, and we assume the $\gamma$ is emitted at angle $\theta$ relative to the boost, we have the relationship
\be
y=\frac{E_\gamma}{m_\chi} = \frac{x}{2} (1 + \cos \theta)
\ee
Then we have a distribution of $\gamma$'s for every $\phi$
\be
\frac{dN}{dy} =\frac{1}{2} \int_{\cos \theta_{min}}^1 d \cos \theta \frac{dN}{dx}\frac{dx}{dy}
\ee
where $\cos \theta_{min} = 2y-1$ is the minimum angle in the $\phi$ frame for which boosted photons can have energy $y m_\chi$.
This gives
\be
\frac{dN}{dy} = \frac{\alpha Q_f^2}{2 \pi} \int_{\cos \theta_{min}}^1 d \cos \theta \frac{1}{y} (1+ (1-x)^2) (\log(m_\phi^2/m_f^2) + \log(1-x)) \frac{2}{1 + \cos \theta}
\ee
The first term, proportional to $\log(m_\phi^2/m_f^2)$, is dominant at small x, and is easily evaluated, while the second is somewhat more complicated. We write this as
\be
\frac{dN}{dy} = \frac{\alpha Q_f^2}{y \pi} (\log(m_\phi^2/m_f^2) P_1 +  P_2),
\ee
where
\be
P_1 = 2-y + 2 y \log y - y^2
\ee
and

\bea
\nonumber P_2 &=& 
\log (2)+\left(3-y-y^2\right) \log(1-y) -\left(1+\frac{\pi ^2}{3}-y\right) y\\&&+y \log (y) (\log
   (y)+2)+{\rm Re}\left[2 y \text{Li}_2\left(\frac{1}{y}\right)\right]
   \eea

A few points to consider: one must of course recall that this is the $\gamma$ distribution per $\phi$, and that there are two $\phi$'s per annihilation. We have taken the approximation that $m_\phi \gg m_f$. (See \cite{Mardon:2009rc} for a further discussion.) Also, we have used the approximate formula \ref{eq:fsr1}, which is only valid in the collinear limit. The hardest part of the spectrum we show would then be expected to be somewhat softer that what we have displayed. A complete analysis is warranted.

\onecolumngrid
\bibliography{atic}
\bibliographystyle{apsrev}

\end{document}